\documentclass[11pt,a4paper]{article} 
\pdfoutput=1
\usepackage{multirow,slashed,jheppub}
\usepackage[utf8]{inputenc}
\usepackage{color}
\usepackage{soul}
\usepackage{eufrak}
\pdfoutput=1

\newcommand{\be}{\begin{equation}}
\newcommand{\ee}{\end{equation}}
\newcommand{\bea}{\begin{eqnarray}}
\newcommand{\eea}{\end{eqnarray}}

\def\bsp#1\esp{\begin{split}#1\end{split}}

\title{Common exotic decays of top partners}

\author[1]{Nicolas Bizot}
\author[1]{\!\!, Giacomo Cacciapaglia}
\author[2]{\!\!, Thomas Flacke}

\affiliation[1]{Univ Lyon, Universit\'e Lyon 1, CNRS/IN2P3, IPNL, F-69622,
   Villeurbanne, France}
\affiliation[2]{Center for Theoretical Physics of the Universe, Institute for
   Basic Science (IBS), Daejeon 34051, Korea}

\emailAdd{bizot@ipnl.in2p3.fr}
\emailAdd{g.cacciapaglia@ipnl.in2p3.fr}
\emailAdd{flacke@ibs.re.kr}

\abstract{
Many standard model extensions that address the hierarchy problem contain Dirac-fermion partners of the top quark, which are typically expected around the TeV scale. Searches for these vector-like quarks mostly focus on their decay into electroweak gauge bosons and Higgs plus a standard model quark. In this article, backed by models of composite Higgs, we propose a set of simplified scenarios, with effective Lagrangians and benchmarks, that include more exotic decay channels, which modify the search strategies and affect the bounds. 
Analysing several classes of underlying models, we show that exotic decays are the norm and commonly appear with significant rates. All these models contain light new scalars that couple to top partners with charge $5/3$, $2/3$, and $-1/3$.
}

\begin{document}

\hspace*{80mm}{\large \tt CTPU-PTC-18-05, LYCEN 2018-03} \\

\maketitle
\flushbottom


\section{Introduction} 

Heavy vector-like quarks (VLQs in the following) are extensively searched for at the LHC due to the important role they play in many models beyond the standard model (BSM). The qualification ``vector-like'' refers to the fact that, contrary to fermions in the standard model (SM), both VLQ chiralities share the same quantum numbers under the SM gauge symmetries. Among the models that predict VLQs, models of composite Higgs have a special stand due to the crucial role the VLQs play for the top quark and Higgs physics. The Higgs is assumed to arise as a composite scalar of a confining and condensing underlying interaction, and its lightness compared to the condensation scale can be accounted for by the Higgs being a pseudo-Nambu-Goldstone boson (pNGB)~\cite{Kaplan:1983fs,Dugan:1984hq}. VLQs materialise as composite fermions, which generate masses for the top (and, eventually, lighter fermions) via the mechanism of partial compositeness (PC)~\cite{Kaplan:1991dc}, i.e. via linear mixing terms between the elementary and the composite fermions. It is also widely accepted that light VLQs (aka top partners) are needed in order to stabilise the loop-induced Higgs potential and keep the Higgs mass light (see, for instance, refs~\cite{Matsedonskyi:2012ym,Redi:2012ha}). We stress that this conclusion is based on the strong assumption of an enhanced calculability present in the effective theory below the condensation scale~\cite{Contino:2011np,Marzocca:2012zn}. This is, however, not always the case for strongly interacting and confining theories (as QCD teaches us). Furthermore, the stabilisation of the Higgs potential can also be achieved without top partners, for instance by tuning a mass term for the underlying fermions~\cite{Galloway:2010bp,Cacciapaglia:2014uja}. 
VLQs also play a useful role in other models, like supersymmetry~\cite{Ibrahim:2008gg,Liu:2009cc,Martin:2009bg}, and their phenomenology can be studied in effective models, independently of the theoretical framework they come from (see, for instance, refs~\cite{delAguila:2000rc,AguilarSaavedra:2009es,Cacciapaglia:2010vn,Okada:2012gy,Garberson:2013jz,Buchkremer:2013bha}).

In this article, we use the framework of partial compositeness and a composite pNGB Higgs as a guide for characterising the phenomenology of VLQs. This has already been the guiding principle behind the current experimental VLQ searches. However, the phenomenological expectations were strongly based on the most minimal model, where the Higgs boson is the only light pNGB in the theory~\cite{DeSimone:2012fs}. The two main assumptions, which have been used for most searches, are first, that the VLQ only decays to a standard boson ($W$, $Z$ and the Higgs $h$) plus a SM quark, and second, that the quarks belong to the third generation, i.e. only top or bottom quarks. 
We will show that, in models that enjoy a simple underlying description in terms of a confining gauge symmetry, the first assumption is not well justified. In fact, generically new decay channels are present that often dominate over the standard ones. The main underlying reason is that the most minimal symmetry breaking pattern $SO(5)\rightarrow SO(4)$ is not realised in any known simple underlying model, and thus additional light pNGBs are present in the spectrum. This is true both in models where only fermions are present, as described in refs~\cite{Barnard:2013zea,Ferretti:2013kya,Ferretti:2014qta,Vecchi:2015fma}, and in models with fermion--scalar bound states~\cite{Sannino:2016sfx,Cacciapaglia:2017cdi}.

Analysing the classes of models in the literature, we identify 4 types of situations that can strongly affect top partner decays, as summarised below:

\begin{enumerate}
	
	\item {\bf Singlet pseudo-scalar, $T \to t\ a$ and $B \to b\ a$.} The presence of a light CP-odd pNGB associated to a non-anomalous $U(1)$ global symmetry is ubiquitous to models of PC with a gauge-fermion underlying description~\cite{Cai:2015bss,Belyaev:2015hgo,DeGrand:2016pgq,Ferretti:2016upr,Belyaev:2016ftv}. We show that the light pseudo-scalar $a$ always couples to the top partners. Thus, a charge $2/3$ VLQ $T$ and a charge $-1/3$ VLQ $B$ also decay to $a$ and a SM quark, as long as the pNGB $a$ is lighter than the VLQ. While the presence of $a$ adds VLQ decay channels, the pair and single production rates of the VLQs are barely modified.
	
	\item {\bf Exclusive pNGB, $\widetilde{T} \to t\ \eta$.} The extended pNGB cosets may also contain additional scalars that couple exclusively with one specific top partner, $\widetilde{T}$. This is the case for a CP-odd singlet $\eta$ present in the $SU(4)/Sp(4) \simeq SO(6)/SO(5)$ coset~\cite{Gripaios:2009pe,Serra:2015xfa,Chala:2017sjk}. The charge $2/3$ top partner $\widetilde{T}$, which is part of a $\bf 5$ of $Sp(4)\simeq SO(5)$,  does not decay to two SM particles but exclusively into $t\ \eta$, and it cannot be singly- but  only pair-produced at colliders.
	
	\item {\bf Coloured pNGB, $X_{5/3} \to \bar{b}\ \pi_6$.} The presence of coloured fermions or scalars in the underlying theory yields potentially light coloured pNGBs. Their couplings to the VLQs imply additional decay channels beyond the standard ones. As an example, we consider a pNGB transforming as a sextet of QCD colour and with charge $4/3$. This state is present in some underlying models~\cite{Cacciapaglia:2015eqa}, and it can couple to the exotic charge $5/3$ top partner $X_{5/3}$. Note that coloured pNGBs can also modify the production rates of the VLQs, especially if heavier than them~\cite{Deandrea:2017rqp}.
	
	\item {\bf Charged pNGB, $X_{5/3} \to t\ \phi^+$.} Some cosets, like $SU(5)/SO(5)$~\cite{Vecchi:2015fma}, also contain additional charged pNGBs which contribute to the decays of the top partners. These decay channels are usually present in addition to the standard ones. 
	
\end{enumerate}

A fifth possibility is that some top partners can decay into a stable (or long-lived) pNGB, which may be identified with a Dark Matter candidate: typically, this leads to exclusive decay modes, as shown in refs~\cite{Anandakrishnan:2015yfa,Balkin:2017aep,Chala:2018qdf}. Such decay modes are efficiently covered by searches focused on supersymmetric final states~\cite{Kraml:2016eti}. Thus, we do not consider this possibility here. 

In Section~\ref{sec:simplified}, we introduce simplified model descriptions and benchmark points for the scenarios listed above. We discuss how the standard searches for VLQs are affected by the new decay modes and which new experimentally promising signatures arise. Several additional decay modes for VLQs have already been considered in the literature, both for composite models~\cite{Serra:2015xfa,Chala:2017xgc} and supersymmetric models~\cite{Aguilar-Saavedra:2017giu} (see also ref.~\cite{Brooijmans:2016vro} for a more general table of allowed final states). Our approach differs, as we identify testable predictions which arise from models with a simple underlying description, where the new modes are predicted and not added by hand.  To better substantiate this, in Section~\ref{sec:models}, we present underlying models and model-parameters that predict the field content of the simplified models of Section~\ref{sec:simplified} as part of their (light) particle spectrum, and that yield the effective couplings used as benchmark points.
Finally, we present our conclusions in Section~\ref{sec:concl}.


\section{Simplified scenarios}
\label{sec:simplified}

\subsection{Singlet pseudo-scalar, $T \to t\ a$ and $B \to b\ a$} 
\label{sec:sm1}

As a first simplified scenario, we consider a model with a charge $2/3$ top partner $T$ and a lighter pseudo-scalar $a$.  
Such a light pseudo-scalar $a$ is genuinely present in  models of PC with a gauge-fermion underlying description~\cite{Cai:2015bss,Belyaev:2015hgo,DeGrand:2016pgq,Ferretti:2016upr,Belyaev:2016ftv}, where it can be associated with the pNGB of a global $U(1)$ symmetry.
We parameterise the interactions of a VLQ with SM particles and the pseudo-scalar $a$ as\footnote{We follow the parametrisation of  ref.~\cite{Buchkremer:2013bha} for the couplings to SM particles.}
\bea
\mathcal{L}_{T} &=& \phantom{+}  \overline{T}\left(i\slashed{D}-M_T\right) T + \left(\kappa^T_{W,L} \frac{g}{\sqrt{2}}  \,\overline{T}\slashed{W}^+P_Lb
+ \kappa^T_{Z,L} \frac{g}{2 c_W}\, \overline{T} \slashed{Z} P_L t \right. \nonumber \\ 
&&
\left. - \kappa^T_{h,L} \frac{M_T}{v}\, \overline{T} h P_L t  + i \kappa^T_{a,L}\, \overline{T} a P_L t + L\leftrightarrow R + \mbox{ h.c. }\right), \label{eq:LTa}
\eea
where $P_{L,R}$ are left- and right-handed projectors, and $T$ denotes the top partner mass eigenstate with mass $M_T$.  The first three interaction terms dictate the partial widths of $T$ decays into $bW$, $tZ$, and $th$ as often considered in VLQ models. In the above parametrisation, the coefficients $\kappa^T_{W/Z/h,L/R}$  are determined by the $SU(2)$ charge and the mixing angles of the top partner with the elementary top.  
If only decays into SM particles are considered, the current bound is of order $M_T\gtrsim 1$ TeV \cite{Sirunyan:2017pks,CMS:2016dmr,Aaboud:2017zfn,Aaboud:2017qpr,TheATLAScollaboration:2016gxs,Aaboud:2016lwz,ATLAS:2016btu}.\footnote{Bounds on $M_T$ from QCD produced $T$-pairs depend on the $T$ branching ratios (BRs) into $bW,tZ, th$. The strongest reported bound is for  100\% BR $T\rightarrow bW$ ($M_T\gtrsim 1.3$~TeV) \cite{Sirunyan:2017pks,Aaboud:2017zfn}, while bounds on 100\% BR  $T\rightarrow tZ$ or $T\rightarrow th$ are around 1 TeV. Bounds on $M_T$ from electroweak single-production \cite{Sirunyan:2017ynj,Sirunyan:2017tfc,Sirunyan:2017ezy,Sirunyan:2016ipo,Khachatryan:2016vph,ATLAS:2016ovj,Aad:2016qpo} are even more model-dependent as the production cross section depends on additional BSM couplings.}
The last term in Eq.~(\ref{eq:LTa}) parameterises the coupling of $T$ to the pseudo-scalar $a$. This term does not significantly affect the top-partner production, which occurs through QCD pair production, or through single-production dictated by the first three terms ({\it Cf. e.g.} refs~\cite{DeSimone:2012fs,Buchkremer:2013bha} for top partner single- and pair production rates). If $M_T > m_a + m_t$, the last term in Eq.~(\ref{eq:LTa}) adds an additional decay channel of $T\rightarrow t\ a$.  
Explicit expressions for the tree-level decay widths can be found in ref.~\cite{Atre:2011ae}.

In analogy, as a second simplified model, we introduce the VLQ $B$ with charge $-1/3$, with the simplified Lagrangian 
\bea
\mathcal{L}_{B} &=& \phantom{+}  \overline{B}\left(i\slashed{D}-M_B\right) B + \left(\kappa^B_{W,L} \frac{g}{\sqrt{2}} \, \overline{B} \slashed{W}^-P_L t   + \kappa^B_{Z,L} \frac{g}{2 c_W} \, \overline{B}\slashed{Z}^+P_L b  \right. \label{eq:LBa} \\ &&
\left. - \kappa^B_{h,L} \frac{M_B}{v}\, \overline{B} h P_L b   + i \kappa^B_{a,L} \, \overline{B} a P_L b  
+L\leftrightarrow R + \mbox{ h.c. } \right) .
\nonumber
\eea

\begin{figure}[tbh]
	\centering
	\begin{tabular}{cc}
		\includegraphics[width=0.48\textwidth]{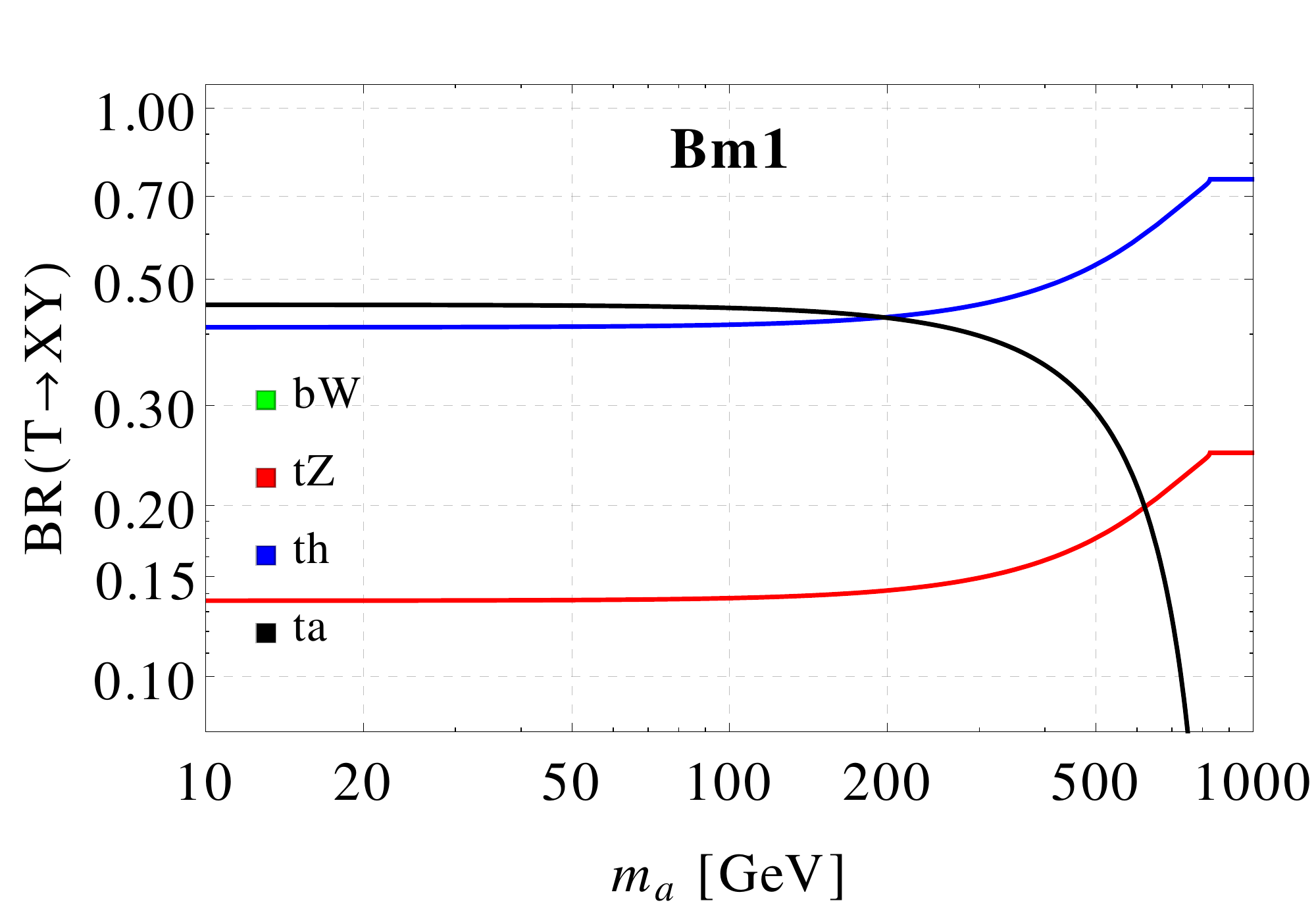} & \includegraphics[width=0.48\textwidth]{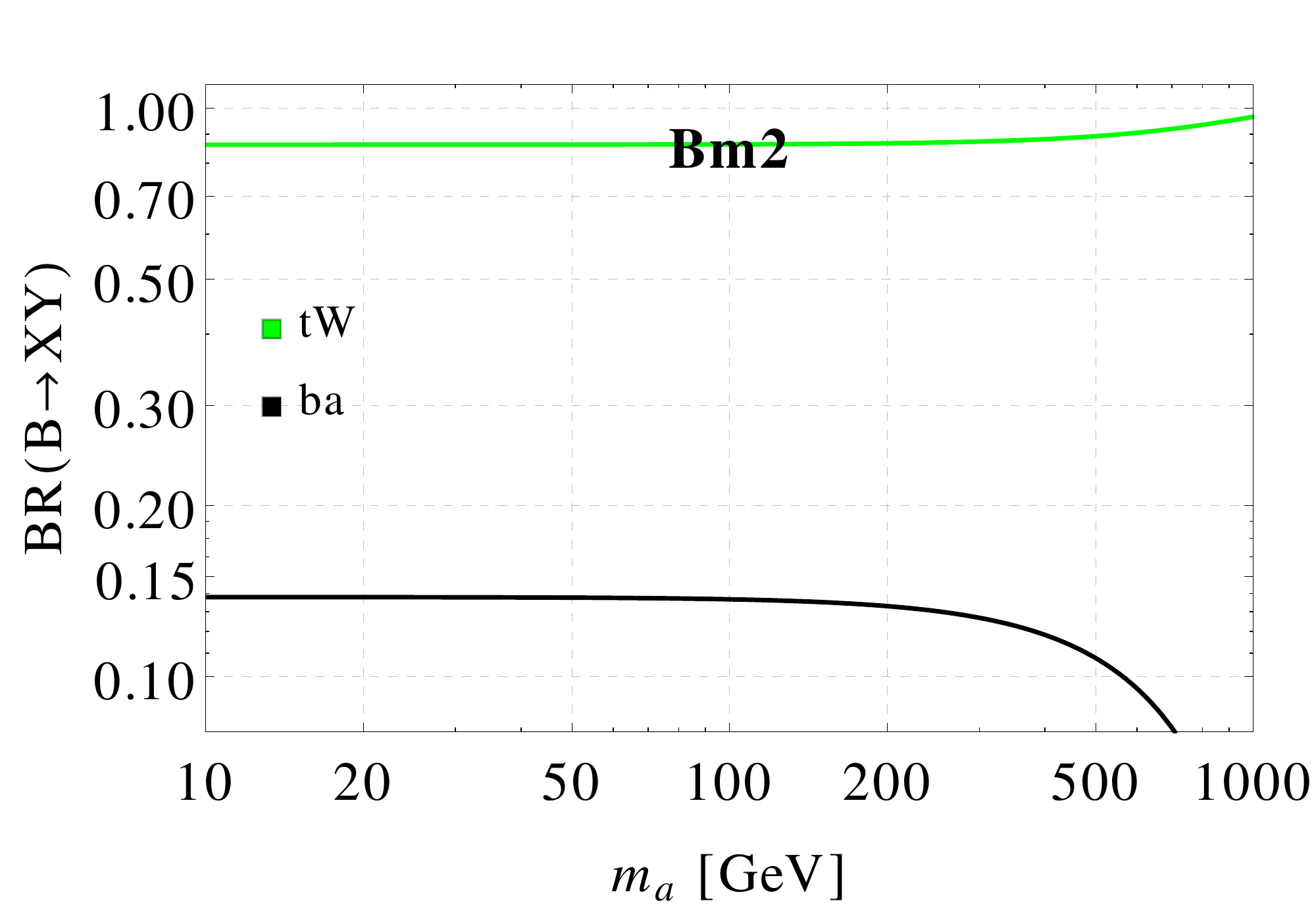}
	\end{tabular}
	\caption{BRs of $T$ (left) and $B$ (right) as a function of mass $m_a$ respectively in two different benchmark models Bm1 and Bm2 introduced in Sec \ref{sec:models}. For $T$, a decay into $bW$ is allowed, but suppressed in the chosen benchmark point Bm1.
	}
	\label{fig:TaBRs}
\end{figure}

To illustrate the relevance of the new decay channels, we consider two benchmark models, ``Bm1'' and ``Bm2'', arising from an underlying UV embedding of composite Higgs models with  $SU(4)/Sp(4)$  breaking, which are discussed more in detail in Sec.~\ref{sec:models}.
In Fig.~\ref{fig:TaBRs} we show the BRs in the two benchmarks as a function of the $a$ mass. 
Each scenario focuses on one VLQ, either $T$ or $B$.
The two benchmark models are respectively characterised by the following couplings:
\bea
&&
\mbox{Bm1}: ~~ M_T = 1~\rm{TeV}~,  
\quad
\kappa^T_{Z,R} =-0.03~,
~~ \kappa^T_{h,R} =0.06 ~,  
~~ \kappa^T_{a,R} =-0.24 ~, 
~~\kappa^T_{a,L} =-0.07 ~; 
\nonumber
\\
&&
\mbox{Bm2}: ~~M_B = 1.38~\rm{TeV}~,
~~
\kappa^{B}_{W,L} = 0.02~, 
~~\kappa^{B}_{W,R} =-0.08~,  
~~\kappa^{B}_{a,L} =-0.25~,  
\eea
while the ones that are not reported are suppressed and thus negligible.  
The BRs of $T\rightarrow t\ a$ and $B\rightarrow b\ a$ are model dependent. However, the benchmarks we present in Fig.~\ref{fig:TaBRs}, which are fairly generic and not tuned to maximise the new channels, clearly show that, in fully realistic models, they can be sizeable and even comparable to the BRs into SM particles, which are considered in standard searches at the LHC. 

\begin{figure}[tbh]
	\centering
	\begin{tabular}{cc}
		\includegraphics[width=0.48\textwidth]{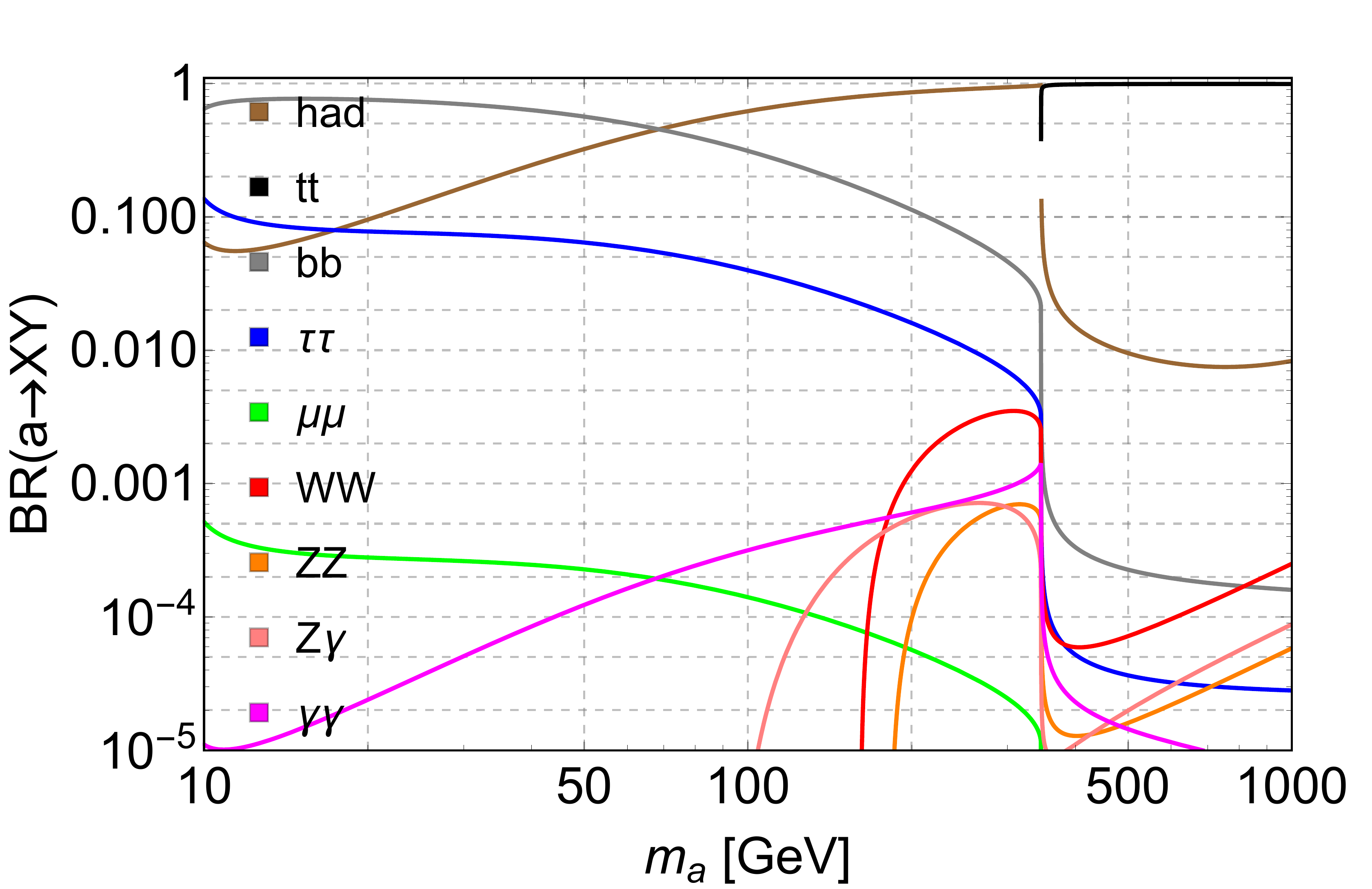} & \includegraphics[width=0.48\textwidth]{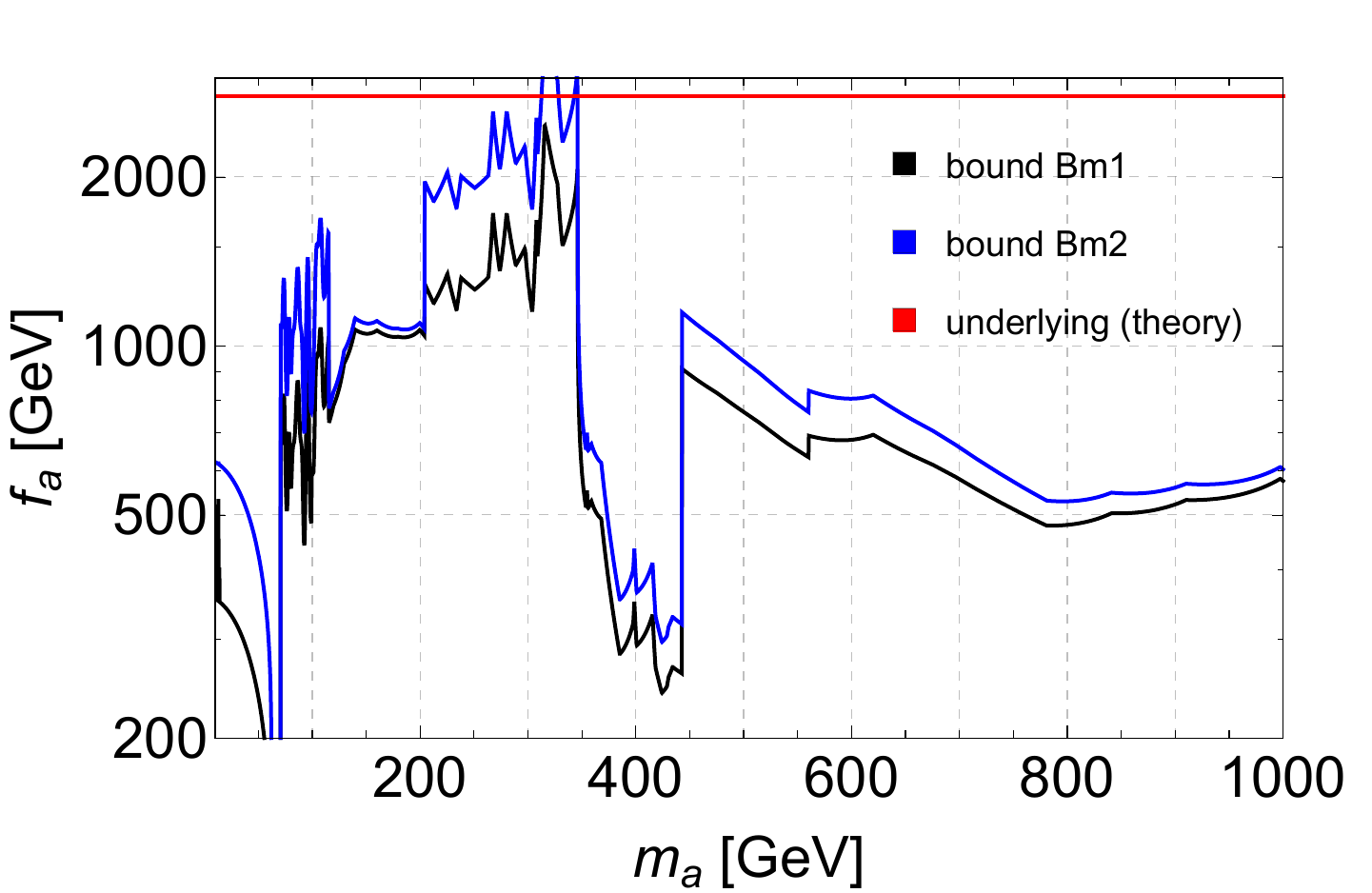}
	\end{tabular}
	\caption{Left: Branching ratios of $a$ as a function of the mass $m_a$ in the benchmark model Bm1 characterised by the parameters in Eq.~(\ref{eq:apara}). Right: Experimental lower bounds on $f_a$ from ATLAS and CMS resonance searches on the benchmark models Bm1 (blue) and Bm2 (black) as compared to the underlying theory value (red). 
	}
	\label{fig:aBRs}
\end{figure}

To determine new possible final states that can occur from the $T\rightarrow t\ a$ (or $B \to b\ a$) decay, we briefly review the properties of, and constraints on, the pseudo-scalar $a$.  
The interactions of the pseudo-scalar $a$ with SM particles can be parameterised as \footnote{We give the effective Lagrangian up to dimension 5 operators. Additional interactions can be generated at higher order. See ref.~\cite{Cacciapaglia:2017iws} for couplings $haa$ and $hZa$.}
\begin{multline}
{\mathcal{L}}_a = \frac{1}{2}(\partial_\mu a)(\partial^\mu a) -\frac{1}{2} m_a^2 a^2 - \sum_f \frac{i C_f^a m_f}{f_a} a \bar f \gamma^5 f +\!\frac{g_s^2 K_g^a }{16\pi^2 f_a}a G^a_{\mu\nu}\tilde G^{a\mu\nu}\!+\! \frac{g^2 K_W^a }{8\pi^2 f_a} a W^+_{\mu\nu}\tilde W^{-,\mu\nu}
\label{eq:aLag}
\\
\!+\!\frac{e^2 K_\gamma^a}{16\pi^2 f_a} a A_{\mu\nu}\tilde A^{\mu\nu}  \!+\!\frac{g^2 c_W^2 K_Z^a}{16\pi^2 f_a} a Z_{\mu\nu}\tilde Z^{\mu\nu} \!+\!\frac{e g c_W K_{Z\gamma}^a}{8\pi^2 f_a} a A_{\mu\nu}\tilde Z^{\mu\nu}  \, .
\end{multline}
Note that we have written the couplings to the gauge bosons in the mass eigenstate basis because the mass of $a$ can well be around or below the electroweak (EW) scale, and we define $V_{\mu \nu} = \partial_\mu V_\nu - \partial_\nu V_\mu$ and $\tilde{V}_{\mu \nu} = \epsilon_{\mu \nu \rho \sigma} V^{\rho \sigma}$. The  couplings are, in general, independent, but relations among them may exist depending on the origin of the pseudo-scalar $a$.
In underlying models where $a$ is associated to a $U(1)$ symmetry~\cite{Belyaev:2016ftv, Cacciapaglia:2017iws}, the couplings to gauge bosons can be determined in terms of two parameters: one coupling, $K_W^a$, to the SU(2) bosons and one, $K_B^a$, to hypercharge. Thus, we have the relations
\bea \label{eq:kappas}
K_\gamma^a = K_W^a + K_B^a\,, \quad K_Z^a = K_W^a + K_B^a t_W^4\,, \quad K_{Z\gamma}^a = K_W^a - K_B^a t_W^2\,,
\eea
where $t_W=\tan(\theta_W)$ is the tangent of the Weinberg angle. The parameters $K_{W,B}^a$ are fully determined in terms of the underlying theory.
On the other hand, the couplings to the fermions depend on the origin of the mass terms, i.e. the choice of top partners and the values of the mixing couplings (more details will be provided in Sec.~\ref{sec:models}).
For a given model and choice of the fermion couplings, the only other remaining parameters are the decay constant $f_a$, which controls the overall width and  coupling strength, and the mass $m_a$, on which the BRs depend. 
The latter can be potentially very small, of the order of few GeV~\cite{Belyaev:2016ftv}. 
As an example, the couplings resulting from the $SU(4)/Sp(4)$ model discussed in Sec.~\ref{sec:SU4} are
\be
K_g^a =-1.6  ~,  \quad
K_W^a = 1.9~,  \quad
K_B^a = -2.3~,  \qquad 
C_{f\neq t}^a = 1.9~,\quad 
C_t^a = \left\{
\begin{array}{l}
	1.46 ~\rm{for} ~\rm{Bm1}
	\\
	2.33 ~\rm{for} ~\rm{Bm2}
\end{array}
\right.~,  
\\
\label{eq:apara}
\ee
where from the UV model, we also fix $f_a = 2.8$ TeV, and only leave the mass $m_a$ as a free parameter.  
They correspond to the underlying model M8, discussed in ref.~\cite{Belyaev:2016ftv}.
The BRs of $a$ for this parameter choice (for Bm1) are shown in Fig.~\ref{fig:aBRs} (left) as a function of $m_a$. The BRs do not depend on the decay constant $f_a$.

The pseudo-scalar $a$ can be directly produced at the LHC in gluon fusion.\footnote{Top-associated production ($t\bar{t}a$) is also possible -- in particular for light $a$ \cite{Casolino:2015cza}. Furthermore, light $a$ can result from $h\rightarrow a a $ or $h\rightarrow Z a$ decays \cite{Bauer:2017ris}. For the benchmark models considered in Sec.~\ref{sec:models}, these processes only yield weak bounds, however, as shown in ref.  \cite{Cacciapaglia:2017iws}.} 
As the BRs of $a$ are determined as given in Fig.~\ref{fig:aBRs}, bounds from ATLAS and CMS resonance searches in the  channels $jj$ \cite{Sirunyan:2016iap,CMS:2017xrr,Aaboud:2017yvp}, $t\bar{t}$  \cite{Sirunyan:2017uhk,Khachatryan:2015sma,Aad:2015fna}, $b\bar{b}$ \cite{ATLAS:2016fol}, $\tau^+\tau^-$ \cite{Khachatryan:2016qkc,CMS:2017epy,Aaboud:2016cre,Aaboud:2017sjh}, $\mu^+\mu^-$ \cite{Chatrchyan:2012am}, $W^+W^-$ \cite{Sirunyan:2017acf,Aaboud:2017eta,Sirunyan:2016cao,CMS:2017mjm,CMS:2016pfl,Aaboud:2017gsl,Aaboud:2017fgj}, $ZZ$ \cite{Sirunyan:2017acf,Aaboud:2017eta,CMS:2017xyz,CMS:2017sbi,Sirunyan:2017jtu,CMS:2017pfj,Aaboud:2017rel,Aaboud:2017itg}, $Z\gamma$ \cite{Sirunyan:2017hsb,Khachatryan:2016odk,Aaboud:2017uhw} and $\gamma\gamma$ \cite{Khachatryan:2016yec,Khachatryan:2016hje,CMS:2017yta,Aaboud:2017yyg} can be translated into bounds on the decay constant $f_a$, which controls the direct production cross section \cite{Belyaev:2016ftv, Cacciapaglia:2017iws}.  
The right panel in Fig.~\ref{fig:aBRs} shows the resulting bounds on $f_a$ for the  benchmarks Bm1 and Bm2 characterised by the couplings given in Eq.~(\ref{eq:apara}).\footnote{The minor difference in bounds between Bm1 and Bm2 arises mainly from $C_t$-dependent one-loop contributions to $a$ production in gluon fusion, and to a lesser extent from $C_t$-dependent one-loop corrections to the decay partial width of $a$ into $gg$, $Z\gamma$, and $\gamma\gamma$. For expressions of the BRs at one-loop, see ref.~\cite{Belyaev:2016ftv}.}   The red line corresponds to the estimated value of $f_a$ from the underlying model (see Sec.\ref{sec:SU4}). 
As it can be seen, this value is not excluded by current searches for almost all masses of $m_a$ (see refs~\cite{Belyaev:2016ftv, Cacciapaglia:2017iws} for bounds on $f_a$ in other models).

Indirect production of $a$, via  the decay $T\rightarrow t\ a$ or $B\rightarrow b\ a$, is therefore an independent test of the models. A dedicated collider study of the signatures is beyond the scope of this article, so we briefly comment on the new allowed final states in a way to relate them to the standard top partner searches. 
While the BRs shown in Fig.~\ref{fig:aBRs} apply to a specific model, the plot shows some typical behaviours which are generic for most models. We can thus infer a general trend for the dominant final states, depending on the mass range of the $a$:
\begin{itemize}
	\item For $m_a > 2 m_t$, the channel $T\rightarrow t\ a \rightarrow t t \bar{t}$ has the dominant BR. This is a generic expectation as long as a sizeable coupling to the tops is present. The resulting ``tri-top'' and ``esa-top'' final states are not searched for, but they will be efficiently covered by existing 4-top searches, as shown in ref.~\cite{Deandrea:2014raa}.  
	\item Below the $t\bar{t}$ threshold, the dominant $a$ decays are into hadrons ($gg$) or $b\bar{b}$. Thus, $T\rightarrow t\ a$ decays provide similar final states as standard channels like $T\rightarrow t\ V_{\rm had}$ and $T\rightarrow t\ h \rightarrow t b \bar{b}$.
	However the mass reconstructions applied in current fully hadronic decay searches typically focus on invariant mass reconstructions in the $W$/$Z$ and Higgs mass ranges. Thus the signal generated by $T \rightarrow t a$ is potentially being rejected unless its mass is close to the one of the standard bosons. Note that a simple recast was possible for some Run-I searches where the Higgs mass reconstruction was not imposed~\cite{Chala:2017xgc}.

	\item  The final state $T\rightarrow t\ a \rightarrow t \tau^+ \tau^-$ could also arise from $T\rightarrow t\ h / t\ Z$ but is to our knowledge currently not covered by any top partner searches. For VLQ masses much larger than $m_t + m_a$, boosted di-tau systems may arise, thus offering interesting final states at the LHC~\cite{Katz:2010iq,Conte:2016zjp,Cacciapaglia:2017iws}.
	\item Decays of $a$ to vector bosons (if kinematically allowed) can yield $t\gamma\gamma$, $tZ\gamma$, $tWW$, or $tZZ$ resonances. In our benchmark model(s), these $a$ decays do not have large BRs. Nevertheless, the final states (and the kinematics with a boosted top and a di-boson resonance) offer many handles for excellent SM background rejection.
\end{itemize}
Similar considerations hold for the VLQ partner $B$.

\subsection{Exclusive pNGB, $\widetilde{T} \to t\ \eta$ } 
\label{sec:sm2}

As a second simplified scenario, we consider a model with a top partner $\widetilde{T}$ with charge $2/3$
that does not mix with the SM top, and a lighter pseudo-scalar $\eta$. This situation is realised, for example, in composite Higgs models based on $SU(4)/Sp(4)$ breaking, where $\eta$ is the additional singlet and the top partner couplings respect a parity associated with $\eta$. A concrete realisation will be discussed in Sec. \ref{sec:SU4}.
The model is described by the Lagrangian
\bea
\mathcal{L}_{\widetilde{T}} &=&
\phantom{+} \overline{\widetilde{T}} \left(i \slashed{D} - M_{\widetilde{T}}\right)\widetilde{T}  - \left(i \kappa^{\widetilde{T}}_{\eta,L} \, \overline{\widetilde{T}} \eta P_L t + L\leftrightarrow R + \mbox{ h.c. } \right)\, , 
\label{eq:TetaLag}
\eea
for the interactions involving $\widetilde{T}$, which differs from Eq.(\ref{eq:LTa}) by the absence of couplings to the SM bosons.
For the pseudo-scalar $\eta$, in principle, one can write an effective Lagrangian similar to Eq.(\ref{eq:aLag}).  However, in this specific case, not all couplings arise on the same footing. If the couplings of the top respect $\eta$-parity, no couplings of  $\eta$ to tops are generated at leading order~\cite{Alanne:2018wtp}.  The couplings to light fermions are model dependent, but they may also be suppressed: for instance, if they are generated by bilinear couplings, they are absent at the leading order~\cite{Arbey:2015exa}.
Thus, to keep the scenario minimal, we will only consider couplings to gauge bosons:  
\begin{multline}
{\mathcal{L}}_\eta = \frac{1}{2}(\partial_\mu \eta)(\partial^\mu \eta) -\frac{1}{2} m_\eta^2 \eta^2 + \!\frac{g_s^2 K_g^\eta }{16\pi^2 f_\eta} \eta G^a_{\mu\nu}\tilde G^{a\mu\nu}\!+\! \frac{g^2 K_W^a }{8\pi^2 f_\eta} \eta W^+_{\mu\nu}\tilde W^{-,\mu\nu}
\label{eq:etaLag}
\\
\!+\!\frac{e^2 K_\gamma^\eta}{16\pi^2 f_\eta} \eta A_{\mu\nu}\tilde A^{\mu\nu}  \!+\!\frac{g^2 c_W^2 K_Z^\eta}{16\pi^2 f_\eta} \eta Z_{\mu\nu}\tilde Z^{\mu\nu} \!+\!\frac{e g c_W K_{Z\gamma}^\eta}{8\pi^2 f_\eta} \eta A_{\mu\nu}\tilde Z^{\mu\nu}  \, ,
\end{multline}
for the interactions of $\eta$ with the SM particles.

\begin{figure}[tbh]
	\centering
	\begin{tabular}{cc}
		\includegraphics[width=0.52\textwidth]{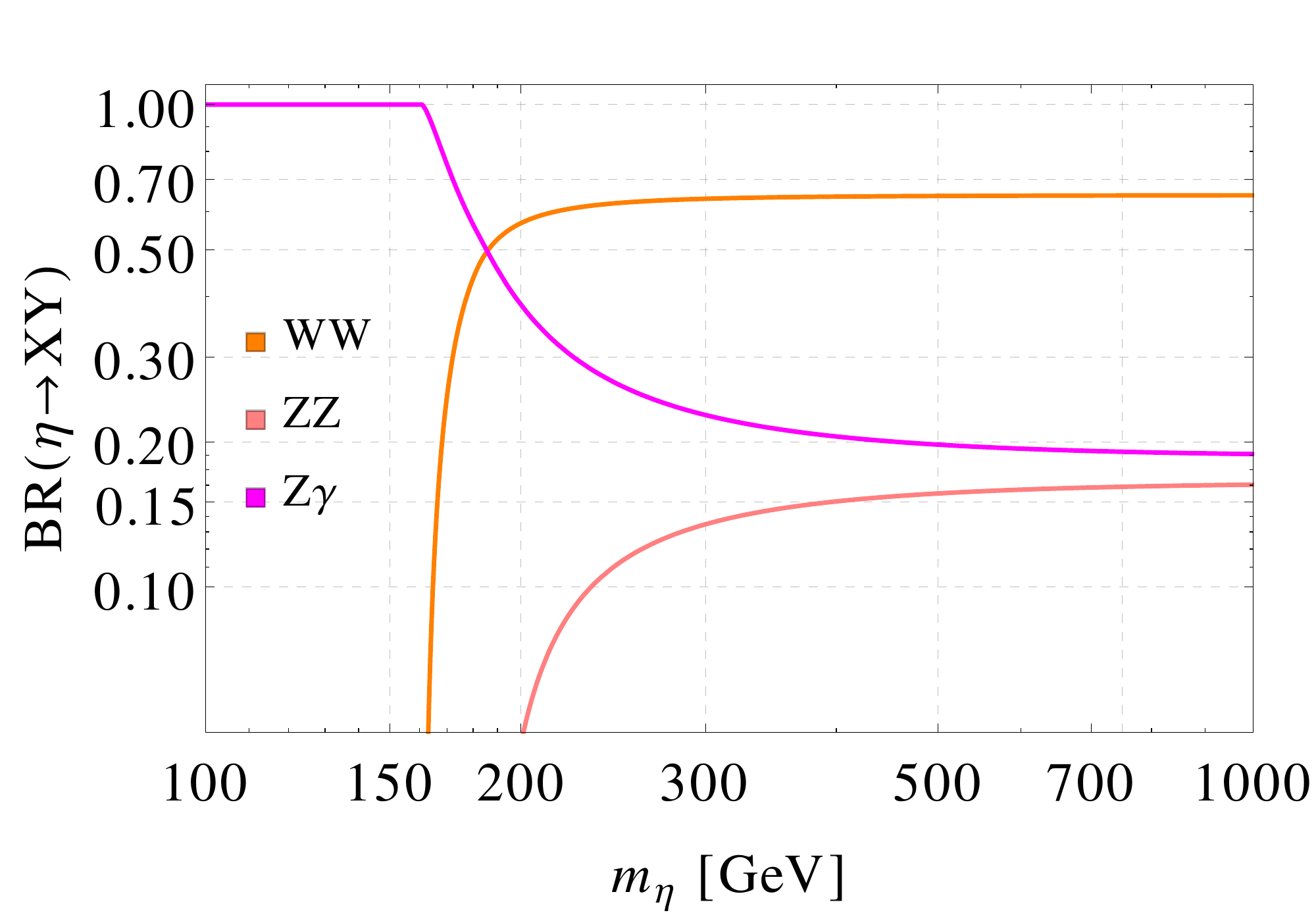}
	\end{tabular}
	\caption{Branching ratio of 
		$\eta$ 
		as a function of $m_{\eta}$  in the benchmark $SU(4)/Sp(4)$ models  introduced in Sec \ref{sec:SU4}.}
	\label{fig:etaBRs}
\end{figure}

In the benchmark model we are interested in, as detailed in Sec.~\ref{sec:SU4}, $\eta$ arises as a singlet from the coset $SU(4)/Sp(4)$ in the EW sector. As a consequence, $K_g^\eta = 0$, and the couplings to the EW bosons can be expressed in terms of two parameters, as in Eq.(\ref{eq:kappas}), with the further constraint $K_B^\eta = - K_W^\eta$.
Thus, the coupling to photons vanishes, and the BRs are fixed in terms of gauge couplings, as shown in Fig.~\ref{fig:etaBRs}. To be concrete, we report here the specific values of the couplings in the benchmark model Bm2  discussed in Sec.~\ref{sec:SU4}:
\begin{equation}
M_{\widetilde{T}}=1.3 ~\rm{TeV}~,
\qquad
\kappa_{\eta,L}^{\widetilde{T}}=-0.08~,
\quad
\kappa_{\eta,R}^{\widetilde{T}}=0.89~,
\end{equation}
while for the decay constant, we fix  $f_\eta=1$ TeV and we leave the mass $m_\eta$ as a free parameter.
Due to the absence of couplings to gluons and the smallness of the anomaly-induced couplings to EW gauge bosons, $\eta$ by itself is not very visible at the LHC nor at lepton colliders: production cross sections have been studied in ref.~\cite{Arbey:2015exa} and give very small yields. At loop level, a coupling $gg \to \eta \eta$ is generated by top-$\widetilde{T}$ loops, and it may give sizeable production rates for small $\eta$ masses. However, below the $Z$ mass, the decay rates of $\eta$ are very model dependent: besides the 3-body decays $Z^\ast \gamma$, competitive rates may be due to sub-leading couplings to light quarks or even 3-photon final states generated by anomalous couplings. A detailed study would, however, be required to establish the precise bounds, thus here we will focus on the $m_\eta > m_Z$ range.

While direct production appears to be negligible, the singlet $\eta$ will be produced via decays of the top partner $\widetilde{T}$ that can only be pair produced.
As in this scenario $\eta$ decays into EW gauge bosons, the signatures resulting from $\widetilde{T}\overline{\widetilde{T}}$ pair production contain fully-reconstructable 3-body resonances with very low SM backgrounds:

\begin{itemize} 
	\item For $m_\eta > 2 m_W$, $\eta$ dominantly decays into $W^+ W^-$ which yields a final state of $pp \rightarrow \widetilde{T}\overline{\widetilde{T}}\rightarrow (tW^+ W^-)(\bar{t}W^+ W^-)$. Decays into $ZZ$ and $Z\gamma$ provide subleading channels.
	\item Below the $2 m_W$ threshold, $\eta$ decays almost exclusively into $Z\gamma$, providing the interesting final state $pp \rightarrow \widetilde{T}\overline{\widetilde{T}}\rightarrow (tZ\gamma)(\bar{t}Z\gamma)$.
\end{itemize}

\subsection{Coloured pNGBs: the case $X_{5/3} \to \bar{b}\ \pi_6$}
\label{sec:sm3}

Models of PC for quarks necessarily contain coloured bound states, as some of the confining underlying fermions need to be charged under $SU(3)_c$ in order to give colour to the composite top partners. In models with a fermionic underlying description, this implies the presence of coloured pNGBs, which may be lighter than the top partners and can thus appear in top partner decays.\footnote{If a coloured pNGB is heavier than top partners, it can affect their production rates~\cite{Deandrea:2017rqp}.}

A colour octet pseudoscalar $\pi_8$, neutral under the EW interactions, is ubiquitous in models with a fermionic underlying description~\cite{Belyaev:2016ftv}. It can couple to a quark and quark-partner and therefore appear in quark partner decays, and itself decays into $t\bar{t}$, $gg$ or $g\gamma$. The presence of $\pi_8$ thus gives rise to final states similar to the ones described in Sec.~\ref{sec:sm1} (with the addition of the $g\gamma$ channel). 

Other colour charged pNGBs are present in some of the models. Here, we focus on the  charge-$4/3$ colour sextet $\pi_6$~\cite{Cacciapaglia:2015eqa}. The main reason behind this choice is that it can modify the decays of a charge $5/3$ top partner $X_{5/3}$. 
The latter is a commonly considered state which is present in top partner multiplets in an $SU(2)_L\times SU(2)_R$ bi-doublet. It is normally assumed to decay exclusively into $t\ W^+$, which yields a same-sign lepton (SSL) signature from leptonic $W$ decays~\cite{Contino:2008hi}, with low SM background and thus very high sensitivity. $X_{5/3}$ is therefore an ideal target for searches at hadron colliders. Semi-leptonic decays of $t\ W^+$ have higher background but also a higher BR and provide another attractive channel.  
For pair-produced $X_{5/3}$, the current bound on its mass is $M_{X_{5/3}} > 1.3$ TeV \cite{CMS:2017wwc,CMS:2017jfv,Sirunyan:2017jin,Aaboud:2017zfn}, while higher sensitivity for single-produced $X_{5/3}$ is possible, but model-dependent \cite{Backovic:2014uma}.
However, all these  bounds assume the absence of ``exotic'' $X_{5/3}$ decays.

The effective Lagrangian for the $X_{5/3}$ couplings, including the sextet, reads
\begin{multline}
\mathcal{L}_{X_{5/3}}^{\pi_6} =
\phantom{+} \overline{X}_{5/3} \left(i \slashed{D} - M_{X_{5/3}}\right)X_{5/3} 
\label{eq:XpiLag}\\ 
+ \left(\kappa^X_{W,L} \frac{g}{\sqrt{2}} \, \overline{X}_{5/3} \slashed{W}^+P_L t  + i \kappa^{X}_{\pi_6,L}  \, \overline{X}_{5/3} \pi_6 P_L b^c  + L\leftrightarrow R + \mbox{ h.c. } \right)~,
\end{multline}
while the one associated to the $\pi_6$  couplings to SM particles is
\bea
\mathcal{L}_{\pi_6} &=&
\phantom{+}   \left| D_\mu \pi_6 \right|^2 - m_{\pi_6}^2 \left|\pi_6 \right|^2 + \left( i \kappa^{\pi_6}_{tt,R} \,   \overline{t} \pi_6 (P_R t)^c + L\leftrightarrow R + \mbox{ h.c. }\right)~,
\nonumber
\label{eq:XpiLag1}
\eea
where $b^c$ and $t^c$ denote the charge conjugate of the bottom and the top quark fields.  Note that, in the model we consider, $\pi_6$ is a singlet of $SU(2)_L$. The coupling $\kappa^{\pi_6}_{tt,L}$ to left handed tops are thus suppressed by $m_t^2/f_{\pi_6}^2$ with respect to $\kappa^{\pi_6}_{tt,R}$.  The sextet decays as $\pi_6 \to tt$, with large dominance to right-handed tops. 

\begin{figure}[tbh]
	\centering
	\begin{tabular}{cc}
		\includegraphics[width=0.48\textwidth]{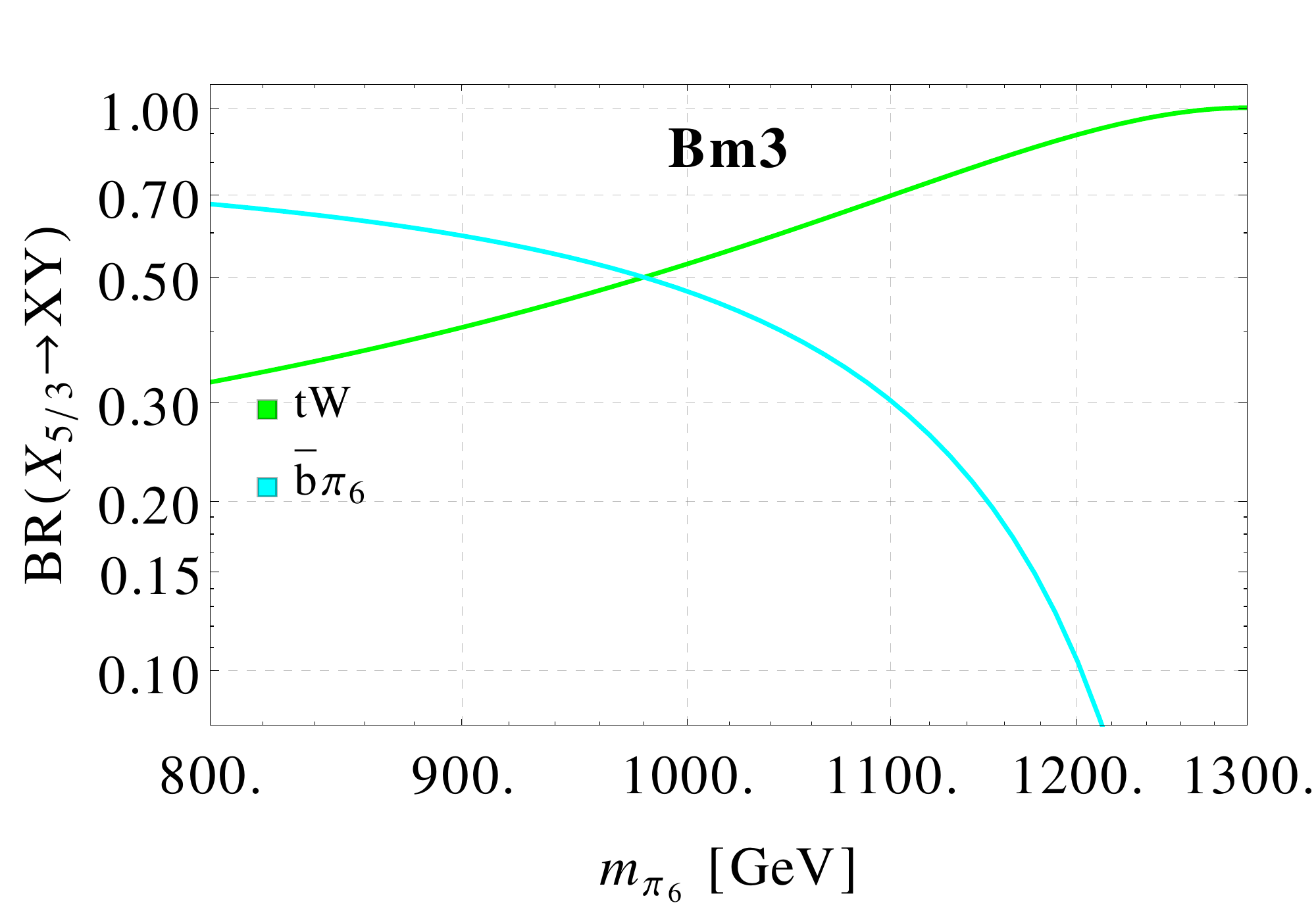}  &\includegraphics[width=0.48\textwidth]{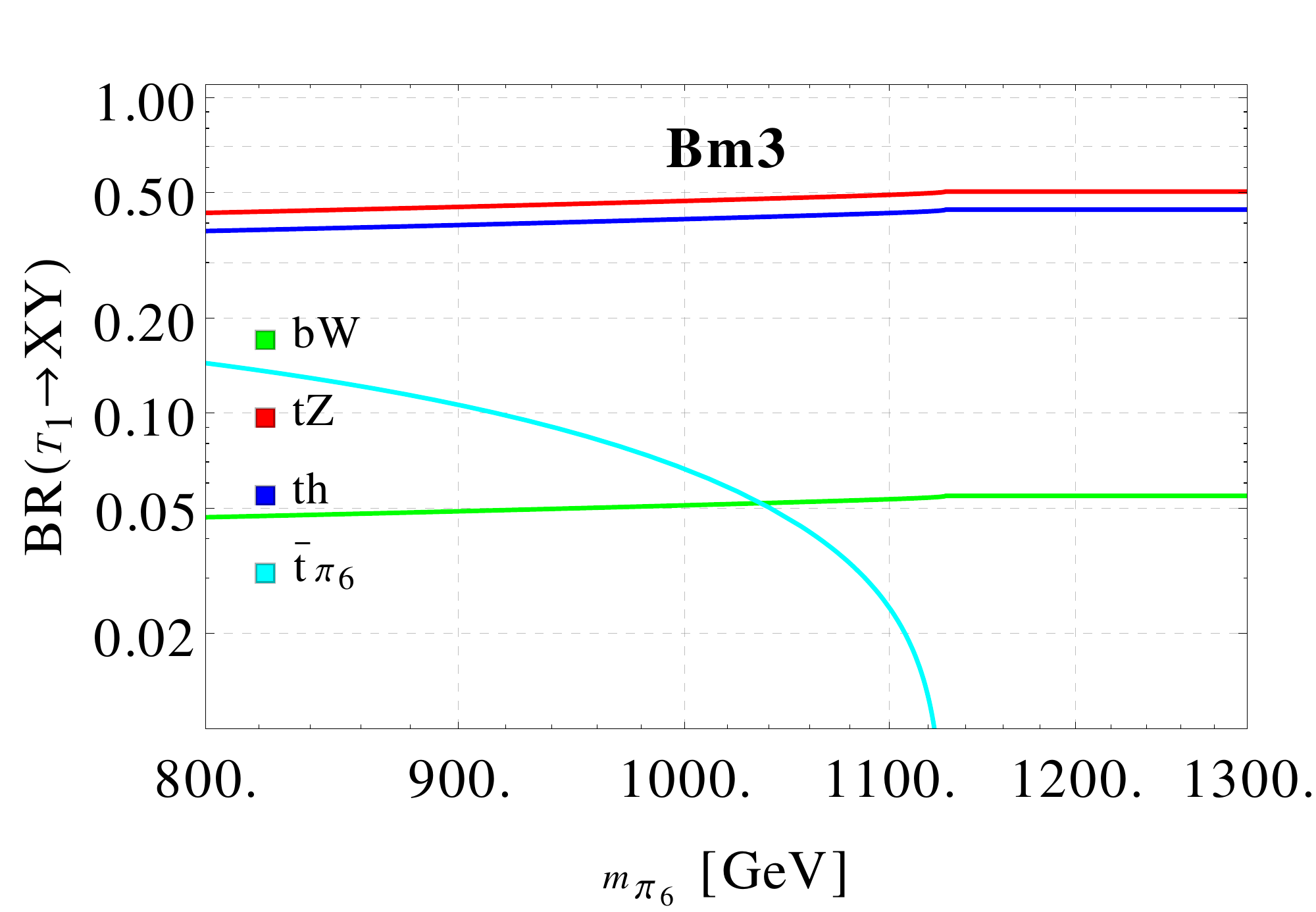}
	\end{tabular}
	\caption{Branching ratios of $X_{5/3}$ and the lightest $T_1$ 
		as a function of $m_{\pi_6}$  in the  benchmark model Bm3  introduced in Sec. \ref{sec:SU6}.}
	\label{fig:XpiBRs}
\end{figure}

The sextet arises, for example, as part of the pNGB spectrum in UV embeddings of composite Higgs models with $SU(4)/Sp(4)$ breaking \cite{Cacciapaglia:2015eqa} (see Sec.\ref{sec:SU4}). 
For illustration purposes, we again use this underlying model to define a benchmark model, Bm3, in Sec.~\ref{sec:SU6}. The values of the couplings are
\begin{equation}
\mbox{Bm3}: ~~M_{X_{5/3}}=1.3 ~\rm{TeV}~,
\quad
\kappa_{W,L}^X=0.03~,
~~
\kappa_{W,R}^X=-0.11~,
~~
\kappa_{\pi_6,L}^X=1.95~,
~~
\kappa_{tt,R}^{\pi_6}=-0.56~,
\end{equation} 
while the other couplings are suppressed, and $f_{\pi_6}=430$~GeV (note that $f_{\pi_6}$ is not directly related to the compositeness scale for the Higgs, as it comes from a different sector of the theory, and we use here an estimate with respect to a decay constant $f = 1$~TeV in the Higgs sector). The BRs are shown in the left panel of Fig.~\ref{fig:XpiBRs}, demonstrating that sizeable rates into the colour sextet are possible in realistic models.

The phenomenology of  $\pi_6$ (in absence of VLQs) has been studied in ref.~\cite{Cacciapaglia:2015eqa}. It is pair-produced via  QCD interactions or singly produced via top-fusion and,
following the decay into two top quarks, leads to 4-top final states. At LHC Run I, SSL searches  imply a bound of $m_{\pi_6} \gtrsim 800$~GeV  \cite{Cacciapaglia:2015eqa}. Additional indirect constraints may apply, however they are more model dependent so we conservatively rely on the direct production bound only. 
The signatures from production via $X_{5/3}$ decays depend on the production mode for the VLQ:

\begin{itemize}
	
	\item For pair produced $X_{5/3}$, the final state contains $tt\bar{t}\bar{t}$ + $b \bar{b}$, thus it will be efficiently covered by 4-top searches. Additionally, one can have different decays on the two legs, yielding $\bar{b} t t \bar{t} W^- + b t \bar{t} \bar{t} W^+$, which again matches 4-top searches.
	
	\item For singly-produced $X_{5/3}$, the final state will always contain two tops, thus this final state can again be searched for in SSL final states.
	
\end{itemize}
While SSL searches seem to efficiently cover this channel, the precise bounds will depend on the different kinematics of the final states. Furthermore, additional requirements, like for instance tagging the b-jets, may improve the reach with respect to standard searches. 

Finally, we  remark that  $\pi_6$ can also couple to other top partners, which can thus decay into it. As $X_{5/3}$ is embedded in an electroweak multiplet (in our example, in an $SU(2)_L\times SU(2)_R$ bi-doublet), additional top partners with mass comparable to $M_{X_{5/3}}$ are generically present. A charge $2/3$ top partner can couple to $\bar{t}\ \pi_6$, thus adding final states with a  $T \to \bar{t}\ \pi_6 \to \bar{t} t t$ decay. The same final states already occurred in the  simplified models in Sec.~\ref{sec:sm1}, although the kinematics differs, as $\pi_6$ decays into two tops (and not $t\bar{t}$). Decays of the individual states of the top partner multiplet in our  benchmark model Bm3 are discussed in more detail in Sec.~\ref{sec:SU6}.  
Here we just give a brief example. In the right panel of Fig.~\ref{fig:XpiBRs} we show the BRs of the lightest charge $2/3$ partner, $T_1$ (with mass $M_{T_1}= 1.3$~TeV) of Bm3 which has a sizeable branching fraction into $\bar{t} \pi_6$, if $\pi_6$ is sufficiently light.

\subsection{Charged pNGB, $X_{5/3} \to t\ \phi^+$}
\label{sec:sm4}

As a second example for exotic decays of  a charge $5/3$ top partner, we consider a model with a colour-neutral, electrically charged scalar $\phi^+$. 
The latter arises for example as part of the pNGB spectrum in  composite Higgs models with $SU(5)/SO(5)$ breaking~\cite{Dugan:1984hq} (see Sec.~\ref{sec:SU5}), where it is accompanied by a 
doubly-charged scalar.
The effective Lagrangians for the VLQ $X_{5/3}$ and the charged scalar couplings, respectively read
\bea
\mathcal{L}_{X_{5/3}}^{\phi} &=&
\phantom{+} \overline{X}_{5/3} \left(i \slashed{D} - M_{X_{5/3}}\right)X_{5/3} 
+\biggl(  \kappa^X_{W,L} \frac{g}{\sqrt{2}} \, \overline{X}_{5/3} \slashed{W}^+ P_L t 
\label{eq:XphiLag}
\\ 
&&
+ i \kappa^{X}_{\phi^+,L}  \, \overline{X}_{5/3} \phi^+ P_L t 
+ i \kappa^{X}_{\phi^{++},L}  \, \overline{X}_{5/3} \phi^{++} P_L b + L \leftrightarrow R +\mbox{h.c.} \biggr) ~,
\nonumber
\eea
and \footnote{We neglect couplings of pNGBs to leptons and light quarks, which are analogous to the last term of Eq.(\ref{eq:XphiLag1}) and suppressed by $m_f/f_\phi$. Additional couplings may also arise, like lepton number violating ones with the triplet, however their presence is model dependent. As they are not required by the lepton mass generation, we can consistently assume their absence.}
\bea
\mathcal{L}_{\phi} &=&
\sum\limits_{\phi=\phi^+,\phi^{++}} \left( \left|D_\mu \phi \right|^2 - m_{\phi}^2 \left|\phi \right|^2 \right)
+\biggl(\frac{e g K^\phi_{W\gamma}}{8\pi^2 f_\phi} \phi^+W^-_{\mu\nu}\tilde{B}^{\mu\nu} 
+\frac{g^2 c_w K^\phi_{WZ}}{8\pi^2 f_\phi} \phi^+W^-_{\mu\nu}\tilde{B}^{\mu\nu}
\nonumber
\\  
&& ~~~~~~~~ +\frac{g^2 K^\phi_{W}}{8 \pi^2 f_\phi} \phi^{++} W^-_{\mu\nu} \tilde{W}^{\mu\nu,-}  
+ i \kappa^\phi_{tb,L} \frac{m_t}{f_\phi} \,  \overline{t} \phi^+ P_L b + L\leftrightarrow R  + \mbox{h.c.} \biggr) ~.
\label{eq:XphiLag1}
\eea
Note that we have defined a unique decay constant, $f_\phi$, for both charged scalars, as they usually originate from the same coset. 
In models based on the $SU(5)/SO(5)$ breaking pattern (minimal coset with charged pNGBs), the charged scalar $\phi^\pm$ belongs to $SU(2)_L$-triplets.   
Thus, in the non-zero hypercharge triplet, a doubly charged scalar $\phi^{\pm\pm}$ is present and has been added to the previous Lagrangians.
The latter can not be neglected, even in this simplified scenario, as it affects the decays of $X_{5/3}$.
Thus, the new exotic channels in this scenario are  $X_{5/3}\rightarrow t\ \phi^+$ and $X_{5/3}\rightarrow b\ \phi^{++}$.

\begin{figure}[tbh]
	\centering
	\includegraphics[width=0.48\textwidth]{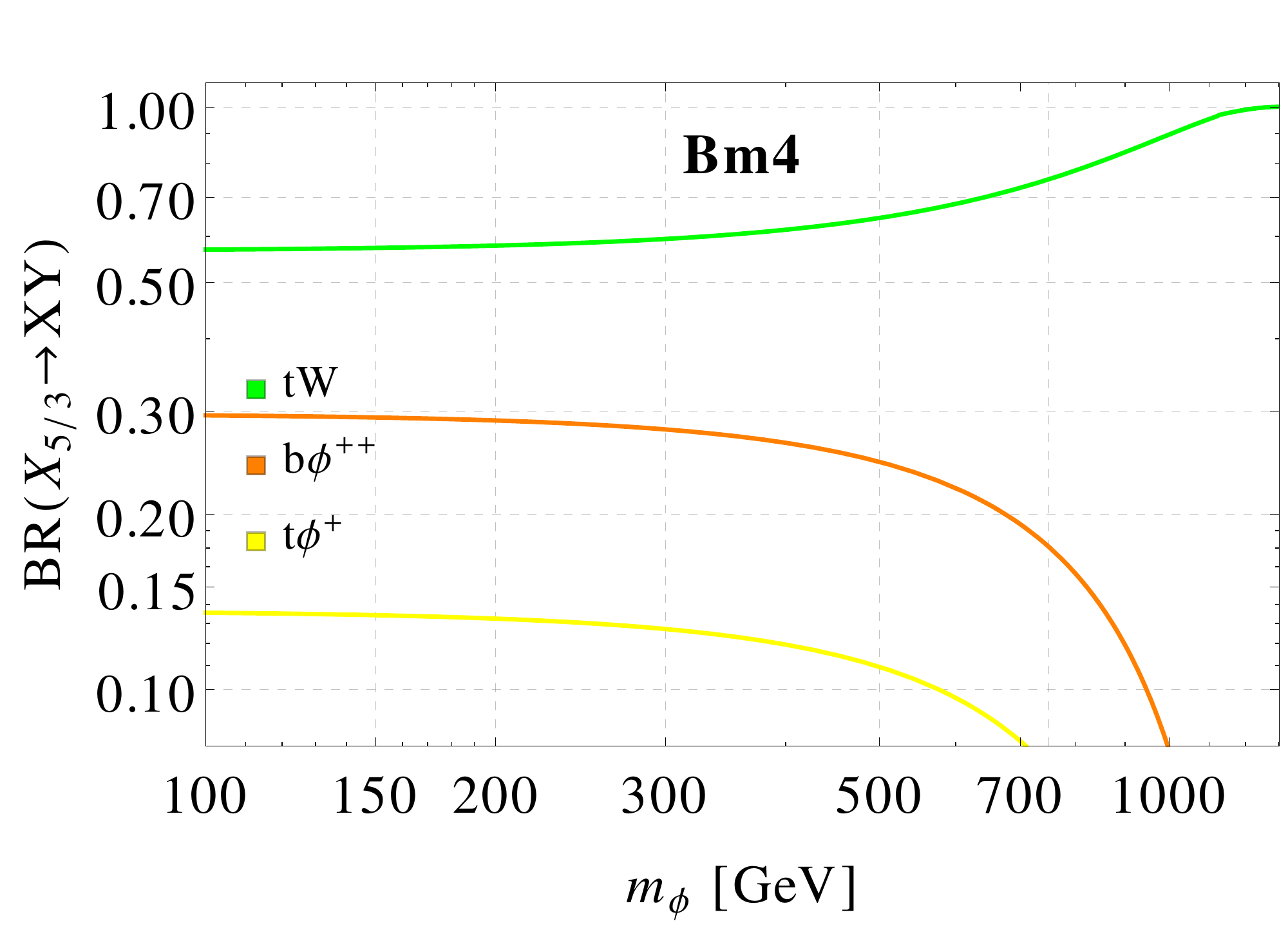}  
	\caption{Branching ratios of $X_{5/3}$ as a function of the mass of the charged pNGBs $m_\phi = m_{\phi^+}=m_{\phi^{++}}$  for the benchmark model Bm4 introduced in Sec. \ref{sec:SU5}.
	}
	\label{fig:XphiBRs}
\end{figure}

To illustrate these exotic decay modes, we define  another benchmark model, Bm4, in Sec.~\ref{sec:SU5}.
The corresponding values of the couplings are given by
\bea
&&
\mbox{Bm4}:~~M_{X_{5/3}}= 1.3 ~\rm{TeV}~,
\quad
\kappa_{W,L}^X=0.03~,
~~
\kappa_{W,R}^X=0.13~,
~~
\kappa_{\phi^+,L}^X=0.49~,
~~
\kappa_{\phi^+,R}^X=0.12~,
\nonumber
\\
&&
\hspace*{5. cm}
\kappa_{\phi^{++},L}^X=-0.69~,
~~
\kappa_{tb,L}^{\phi}=0.53~,
\eea
while the other couplings are suppressed, and $f_\phi=1$ TeV.
The BRs are displayed in Fig.~\ref{fig:XphiBRs}, showing that non-negligible rates into the charged pNGBs $\phi^\pm$ and $\phi^{\pm\pm}$ are present in realistic models.
Note that we assume for simplicity a common mass $m_\phi$ for the two charged pNGBs.

Due to its anomalous couplings in Eq.~(\ref{eq:XphiLag1}), the charged pNGB $\phi^+$ can decay into a pair of SM gauge bosons, either $W^+ \gamma$ or  $W^+ Z$.
A coupling to $t \overline{b}$ is also generated from PC. 
Couplings to light fermions are model dependent, as they vary according to the mechanism generating their mass: here, for simplicity, we will neglect them. 
For the doubly charged pNGB $\phi^{++}$, the only available channel arises from an anomalous couplings to $W^+ W^+$. 
In the underlying models based on SU(5)/SO(5), the anomalous couplings of $\phi^+$ are related by gauge couplings, as they both originate from the coupling $K_{WB}^\phi$ of the triplet to an SU(2)$_L$ and a U(1)$_Y$ gauge boson.
This leads to the relations
\begin{equation}
K_{W\gamma}^\phi=K_{WB}^\phi~,
\qquad
K_{WZ}^\phi=-K_{WB}^\phi t_w^2~.
\end{equation}
Below the $t\bar{b}$ threshold, $\phi^+$ mostly decays into $W^+ \gamma$: this is due  both to the suppression of the coupling to $W^+ Z$ (shown above) and to the fact that the mass threshold for the $WZ$ channel is very close to the $t\bar{b}$ one. Above the $t \overline{b}$ threshold, the fermionic channel typically dominates.
Note that below the $W$ mass, the decays into a virtual $W$ boson (i.e., three body decays) may be competitive with more model dependent decays into light fermions, thus we will not consider this mass region here. It should also be noted that, while dedicated searches are not available, collider bounds on direct production of the charged scalars are very mild: bounds on similar models, which should be applied with a pinch of salt, point towards mass bounds below the $W$ mass~\cite{Kanemura:2014ipa,Degrande:2017naf}, so no direct bounds should apply to the mass region we chose.
The above scenario leads to different signatures depending on the masses of the charged pNGBs:
\begin{itemize}
	\item For $m_{\phi^+}$ below the $t\overline{b}$ threshold, the channel $X_{5/3}\to t\ \phi^+ \to t W^+ \gamma$ leads to extra hard photons in addition to the standard final states. 
	\item Above the $t\bar{b}$ threshold, $\phi^+$ decays almost exclusively into $t\overline{b}$, thus offering an interesting final state $X_{5/3}\to t t \bar{b}$ that  will be easily covered by the existing 4-top searches when $X_{5/3}$ is pair-produced and both decay into this exotic channel.  Different decays on the two legs  produce final states similar to four tops, i.e. $tt\bar{b}\bar{t}W^-$ (for one decay through $\phi^+$ and one standard) or $tt\bar{b}\bar{b}W^-W^-$ (for one decay through $\phi^+$ and one through $\phi^{--}$).
	\item The channel $X_{5/3}\rightarrow b\ \phi^{++} \rightarrow b W^+ W^+$ 
	leads to a signature similar to the standard $X_{5/3}\rightarrow tW$ (with subsequent top decay to $bW^+$), but with  different kinematics.
\end{itemize}

Finally, let us remark that the charged pNGBs couple in general to the other top partners.
The resulting new decay modes are discussed in more details in Sec.~\ref{sec:SU5}.
One interesting final state that we want to mention is due to decays of a charge $2/3$ partner in the charged scalar leading to $T \to b\, \phi^+ \to b W^+ \gamma$, which is similar to a top final state with the addition of a hard photon.


\section{Exotica in minimal composite Higgs models}
\label{sec:models}

The simplified models that describe the new decay modes arise quite naturally in models of a composite Higgs with partially composite fermions.
In this section we provide some explicit examples to illustrate the origin of the new channels.
We start by providing a simplified scenario where only the minimal matter content is introduced, before analysing two realistic scenarios based on the symmetry breaking patterns SU(4)/Sp(4) and SU(5)/SO(5).

\subsection{ Simple model of partial compositeness}
\label{Minimal scenario}

 The main principle behind PC is that elementary fermions linearly couple to vector-like fermion partner states such that they mix, and the lighter eigenstate -- which is to be identified with the SM fermion -- obtains a mass through electroweak symmetry breaking. This structure appears naturally in composite Higgs models, as the VLQs are identified with composite states themselves. The coupling to the Higgs (as a pNGB) thus arises via the linear mixing operators that connect the elementary fields to the composite fermions.

At an effective model level, the most minimal field content involves two VLQs that have exactly the same quantum numbers as the elementary top fields: an $\rm{SU(2)}_L$ doublet $Q = \begin{pmatrix} U \\ D \end{pmatrix}$ with hypercharge $1/6$, and a singlet $S$ with hypercharge $2/3$. This simple PC Lagrangian, including only linear interactions of the Higgs doublet scalar $\phi_H$, is given by
\begin{multline} \label{eq:Lag_minimal}
- \mathcal{L}_{\rm PC} = M_Q\ \bar{Q} Q + M_S\ \bar{S} S +  \left( y_L f\ e^{i \xi_Q \frac{a}{f_a}} \ \bar{Q} P_L q + y_R f\ e^{i \xi_S \frac{a}{f_a}}\ \bar{t} P_L S \right. \\
- \left. y'_L \phi_H^\dagger\ e^{- i \xi_S \frac{a}{f_a}}\ \bar{S} P_L q - y'_R \phi_H^\dagger\ e^{- i \xi_Q \frac{a}{f_a}}\ \bar{t} P_L Q +  \mbox{h.c.}\right)\,,
\end{multline}
where we have also included the couplings of a pseudo-scalar $a$ associated to a spontaneously broken global U(1) symmetry with charges $\xi_{Q,S}$ assigned to the VLQs. Such a pNGB arises naturally in underlying models of PC~\cite{Ferretti:2016upr,Belyaev:2016ftv}, where an anomaly-free $U(1)$ global symmetry is spontaneously broken by the condensation of underlying fermions.
The couplings $y_{L/R}$ parametrise the linear mixing of the elementary fields with the composite ones, following the PC prescription, while the couplings $y'_{L/R}$ generate additional mixing terms once the Higgs acquires its vacuum expectation value (VEV).
Note that $y_{L/R}$ generate a mass mixing between fermions with the same SM quantum numbers. They can thus be rotated away to bring Eq.(\ref{eq:Lag_minimal}) to a basis which is more familiar to the VLQ literature~\cite{delAguila:2000rc,AguilarSaavedra:2009es,Cacciapaglia:2010vn} by defining
\begin{equation}
\begin{array}{ccc}
& M_{Q'} = \sqrt{M_Q^2 + y_L^2 f^2}\,, \quad s_L \equiv \sin \theta_L = \frac{y_L f}{M_Q'}\,,& \\
& M_{S'} = \sqrt{M_S^2 + y_R^2 f^2}\,, \quad s_R \equiv \sin \theta_R = \frac{y_R f}{M_S'}\, , &
\end{array}
\label{eq:QSmass}
\end{equation}
where we remark that the new doublet $Q'$ and singlet $S'$ are the genuine VLQs. The couplings of the $W$, $Z$ and Higgs can then be obtained  diagonalising the full mass matrix, including the electroweak symmetry breaking contributions~\cite{Buchkremer:2013bha}, while the new couplings of the singlet $a$, following the notation of Eq.(\ref{eq:LTa}), are given by
\begin{equation} \label{eq:couplingsTa}
\begin{array}{ccc}
& \kappa^T_{a,R} = \frac{\xi_S M'_S}{f_a} s_R c_R  + \mathcal{O} (v/M'_S)\,, \qquad \mbox{for the singlet}\,, & \\
& \kappa^T_{a,L} = - \frac{\xi_Q M'_Q}{f_a} s_L c_L  + \mathcal{O} (v/M'_Q)\,, \qquad \mbox{for the doublet}\,. &
\end{array}
\end{equation}
Furthermore, a coupling of the pseudo-scalar $a$ to two tops is also generated by the diagonalisation of the mass matrix, leading to
\begin{equation}
g_{att} = - i \frac{m_t}{f_a} (\xi_Q s_L^2 + \xi_S s_R^2 )\,.
\label{eq:effopp}
\end{equation}
Interestingly, this result is different from what we would obtain if we wrote an effective operator generating the mass of the top, as it was done in ref.~\cite{Belyaev:2016ftv}, which would give $g_{att} = - i \frac{m_t}{f_a} (\xi_Q + \xi_S)$. This difference is due to the effect of the mixing induced by the VLQs. The fact that the results are truly different can be appreciated if we expand for small Yukawas ($y_{L/R} \sim y'_{L/R} \sim y \ll 1$): the coupling we obtained in Eq.(\ref{eq:effopp}) scales like $y^4$ (we recall that $m_t \sim y^2$), while the prediction from the effective operator scales like $m_t \sim y^2$.

To understand the physics entailed by the above minimal scenario, it is instructive  to study the theory before the Higgs develops its VEV. This is justified as the mass of the VLQs is expected to be much larger than the EW scale (Higgs mass), and we can thus use the equivalence principle to study the couplings of the Goldstone bosons instead of the vector bosons.
Calling $t,\ T_1,\ T_2$ and $b,\ B$ the mass eigenstates (without including the Higgs effects), the Lagrangian in Eq.~(\ref{eq:Lag_minimal}) can be rewritten as
\begin{multline}
- \mathcal{L}_{\rm PC} = M_Q' \left( c_L^2 + s_L^2 e^{i \xi_Q \frac{a}{f_a}} \right)\ (\bar{T}_{1} P_L T_{1} + \bar{B} P_L B) + M_S' \left( c_R^2 + s_R^2 e^{i \xi_S \frac{a}{f_a}} \right)\ \bar{T}_{2} P_L T_{2} + \\
M_Q' s_L c_L \left( e^{i \xi_Q \frac{a}{f_a}} -1\right)\ (\bar{T}_1 P_L t + \bar{B} P_L b ) + M_S' s_R c_R \left( e^{i \xi_S \frac{a}{f_a}} -1\right)\  \bar{t} P_L T_{2} +  \\ 
+ \left( y'_L c_L s_R e^{-i \xi_S \frac{a}{f_a}} + y'_R s_L c_R e^{-i \xi_Q \frac{a}{f_a}} \right)\ (\phi_0^\dagger \bar{t} P_L t - \phi^- \bar{t} P_L b) +  \\
-  \left( y'_L s_L c_R e^{-i \xi_S \frac{a}{f_a}} + y'_R c_L s_R e^{-i \xi_Q \frac{a}{f_a}} \right)\ (\phi_0^\dagger \bar{T}_2 P_L T_{1} - \phi^- \bar{T}_2 P_L B) +  \\
- \left( y'_L c_L c_R e^{-i \xi_S \frac{a}{f_a}} - y'_R s_L s_R e^{-i \xi_Q \frac{a}{f_a}} \right)\ (\phi_0^\dagger \bar{T}_2 P_L t - \phi^- \bar{T}_2 P_L b) +  \\
+ \left( y'_L s_L s_R e^{-i \xi_S \frac{a}{f_a}} - y'_R c_L c_R e^{-i \xi_Q \frac{a}{f_a}} \right)\ (\phi_0^\dagger \bar{t} P_L T_{1} - \phi^- \bar{t} P_L B) + \mbox{h.c.} \,. \label{eq:lagr}
\end{multline}
The mass of the top quark is generated by the interaction on the third line of the above equation, allowing us to identify the top Yukawa with
\begin{equation}
y_{\rm top} = \left( y'_L c_L s_R  + y'_R s_L c_R \right)\,.
\end{equation}
The couplings of the VLQs to $a$ and a SM quark, on the other hand, are obtained from the second line after expanding the exponential, thus yielding the results in Eq.(\ref{eq:couplingsTa}).
Relying on the equivalence principle, the decay rate of the VLQs into $a$ can be estimated as follows:
\begin{eqnarray}
& \frac{\Gamma (T_1 \to t\ a)}{\Gamma (T_1 \to t\ \phi_0) } = \frac{\Gamma (B \to b\ a)}{\Gamma (B \to t \ \phi^-)} =  \xi_Q^2 \left( \frac{M_Q'}{f_a} \right)^2 \frac{s_L^2 c_L^2}{(y'_L s_L s_R  - y'_R c_L c_R )^2}\,, & \\
& \frac{\Gamma (T_2 \to t\ a)}{\Gamma (T_2 \to t\ \phi_0) + \Gamma (T_2 \to b\ \phi^+)}  = \xi_S^2 \left( \frac{M_S'}{f_a} \right)^2 \frac{s_R^2 c_R^2}{2 (y'_L c_L c_R  - y'_R s_L s_R )^2}\,. &
\end{eqnarray}
This result clearly shows that the decay rates in the new pseudo-scalar can be substantial, as there is no parametric suppression in their couplings as compared to the couplings to the Higgs field. Furthermore, as long as the charges $\xi_{Q/S}$ are non-vanishing, it is not possible to remove the couplings without affecting the mass of the top.
To clarify this statement, we can check the result in the limit where the singlet is much lighter than the doublet, i.e. for $s_L \ll 1$:
\begin{equation}
\frac{\Gamma (T_2 \to t\ a)}{\Gamma (T_2 \to t\ \phi_0) + \Gamma (T_2 \to b\ \phi^+)} \stackrel{s_L\rightarrow 0}{\,\,\,\,\,=\,\,\,\,\,} \frac{1}{2} \xi_S^2 \left( \frac{M_S'}{f_a} \right)^2 \left(\frac{v}{\sqrt{2}  m_{\rm top}} \right)^2 s_R^4\,,
\label{eq:bflim}
\end{equation}
which is substantial as long as $s_R \sim 1$.


\subsection{The SU(4)/Sp(4) scenario}
\label{sec:SU4}

We now analyse explicit models of composite Higgs: we first consider  the coset $SU(4)/Sp(4)$, which is the minimal one to enjoy a simple gauge-fermion underlying realisation~\cite{Cacciapaglia:2014uja}.
The composite VLQs as well as the pNGBs (including the Higgs boson) now originate from a composite sector which is globally invariant under an $SU(4)$ flavour symmetry that is spontaneously broken down to $Sp(4)$.
As a consequence, the SM Higgs doublet is accompanied by a pseudo-scalar singlet $\eta$ in order to form a complete representation of the unbroken flavour symmetry.
In the same way, the VLQ multiplets must contain additional top partners, whose quantum numbers depend on the choice of the $Sp(4)$ representations.

As a concrete example, we consider two multiplets: one transforming as a $5$-plet of $Sp(4)$  and one in the singlet representation. Together they may form a $6$-plet of $SU(4)$~\cite{Gripaios:2009pe}, and such a top partner easily arises as a ``chimera baryon''~\footnote{The name ``chimera baryon'' was first coined in ref.~\cite{DeGrand:2016mxr}.} in underlying models with two species of fermions~\cite{Barnard:2013zea,Ferretti:2013kya}.
Under the $SU(2)_L \times U(1)_Y$ symmetry, the 5-plet decomposes as $2_{7/6}+2_{1/6}+1_{2/3}$. It thus contains an additional exotic doublet and a singlet together with the $SU(2)_L$ doublet $Q$ of Eq.~(\ref{eq:Lag_minimal}).
The $Sp(4)$ singlet representation, having hypercharge $2/3$, is trivially identified with the singlet $S$ that couples linearly to the right-handed top.
The various top partners are labelled as follows:
\begin{equation}
\mbox{$5$-plet} \rightarrow \quad
\begin{pmatrix}
X_{5/3}
\\
X_{2/3}
\end{pmatrix}~,
~~
\begin{pmatrix}
T 
\\
B
\end{pmatrix}
~,
~~
\widetilde{T}_5~;
\qquad\qquad
\mbox{singlet} \rightarrow \quad \widetilde{T}_1~.
\end{equation}
We then introduce a linear mixing of the left-handed top (and bottom) with the doublet contained in the $5$-plet and of the right-handed top with the singlet: they are the sources of PC, and their effect can be introduced in the effective Lagrangian in the standard way~\cite{Marzocca:2012zn}.
In this work, we will follow the same procedure and notations as in ref.~\cite{Cacciapaglia:2015eqa} to obtain the mass matrices associated to the top partners and the elementary fermions.
As we study the couplings to the pNGBs other than the Higgs doublet, we will keep them explicitly in the mass matrix.
For the charge $2/3$ fermions, in the basis $\psi_t = \{t,T,X_{2/3},\widetilde{T}_1,\widetilde{T}_5\}$, we obtain the following matrix:  
\begin{equation}
\bar{\psi}_{tR}
\begin{pmatrix}
0 & -\frac{y_{5R}}{\sqrt{2}} e^{i \xi_5 \frac{a}{f_a}} f s_\theta & -\frac{y_{5R}}{\sqrt{2}} e^{i \xi_5 \frac{a}{f_a}} f s_\theta & y_{1R}e^{i \xi_1 \frac{a}{f_a}}  f c_\theta & i y_{5R} c_\theta \eta
\\
y_{5L} e^{i \xi_5 \frac{a}{f_a}} f c^2_{\theta/2}  & M_5 & 0 & 0 & 0
\\
-y_{5L} e^{i \xi_5 \frac{a}{f_a}} f s^2_{\theta/2}  & 0 & M_5 & 0 & 0
\\
-\frac{y_{1L}}{\sqrt{2}} e^{i \xi_1 \frac{a}{f_a}} f s_\theta  & 0 & 0 & M_1 & 0
\\
-i \frac{y_{5L}}{\sqrt{2}} s_\theta \eta & 0 & 0 & 0 & M_5
\end{pmatrix} \psi_{tL}~,
\label{mass-SU4-Sp4}
\end{equation}
where we kept only linear terms in the singlet $\eta$, while the charges $\xi_{1,5}$ can be computed from the underlying theory following ref.~\cite{Belyaev:2016ftv}.\footnote{Note that the decay constant for the $U(1)$ pNGB, $f_a$, that we use here follows the convention of ref.~\cite{Cacciapaglia:2017iws}, and we assume that the pseudo-scalar associated to the anomalous $U(1)$ combination decouples.}
For trigonometric functions we use the shorthand notation $s_\theta = \sin \theta$, etc. . 
The angle $\theta$ is related to the Higgs VEV as $s_\theta \equiv v/f$, and it describes the misalignment of the vacuum in the global flavour space~\cite{Kaplan:1983fs}.
The matrix above matches the simplified Lagrangian in Eq.(\ref{eq:Lag_minimal}) once we expand for small $\theta$ up to linear terms. 
Non-linearities, expressed by higher orders in $\theta$ and due to the non-linear nature of the Higgs boson, affect the couplings as follows: 
\begin{equation}
M_{S,Q}=M_{1,5}~,
\qquad y_{L}=y_{5L} c^2_{\theta/2}~,
\quad
y_R=y_{1R} ~c_\theta~,
\quad
y_L^\prime=y_{1L}~,
\quad
y_{R}^\prime=y_{5R}~,
\end{equation}
while a new ingredient is due to the presence of the exotic doublet and of the top partner $\widetilde{T}_5$ that couples to $\eta$.
For completeness, the matrix for the charge $-1/3$ fermions, in the basis $\psi_b = \{b,B\}$, is given by
\begin{equation}
\bar{\psi}_{bR}
\begin{pmatrix}
0 & 0
\\
y_{5L} e^{i \xi_5 \frac{a}{f_a}} f  & M_5
\end{pmatrix} \psi_{bL}~,
\label{mass-SU4-Sp4-b}
\end{equation}
while for the exotic charge $5/3$ fermion have mass $M_{X_{5/3}}=M_5$.

The matrix in Eq.~(\ref{mass-SU4-Sp4}), which contains both the mass mixings and the couplings to the two singlets $\eta$ and $a$, has several remarkable features.
Firstly, the EW singlet $\widetilde{T}_5$ that belongs to the $5$-plet of $Sp(4)$ does not mix to other fermions but couples to them via the singlet $\eta$. This is due to the fact that the couplings we wrote preserve a parity~\cite{Gripaios:2009pe} under which both $\eta$ and $\widetilde{T}_5$ are odd: this parity can be broken if a mixing of the right-handed top to the $\widetilde{T}_5$ is added, thus inducing mass mixing and couplings of $\eta$ to all top partners~\cite{Serra:2015xfa}. 
This will also induce a coupling of $\eta$ to a pair of tops, which is otherwise absent: for an in-depth discussion of the effect of this mixing, we refer the reader to ref.~\cite{Alanne:2018wtp}, while here we limit ourselves to the simpler case that preserves the parity.
Another interesting feature regards the mass ordering inside each multiplet:
the composite fermions that mix with the elementary tops receive additional mass contributions from symmetry breaking (analogous to Eq.~(\ref{eq:QSmass})).
For this reason, within the $5$-plet, the components of the SM doublet, $T$ and $B$, tend to be the heaviest, followed by $X_{2/3}$ whose mixings are suppressed by $\theta$, while $X_{5/3}$ and $\widetilde{T}_5$ remain degenerate and lighter than the others. The singlet $\widetilde{T}_1$, on the other hand, is the lightest state if $M_1 \ll M_5$.\footnote{We use here the notation of the multiplet components to indicate the mass eigenstates with largest superposition with them.}
It is important to identify the lightest states as they are most likely to be more copiously produced at colliders and thus first discovered (or more strongly constrained).
The final point we want to make regards the bottom quark: its mass is not generated from the matrix in Eq.~(\ref{mass-SU4-Sp4-b}). Thus, the model needs to be completed by the addition of a partner of the right-handed bottom, or via an effective operator coming from the strong dynamics~\cite{Cacciapaglia:2015dsa}. In either case, the coupling of the bottom to the strong dynamics is typically smaller than the ones of the top, and  we thus neglect this effect.
The features we listed here are rather general and also apply for other choices of the top partner representations, and typically even in more general set-ups, as the elementary fields may couple to more than one representation~\cite{Golterman:2017vdj}.

In the following, we will use the case of a $5$-plet and singlet to define benchmark models that can be matched to the simplified models introduced in the previous Section \ref{sec:simplified}.

\begin{itemize}
	\item $\bf  \widetilde{T}_5 \rightarrow t\ \eta$: 
	The top partner $\widetilde{T}_5$ is an ideal candidate for the simplified model in Section~\ref{sec:sm2}, as it only couples to the singlet $\eta$ and it has a 100\% BR in $t\ \eta$. Thus, the decay constant $f_\eta$ defined in Eq.(\ref{eq:etaLag}) is equal to $f$, the decay constant of the Higgs.
	This property is a consequence of our choice to couple the right-handed top to the singlet only, thus preserving a parity associated to the pNGB $\eta$.
	The couplings of $\eta$ to SM particles also depend on this choice: in fact, the pre-Yukawas we write do not generate a coupling to top nor bottom quarks.
	Thus, $\eta$-parity is only broken by the Wess-Zumino-Witten (WZW) term, which provides couplings to the EW gauge bosons \cite{Galloway:2010bp,Arbey:2015exa}:
	\begin{equation} \label{eq:WZWsu(4)}
	\mathcal{L}_{\rm WZW} = \frac{d_\mathcal{F} ~c_\theta}{16 \sqrt{2} \pi^2 f} \eta \left( g^2 W_{\mu \nu} \tilde{W}^{\mu \nu} - {g'}^2 B_{\mu \nu} \tilde{B}^{\mu \nu} \right)\,,
	\end{equation}
	where $d_\mathcal{F}$ is the dimension of the representation of the underlying fermion under the confining gauge group.
	Note that additional couplings to the SM fermions may be generated by higher order operators, examples of which can be found in refs~\cite{Galloway:2010bp,Arbey:2015exa}: however, such couplings are small as they come from operators that do not generate the mass of the light fermions, thus inducing $C_f^\eta \ll 1$ (c.f. Eq.(\ref{eq:etaLag})). From the  WZW term, we see that the singlet $\eta$ can only decay to the final states $W^+ W^-$, $ZZ$ and $Z\gamma$, with rates shown in Fig.~\ref{fig:etaBRs}. \footnote{The WZW term in Eq.(\ref{eq:WZWsu(4)}) applies to all models with coset $SU(4)/Sp(4)$, however it may be different in other cosets.}
	
	Adding an $\eta$-parity violating coupling of the right-handed top to the $5$-plet would both induce couplings $\eta \bar{t} t$ and a mixing of $\widetilde{T}_5$ to the top~\cite{Serra:2015xfa}, thus this case would match the simplified model in Section~\ref{sec:sm1}. Finally, we remark that the presence of top partners that decay exclusively into $\eta$ also appears for other choices of top partner representations: for instance, for right-handed top into a $5$-plet and left-handed top into a $10$-plet \cite{Alanne:2018wtp}. In other cases when a $\eta \bar{t} t$ coupling is inevitable, e.g. when both tops are into a $10$-plet \cite{Alanne:2018wtp}, this state is absent. Finally, we remark that in some models there might also be a bottom partner that decays exclusively in $\widetilde{B} \to b\ \eta$.

	\begin{table}[tb]
		\renewcommand{\arraystretch}{1.}
		\begin{center}
			\begin{tabular}{ c| c c c |c c c |c }
				& $t(b)\ a$ & $T_1\ a$ &$T_2\ a$  & $b(t)\ W$ & $B(T_1)\ W$ & $X_{5/3}(T_2)\ W$ & $\widetilde{T}_5\ \eta$ 
				\\
				\hline
				$T_1$ & $0.45$ & $-$ & $-$   & $2 \cdot 10^{-3}$ & $-$ & $-$ & $-$
				\\
				$T_2$ & $0.03$ & $1 \cdot 10^{-3}$ & $-$  & $0.03$ & $-$ & $-$ & $-$ 
				\\
				$T_3$ & $0.15$ & $2 \cdot 10^{-3}$ & $2 \cdot 10^{-4}$   & $6 \cdot 10^{-3}$  & $-$ & $7 \cdot 10^{-3}$  &$0.04$ 
				\\
				$B$ & $0.19$ & $-$ & $-$   &  $6 \cdot 10^{-3}$ &  $0.76$ &  $0.05$   & $-$  
				\\
				\hline
				\hline
				& $t\ Z$ & $T_1\ Z$ & $T_2\ Z$ & $t\ h$ & $T_1\ h$ & $T_2\ h$ &
				\\
				\hline
				$T_1$ & $0.14$ & $-$ & $-$ & $0.41$ & $-$ &  $-$ &
				\\
				$T_2$ & $0.24$ & $0.29$ & $-$ & $0.23$ & $0.17$ & $-$ &
				\\
				$T_3$ & $2 \cdot 10^{-3}$ & $0.31$ & $0.04$ & $9 \cdot 10^{-3}$ & $0.39$ & $0.04$ &
				\\
				$B$ & $-$ &$-$ & $-$      & $-$ & $-$ & $-$ &
				\\
			\end{tabular}
		\end{center}
		\caption{Branching ratios of the VLQs in the benchmark scenario Bm1 for a fixed value of the $a$ mass, $m_a=15$ GeV, while $m_\eta= 100$ GeV. 
			The absence of number for a given BR indicates that the decay channel is not kinematically allowed or the corresponding tree-level coupling vanishes.
			The parenthesis in the first row refer to the decay channels of the VLQ $B$ while the others channels corresponds to the VLQs $T_{1,2,3}$.}
		\label{table-Bm1}
	\end{table}

	\begin{figure}[tb]
		\centering
		\begin{tabular}{cc}
			\includegraphics[width=0.48\textwidth]{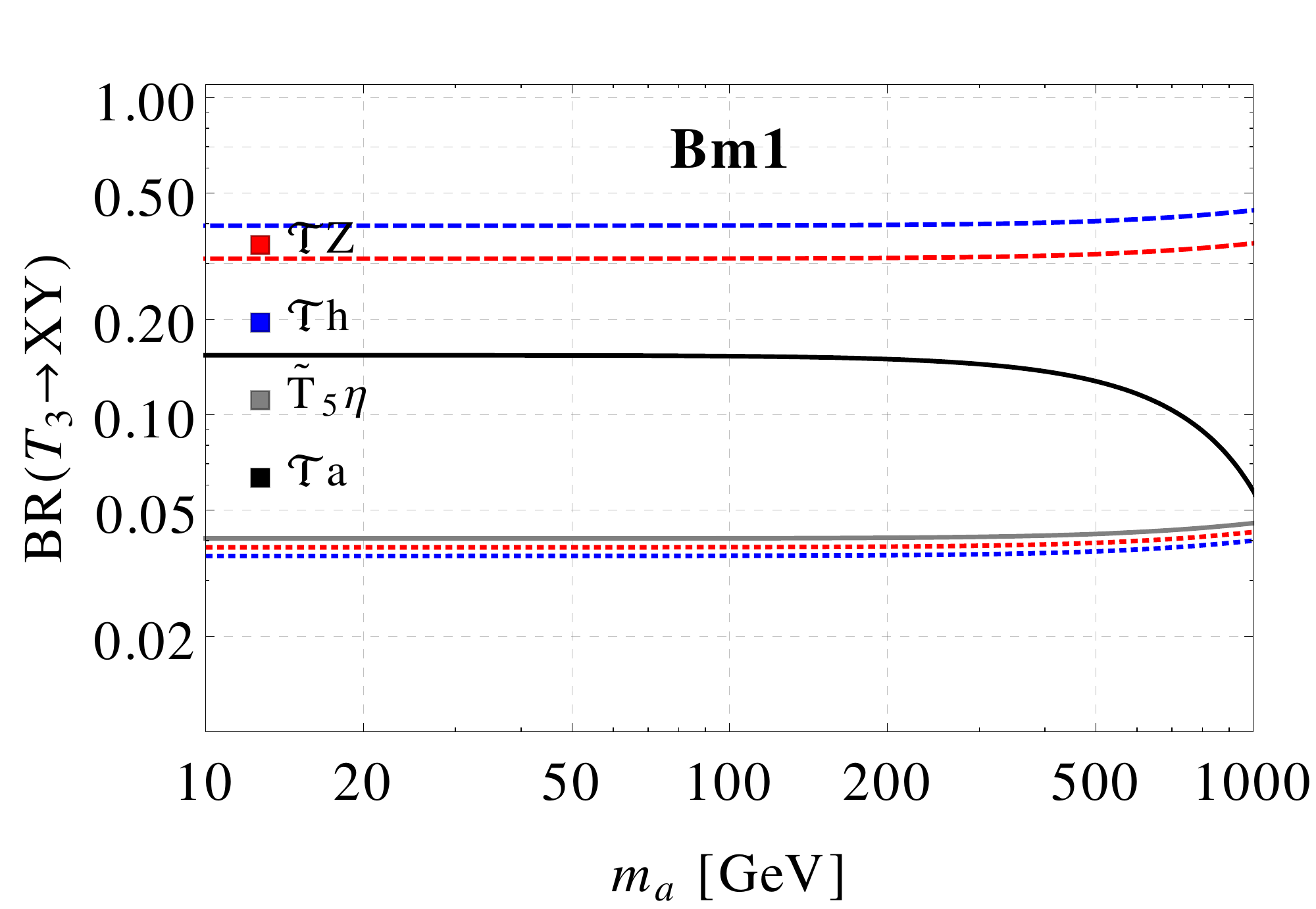} & \includegraphics[width=0.48\textwidth]{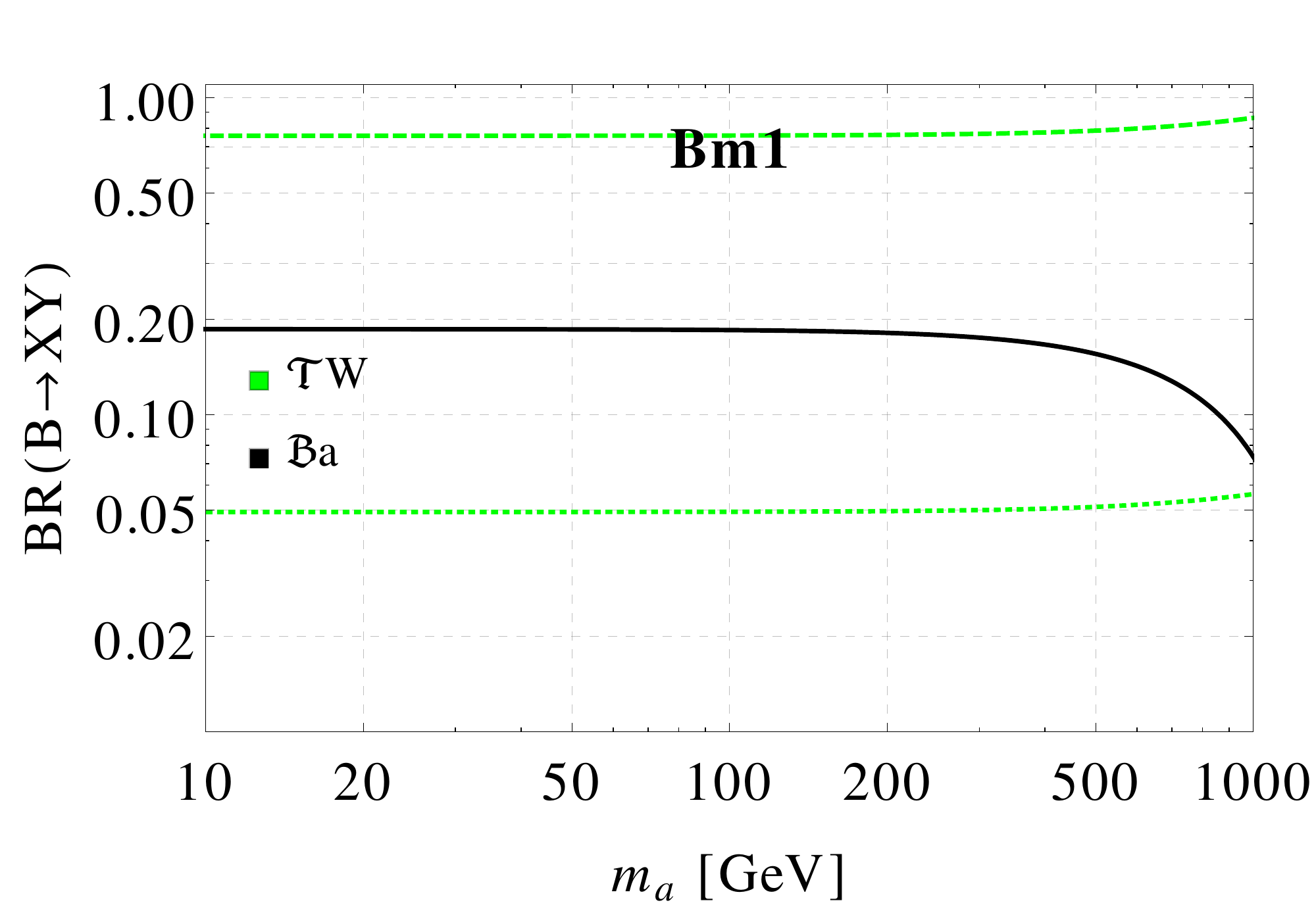}
		\end{tabular} 
		\caption{\label{fig:benchTta} 
			Branching ratios for heavier VLQs as a function of the mass $m_a$ in the benchmark model Bm1 defined in Eqs.~(\ref{Bm1-def-1}) and (\ref{Bm1-def-2}).
			The continuous, dashed and dotted lines correspond respectively to $\mathfrak{T}=\{t,T_1,T_2\}$ or $\mathfrak{B}=\{b,B\}$.
		}
	\end{figure}

	\item $\bf T\rightarrow t\ a$: The singlet pNGB $a$ derives from a spontaneously broken $U(1)$ global symmetry, that is always present in models of fundamental PC with two representations. As it can be seen in Eq.(\ref{mass-SU4-Sp4}), all the top partners that mix with the SM fermions  have a coupling to the singlet $a$. Furthermore, as discussed in ref.~\cite{Belyaev:2016ftv}, the mass of $a$ only comes from explicit mass terms for the underlying fermions, thus it can be as light as possible. In the following, we will use the models M8 and M9, as defined in ref.~\cite{Belyaev:2016ftv}, as benchmark models, referring the reader to that reference for all details about the models (we just recall that we define $f_a$ following the convention of ref.~\cite{Cacciapaglia:2017iws}). The main differences between the two models lie in the decay rates of the singlet $a$, and in the different charges $\xi_{1,5}$ that determine the couplings of $a$ to the top partners.
	
	To guarantee that the lightest partner is one that decays into $a$, we choose benchmark values of the parameters of the model such that the singlet is the lightest, i.e. $M_1 \ll M_5$. We  focus on the model M8 (which has larger couplings to $a$), and define the Benchmark model 1 (Bm1) according to
	the specific values of the input parameters listed below:
	\begin{equation} 
	\mbox{Bm1:}\qquad \begin{array}{c}
	M_1 = 600~\mbox{GeV}\,, \quad M_5 = 1.2~\mbox{TeV}\,, \quad f = 1~\mbox{TeV}\,, \\
	y_{1L} = y_{5L} = 1\,, \quad y_{1R} = 0.87\,, \quad y_{5R} = 1.02\,, \\
	\xi_L = \xi_R = -1.58\,, \quad f_a = 2.8~\mbox{TeV}\,; 
	\end{array}
	\label{Bm1-def-1}
	\end{equation}
	which reproduce the correct value of the top mass.\footnote{While the same masses and pre-Yukawa couplings can be chosen for M9, the values of the charges and decay constant are different: $\xi_L = \xi_R = 0.23\,, \;\; f_a = 1.2~\mbox{TeV}$.}
	The resulting spectrum of VLQs reads:
	\begin{equation} 
	\begin{array}{c}
	M_{T_1} = 1~\mbox{TeV}\,, \quad M_{\widetilde{T}_5} = M_{X_{5/3}} = 1.2~\mbox{TeV}\,, \quad M_{T_2} = 1.23~\mbox{TeV}\,, \\
	M_B = 1.56~\mbox{TeV}\,, \quad M_{T_3} = 1.57~\mbox{TeV}\,.
	\end{array}
	\label{Bm1-def-2} 
	\end{equation}
	The BRs for the lightest VLQ $T_1$ are reported and discussed in Section~\ref{sec:sm1}, so here we will focus on the heavier states. In Fig.~\ref{fig:benchTta} we show the BRs of the heaviest states $T_3$ and $B$ as a function of the mass $m_a$. We see that in both cases sizeable BRs in the final state $t\ a$ are present, of the order of 15-20\%, while the main rates involve the lightest VLQ $T_1$ and a SM boson. This example shows the importance of chain decays for the searches of heavier states together with the final state containing the new pNGB. For the intermediate mass state $T_2$, the channel $ta$ only amounts to a few percent, while the main channels involve equally $T_1$ and the top quark. Table~\ref{table-Bm1} reports the BRs for all the VLQs in the spectrum for fixed value of $m_a = 15$~GeV, as a reference.

	\begin{table}[tb]
		\renewcommand{\arraystretch}{1.}
		\begin{center}
			\begin{tabular}{ c| c c c |c c c | c }
				& $t(b)\ a$ & $T_1\ a$ &$T_2\ a$  & $b(t)\ W$ & $B (T_1)\ W$ & $X_{5/3}(T_2)\ W$  & $\widetilde{T}_5\ \eta$
				\\
				\hline
				$T_1$ & $5 \cdot 10^{-4}$ & $-$ & $-$  & $0.05$ & $-$ & $-$ & $-$
				\\
				$T_2$ & $0.10$ & $2.10^{-8}$ & $-$  & $0.08$ & $-$ & $-$ & $-$ 
				\\
				$T_3$ & $0.08$ & $1 \cdot 10^{-5}$ & $5 \cdot 10^{-5}$    & $0.08$  & $0.19$ & $0.12$  &$0.12$
				\\
				$B$ & $0.14$ & $-$ & $-$       &  $0.86$ &  $-$ &  $-$  & $-$ 
				\\
				\hline
				\hline
				& $Zt$ & $ZT_1$ & $ZT_2$ & $ht$ & $hT_1$ & $hT_2$ &
				\\
				\hline
				$T_1$ & $0.40$ & $-$ & $-$ & $0.54$ & $-$ &  $-$ &
				\\
				$T_2$ & $0.47$ & $-$ & $-$ & $0.35$ & $-$ & $-$
				\\
				$T_3$ & $0.04$ & $0.05$ & $0.11$ & $0.04$ & $0.08$ & $0.08$
				\\
				$B$ & $-$ &$-$ & $-$  & $-$ & $-$ & $-$
				\\
			\end{tabular}
		\end{center}
		\caption{Same as in Tab.~\ref{table-Bm1} but for the benchmark model Bm2.}
		\label{table-Bm2}
	\end{table}

	\begin{figure}[tb]
		\centering
		\begin{tabular}{cc}
			\includegraphics[width=0.48\textwidth]{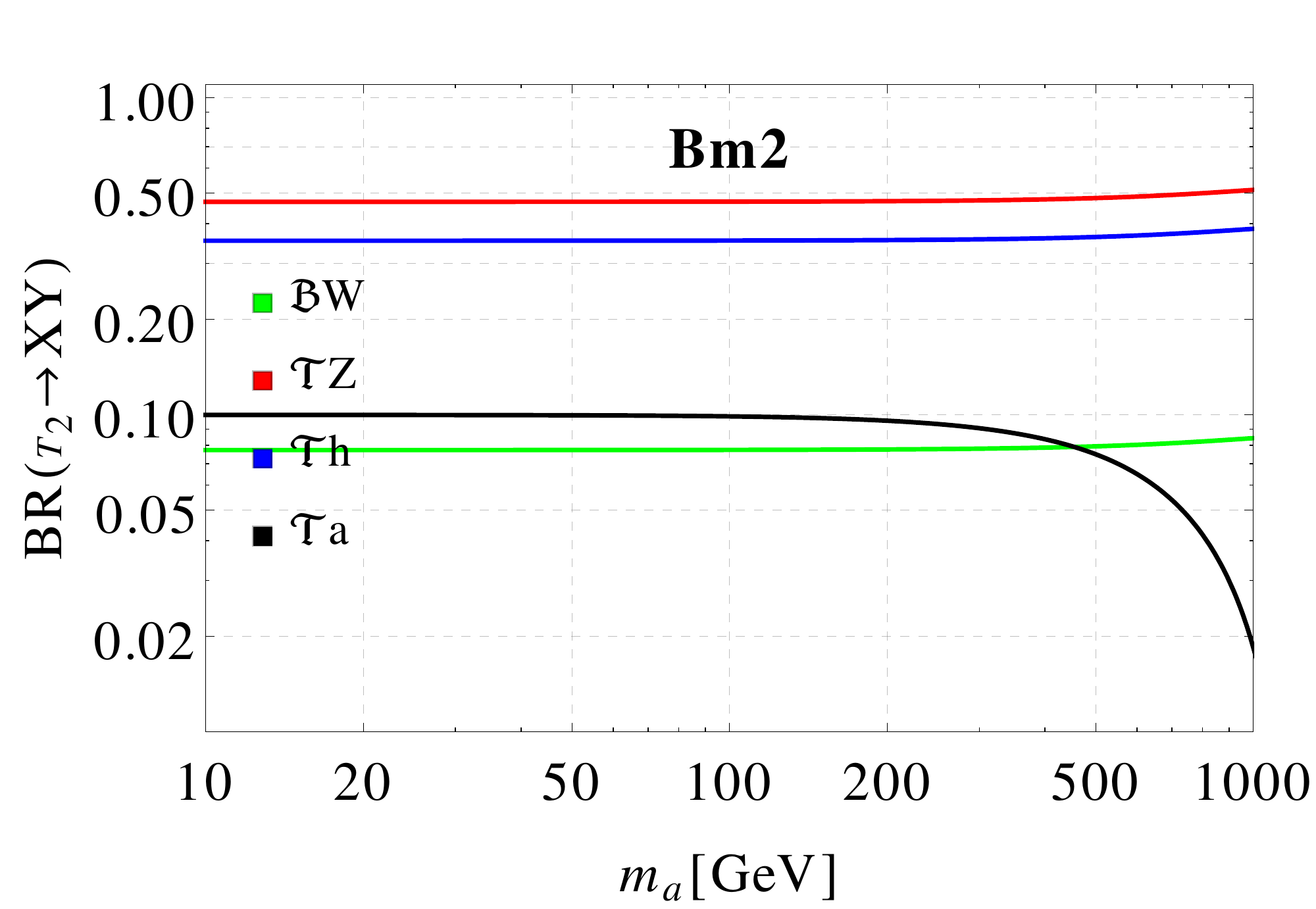} & \includegraphics[width=0.48\textwidth]{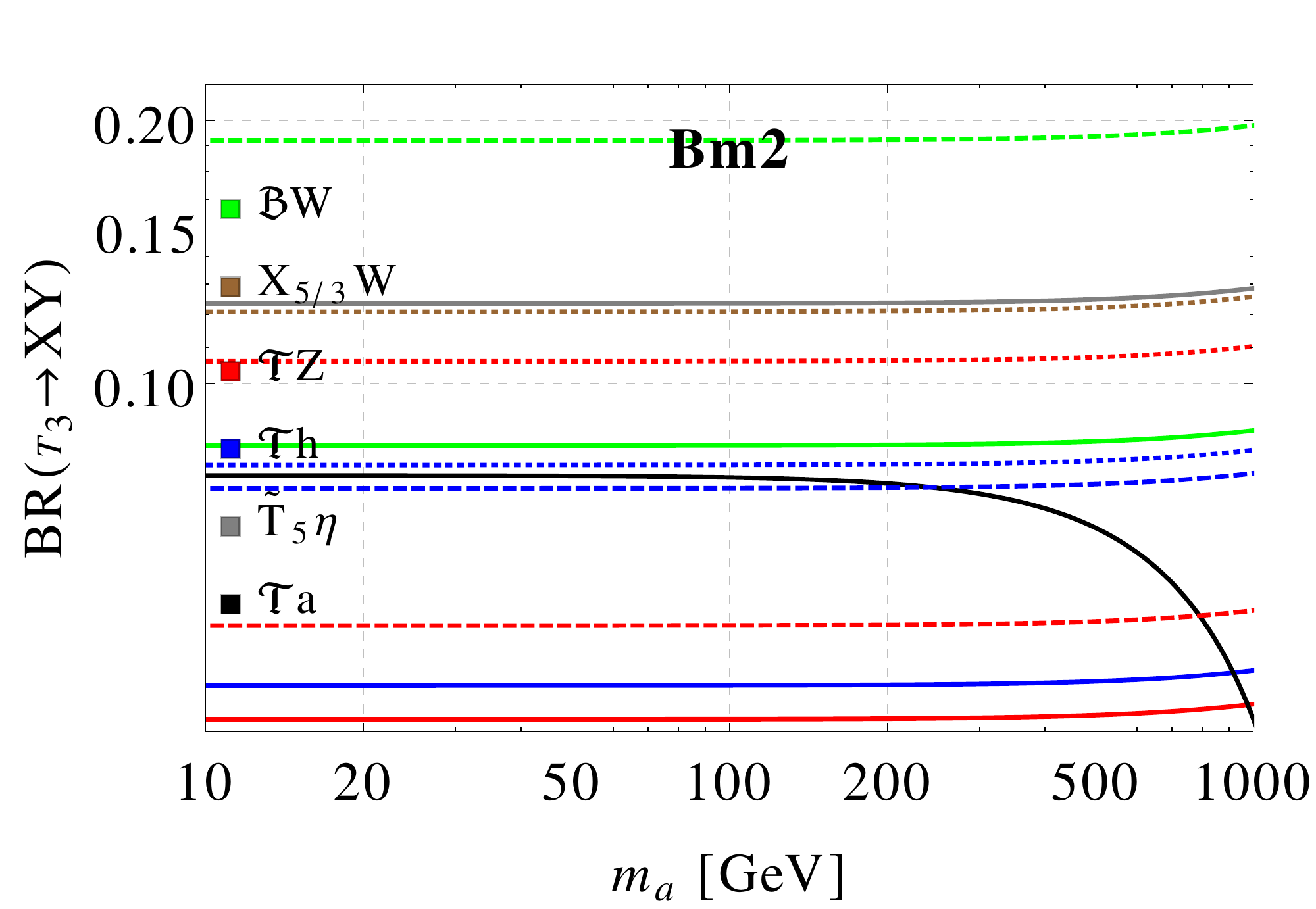}
		\end{tabular} 
		\caption{\label{fig:benchBba} 
			Branching ratios for heavier $T_{2,3}$ VLQs as a function of the mass $m_a$
			in the benchmark model Bm2 defined in Eqs~(\ref{Bm2-def-1}) and (\ref{Bm2-def-2}).
			The continuous, dashed and dotted lines correspond respectively to $\mathfrak{T}=\{t,T_1,T_2\}$ or $\mathfrak{B}=\{b,B\}$.
		}
	\end{figure}

	\item $\bf B\rightarrow b\ a$:  As illustrated in the previous benchmark, if $B$ is heavy it will preferentially decay into a lighter VLQ, thus its phenomenology does not match that of the simplified scenario presented in Sec.~\ref{sec:sm1}. To obtain a new benchmark model, we lower the value of $M_5$ and reduce $y_{5L}$ in order to reduce the mass split between $B$ and the lighter VLQs. The benchmark parameters of Benchmark model 2 are:
	\begin{equation} 
	\mbox{Bm2:}\qquad \begin{array}{c}
	M_1 = 1.4~\mbox{TeV}\,, \quad M_5 = 1.3~\mbox{TeV}\,, \quad f = 1~\mbox{TeV}\,, \\
	y_{1L} =1.17\,, \quad y_{5L} = 0.46\,, \quad y_{1R} = y_{5R} = 1.2\,, 
	\end{array} 
	\label{Bm2-def-1}
	\end{equation}
	which again reproduce the correct value of the top mass, together with the following VLQ spectrum 
	\begin{equation} \begin{array}{c}
	M_{T_1} = 1.30~\mbox{TeV}\,, \quad M_{\widetilde{T}_5} = M_X = 1.3~\mbox{TeV}\,, \quad M_{T_2} = 1.37~\mbox{TeV}\,, \\
	M_B = 1.38~\mbox{TeV}\,, \quad M_{T_3} = 1.85~\mbox{TeV}\,.
	\end{array} 
	\label{Bm2-def-2}
	\end{equation}
	The couplings of $a$ are calculated for model M8 of ref.~\cite{Belyaev:2016ftv}.
	The decays of $B$ are described in Section~\ref{sec:sm1}, while the lightest $T_1$ decays in the standard channels. Decays into $t\ a$ appear for the heavier $T_{2,3}$, whose BRs are shown in Fig.~\ref{fig:benchBba}. Finally in Table~\ref{table-Bm2} we report the BRs of the whole spectrum for $m_a = 15$~GeV, as a reference.

\end{itemize}

\subsection{The SU(4)/Sp(4)$\times$SU(6)/SO(6) scenario} 
\label{sec:SU6}

Top partners are, by definition, charged under QCD such that the underlying theory should contain coloured fundamental fermions.
This leads to a new sector that, upon condensation, contains coloured pNGBs. 
While strong bounds apply  from QCD production at the LHC, they may still be lighter than the top partners and thus appear in their decays.
For models with the Higgs coset $SU(4)/Sp(4)$, it has been shown that the coloured underlying fermions belong to a real representation of the confining gauge group~\cite{Barnard:2013zea,Ferretti:2013kya},
thus an $SU(6)/SO(6)$ pattern of symmetry breaking takes place~\cite{Cacciapaglia:2015eqa}: the theory, therefore, contains 20 additional pNGBs transforming as $8_0+6_{4/3}+\overline{6}_{-4/3}$ under $SU(3)_c \times U(1)_Y$.
The charged colour sextet plays a special role, as it is the only pNGB that can give non-standard decay channels for the exotic charge $X_{5/3}$, thus we will focus on this channel here (matching the simplified model in Section~\ref{sec:sm3}).
The corresponding coupling is given by:
\begin{equation}
{\cal L}_{\pi_6}= i \pi_6 ~\dfrac{y_{5L}f }{2 f_\chi} ~ \overline{X}_{5/3} b_L +h.c.
\label{lagrangian-pi6-couplings}
\end{equation}
where $\pi_6 \equiv \pi_6^{ a} \lambda_S^{a}$ and $f_\chi$ is the decay constant associated to the condensate in the  new sector $SU(6)/SO(6)$.  
Note that $f_\chi$ corresponds to the decay constant $f_{\pi_6}$ defined in Sec.~\ref{sec:sm3} and its value can be determined on the lattice \footnote{See ref.\cite{Ayyar:2017qdf} for results in a different model based on $SU(5)/SO(5)$, and ref.~\cite{Bennett:2017kga} for preliminary results for the model of ref.~\cite{Barnard:2013zea}.}, or be estimated based on the maximally attractive channel hypothesis \cite{Raby:1979my}: the latter gives $f/f_\chi=0.38$ for model $M_8$ and $f/f_\chi=2.3$ for model $M_9$~\cite{Ferretti:2016upr,Belyaev:2016ftv}.

\begin{figure}[tbh]
	\centering
	\begin{tabular}{cc}
		\includegraphics[width=0.48\textwidth]{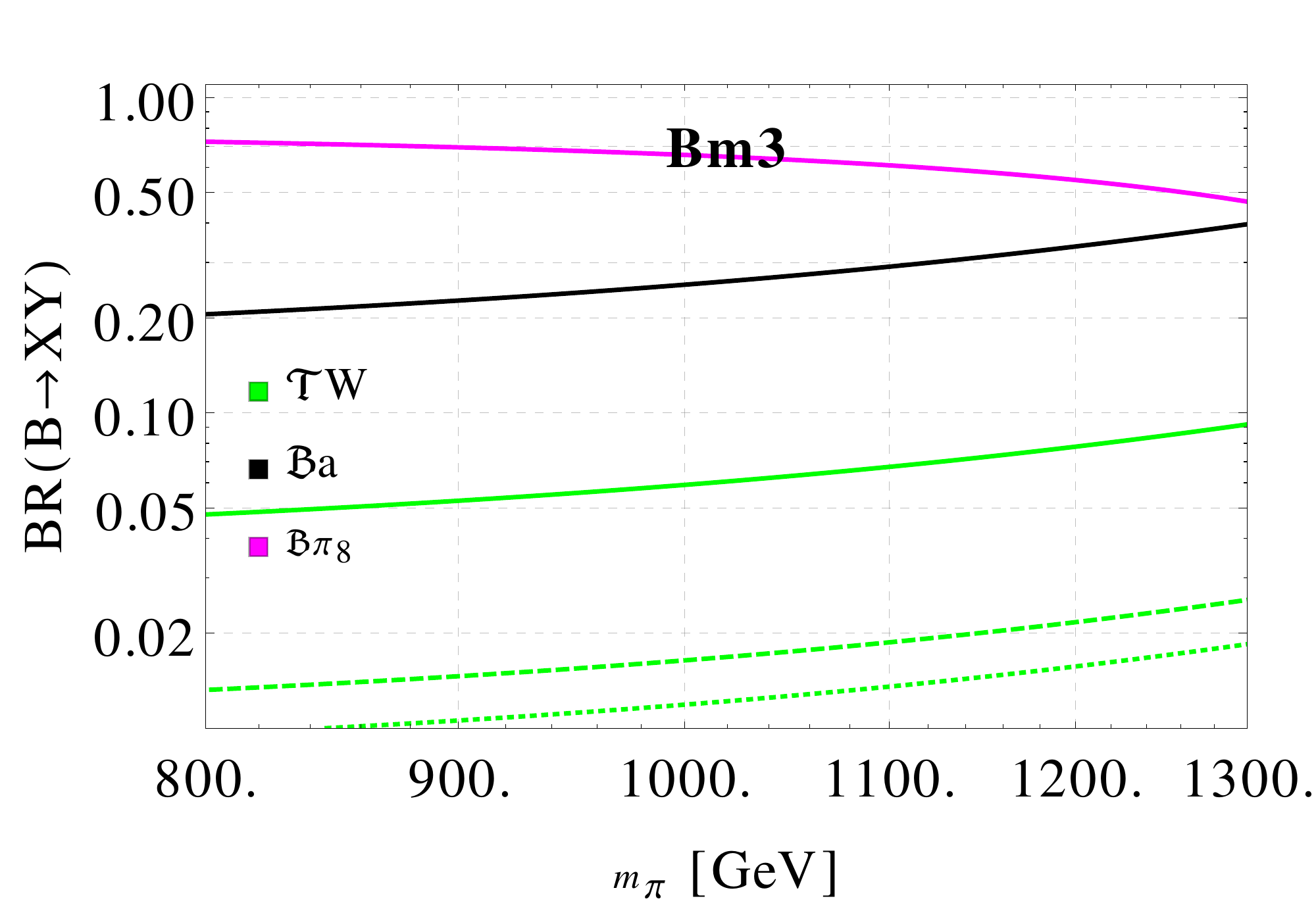} & \includegraphics[width=0.48\textwidth]{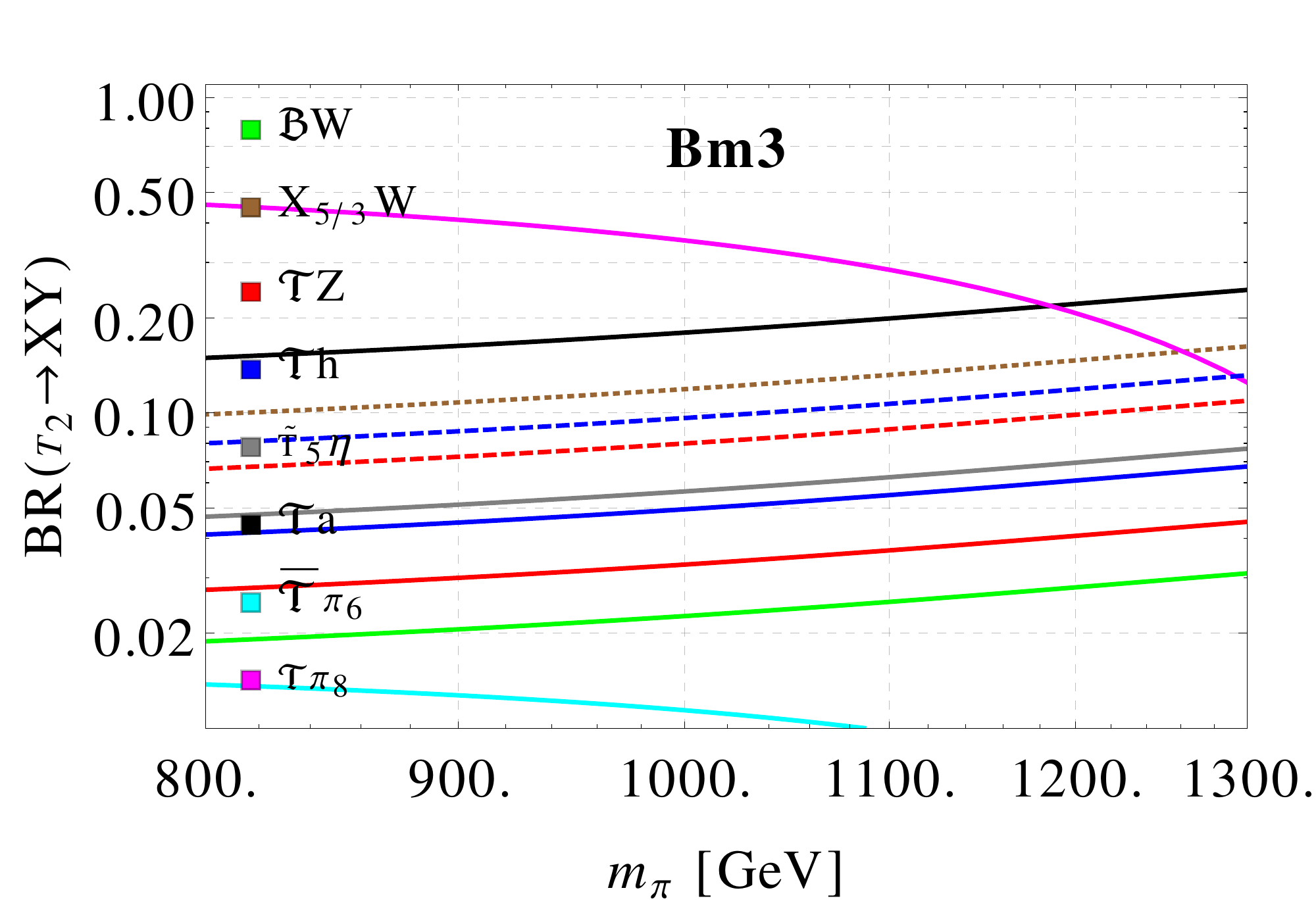}
	\end{tabular}
	\caption{\label{fig:benchXpi6} Branching ratios of $B$ and $T_2$ as a function of the mass of the coloured pNGBs, $m_\pi = m_{\pi_6} = m_{\pi_8}$, for the benchmark model Bm3 detailed in the text.
		The continuous, dashed and dotted lines correspond respectively to $\mathfrak{T}=\{t,T_1,T_2\}$ or $\mathfrak{B}=\{b,B\}$.}
\end{figure}

To ensure that $X_{5/3}$ is one of the lightest top partners, it is enough to consider $M_1>M_5$, without any further assumption on the pre-Yukawas.
We further focus on model $M_9$, which has a smaller value for $f_\chi$, in order to maximise the BR $X_{5/3} \to  \overline{b}\pi_6$ with respect to the standard one  $X_{5/3} \to t W^+$.
The benchmark model we consider here, therefore, is defined by the following choice of parameters:
\begin{equation} 
\mbox{Bm3:} \qquad \begin{array}{c}
M_1 = 1.4~\mbox{GeV}\,, \quad M_5 = 1.3~\mbox{TeV}\,, \quad f = 1~\mbox{TeV}\,, \\
y_{1L} =1\,, \quad y_{5L} = 1.2\,, \quad y_{1R} = 1.1\,, \quad y_{5R} = 1.05\,,  \\
\xi_L = \xi_R = 0.23\,, ~~\quad f_a = 1.2~\mbox{TeV}\,,
\end{array} 
\end{equation}
which again reproduces the correct value of the top mass, together with the following VLQ spectrum 
\begin{equation} 
\begin{array}{c}
M_{T_1} = 1.30~\mbox{TeV}\,, \quad M_{\widetilde{T}_5} = M_X = 1.3~\mbox{TeV}\,, \quad M_{T_2} = 1.68~\mbox{TeV}\,, \\
M_B = 1.77~\mbox{TeV}\,, \quad M_{T_3} = 1.85~\mbox{TeV}\,.
\end{array} 
\end{equation}
The BRs for the lightest $T_1$ and $X_{5/3}$, which only involve the sextet, are described in Section~\ref{sec:sm3}, so here we focus on the heavier states.
For simplicity we assume that the two coloured pNGBs are degenerate, and show the BRs for the $B$ and heavier $T_2$ in Fig.~\ref{fig:benchXpi6}, while numerical values for fixed masses $m_{\pi_6} = m_{\pi_8} = 800$~GeV are show in Table~\ref{table-Bm3}.
Remarkably, the main decay mode involves the colour octet $\pi_8$, which will decay dominantly into $t\bar{t}$, with subleading rates in two gluons and in gluon-photon pair.

\begin{table}[tb]
	\renewcommand{\arraystretch}{1.}
	\begin{center}
		\begin{tabular}{ c| cc| c c c |c c c  }
			& $\overline{t}\ \pi_6$  & $t(b)\ \pi_8$ & $t(b)\ a$ & $T_1\ a$ &$T_2\ a$  & $b(t)\ W$ & $B(T_1)\ W$ & $X_{5/3}(T_2)\ W$  
			\\
			\hline
			$T_1$ & $0.14$ & $5 \cdot 10^{-4}$ & $3 \cdot 10^{-4}$  & $-$ & $-$ & $0.05$ & $-$& $-$
			\\
			$T_2$ & $0.01$ & $0.46$ & $0.15$  & $2 \cdot 10^{-4}$ & $-$ & $0.02$ & $-$ & $0.10$
			\\
			$T_3$ & $1 \cdot 10^{-3}$ & $0.55$ & $0.15$    & $1 \cdot 10^{-4}$  & $1 \cdot 10^{-3}$ & $7 \cdot 10^{-4}$  &$4 \cdot 10^{-4}$ & $0.07$
			\\
			$B$ & $-$ & $0.72$ & $0.21$       &  $-$ &  $-$ &  $0.05$  & $0.01$ & $0.01$
			\\
			\hline
			\hline
			& $\widetilde{T}_5\ \eta$ & & $t\ Z$ & $T_1\ Z$ & $T_2\ Z$ & $t\ h$ & $T_1\ h$ & $T_2\ h$ 
			\\
			\hline
			$T_1$ & $-$ &  & $0.43$ & $-$ & $-$ &  $0.38$ & $-$ & $-$
			\\
			$T_2$ & $0.05$ &  & $0.03$ & $0.07$ & $-$ & $0.04$ & $0.08$ & $-$
			\\
			$T_3$ & $0.13$ &  & $4 \cdot 10^{-3}$ & $0.02$ & $0.05$ & $7 \cdot 10^{-6}$ & $0.02$ & $3 \cdot 10^{-4}$
			\\
			$B$ & $-$ & & $-$  & $-$ & $-$ & $-$ &$-$ & $-$
			\\
		\end{tabular}
	\end{center}
	\caption{Branching ratios of the VLQs in the benchmark scenario Bm3 for  fixed values of the coloured PNGBs masses, $m_{\pi_6}=m_{\pi_8}=800$ GeV while $m_a=m_\eta=100$ GeV. 
		The other conventions are the same as in Tab.~\ref{table-Bm1}.}
	\label{table-Bm3}
\end{table}

\subsection{The SU(5)/SO(5) scenario }
\label{sec:SU5}

A scenario with a larger pNGB sector is based on the $SU(5)/SO(5)$ coset, which is present in many models of PC with an underlying completion~\cite{Ferretti:2013kya} and
yields $14$ pNGBs with quantum numbers $3_{\pm 1}+3_0+2_{\pm 1/2}+1_0$ under $SU(2)_L \times U(1)_Y$.
Thus, in addition to the Higgs doublet and the singlet $\eta$, the model contains three $SU(2)_L$-triplets: 
$\Phi_0=\begin{pmatrix}
\phi_0^0 & \phi_0^\pm
\end{pmatrix} $, 
$\Phi_{+1}=\begin{pmatrix}
\phi_{1}^0 & \phi_1^+ & \phi_1^{++}
\end{pmatrix} $
and $\Phi_{-1} = \Phi_{+1}^\dagger$, with hypercharge $0$ and $+1$ and $-1$ respectively.   Note that, in general, the two charged scalars, $\phi_0^\pm$ and $\phi_1^\pm$, will mix, while the $\phi_0^0$ will mix with the imaginary part of the complex scalar $\phi_1^0$ (and with the singlet $\eta$), which are all pseudo-scalars. 
This scenario provides the same VLQ exotic decays as in the previous sections (i.e., final states containing $a$, $\eta$, and coloured pNGBs), but it also offers new decay channels involving the pNGB triplets.
In this section, we will focus on the decays to the charged scalars $\phi^+_{0,1}$ and $\phi^{++}_1$.

Following the explicit model in ref.~\cite{Ferretti:2014qta}, that corresponds to model M4~\cite{Belyaev:2016ftv}, we assume that the top partners belong to the fundamental representation of $SO(5)$, which decomposes as $2_{7/6}+2_{1/6}+1_{2/3}$  under the EW symmetry
\footnote{This is the same decomposition as for $5$-plet of $Sp(4)$ due to the isomorphism between the two groups.
	However, in the $SU(4)/Sp(4)$ case, a linear coupling alone between the right-handed top quark and the $5$-plet was not possible as it leads to a massless top quark \cite{Alanne:2018wtp}. } 
and may then couple to both the left and right-handed top quark.
In the $\psi_t = \{t,T,X_{2/3},\widetilde{T}_5\}$  and $\psi_b = \{ b, B\}$ bases (we borrow the same notation as from the previous section) we get the following matrices for the top and bottom sectors respectively (where the $U(1)$ singlet $a$ can be introduced in a straightforward way):
\begin{equation}
\bar{\psi}_{tR}
\begin{pmatrix}
0 & \sqrt{2} y_{5R} f s_\theta & \sqrt{2} y_{5R} f s_\theta &  2 y_{5R} f c_\theta
\\
2 y_{5L} f c^2_{\theta/2}   & M_5 & 0 & 0 
\\
-2y_{5L} f s^2_{\theta/2}  & 0 & M_5 & 0 
\\
\sqrt{2} y_{5L} f s_\theta   & 0 & 0 & M_5
\end{pmatrix} \psi_{tL}~,
\quad
\bar{\psi}_{bR}
\begin{pmatrix}
0 & 0
\\
2y_{5L} f  & M_5
\end{pmatrix} \psi_{bL}~,
\label{matrix-SU5-SO5}
\end{equation}
while for the exotic charged state, we have $M_{X_{5/3}}=M_5$.
The lightest top partner, therefore, is always $X_{5/3}$ as it does not receive any contribution to its mass from the Higgs.
Note that considering only one VLQ multiplet coupling to both left and right-handed top quarks leads to fewer parameters: one mass $M_5$ and two pre-Yukawa couplings $y_{5L}$ and $y_{5R}$.
As before, the mixing pattern of the simplified scenario is recovered for a small misalignment angle $\theta$, and we have the following identifications: 
\begin{equation}
M_{Q}=M_S=M_{5}~,
\qquad
y_{L}=2 y_{5L} c^2_{\theta/2}~,
\quad
y_R=2 y_{5R} c_\theta~, 
\qquad
y_{L}^\prime=-2 y_{5L}~,
\quad
y_R^\prime=-2 y_{5R}~.
\end{equation}
We will mainly focus on the exotic decays of the $X_{5/3}$ top partner, because it is the lightest state in the multiplet. The allowed decays are $X_{5/3} \to t\ \phi_{0,1}^+$ and $X_{5/3} \to b\ \phi_1^{++}$, together with the standard $X_{5/3} \to t\ W^+$.
The decay to the doubly-charged scalar is intriguing, but it does not yield truly new final states: the only decay generated by the WZW anomaly is $\phi_1^{++} \to W^+ W^+$, thus the final state of the exotic $X_{5/3}$ decay is $bW^+W^+$ like for the standard channel $X_{5/3} \to t\, W^+$ after the decay of the top quark (although the kinematics differ).
On the other hand, both singly charged scalars can decay to $\phi_{0,1}^+ \to t \bar{b}$ above the $t\bar{b}$ threshold via the PC mixing, and to $\phi_{0,1}^+ \to W^+ \gamma$ below threshold. Decays to $W^+ Z$ are always suppressed because the threshold is very close to the $t\bar{b}$ one, while below the $W$ mass the decays become more model dependent (a more detailed discussion can be found in Sec.~\ref{sec:sm4}).

We now discuss some numerical results to show if sizeable BRs to $X_{5/3} \to t\ \phi_{0,1}^+$ and $X_{5/3} \to b\ \phi_1^{++}$ can be achieved.

\begin{itemize}
	\item $\bf X_{5/3} \rightarrow t\ \phi^+ / b\ \phi^{++}$: 
	The exotic charge state $X_{5/3}$ couples to the charged pNGBs $\phi_{0,1}^+$ as well as to $\phi_1^{++}$.
	The corresponding couplings are given by:
	\bea
	{\cal L}_{\phi}&=&i \sqrt{2} y_{5L}  ~(\phi_0^+ s^2_{\theta/2}+\phi_1^+ c^2_{\theta/2})~\overline{X}_{5/3} t_L 
	+i y_{5R} s_\theta~  \phi_1^- ~ \overline{t}_R  X_{5/3} 
	\nonumber
	\\
	&&+i 2 y_{5L} \phi_1^{++} ~\overline{X}_{5/3} b_L  +h.c.
	\eea
	As the final states of the decays of the two charged scalars are the same, they cannot be distinguished except for the different kinematics due to their mass. In the following, we work under the assumption that all the non-Higgs pNGBs are degenerate, so it makes sense to consider them as a single particle. The couplings above thus match to the simplified model of Sec.~\ref{sec:sm4}, defined in Eqs~(\ref{eq:XphiLag}) and (\ref{eq:XphiLag1}), provided the identification $\phi^+ \to \phi^+_{0,1}$ and $\phi^{++} \to \phi^{++}_1$.
	We chose to focus on model M4 which has the smallest couplings to $a$, thus maximising the BRs into the pNGB triplets.
	The benchmark model we consider here, is defined by the following choice of parameters:
	\begin{equation} 
	\mbox{Bm4}:
	\qquad \begin{array}{c}
	M_5 = 1.3~\mbox{TeV}~, \quad f = 1~\mbox{TeV}\,, \quad y_{5L} =0.41\,, \quad y_{5R} = 0.7\,, \\
	\xi_L = \xi_R = -0.17~, \quad f_a = 2.0~\mbox{TeV}~,
	\end{array}
	\label{Bm5-def-1} 
	\end{equation}
	yielding the following VLQ spectrum 
	\begin{equation} 
	\begin{array}{c}
	M_{T_1} = 1.30~\mbox{TeV}\,, \quad  M_{X_{5/3}} = 1.3~\mbox{TeV}\,, \quad M_{T_2} = 1.51~\mbox{TeV}\,, \\
	M_B = 1.54~\mbox{TeV}\,, \quad M_{T_3} = 1.92~\mbox{TeV}\,.
	\end{array} 
	\label{Bm5-def-2} 
	\end{equation}
	The BRs for $X_{5/3}$ are characterised in Sec.~\ref{sec:sm4}. 
	Here we discuss in more details the benchmark scenario.
	As already mentioned, $X_{5/3}$ is always the lightest top partner, and we fix its mass (and consequently $M_5$) to the most conservative experimental bound of $1.3$~TeV.
	The two pre-Yukawa couplings $y_{5L}$ and $y_{5R}$ determine the correct value of the top mass and the BRs into the charged pNGBs:
	in order to maximise the latter, it is not enough to increase $y_{5L}$  as the mixing in Eq.~(\ref{matrix-SU5-SO5}) changes as well. 
	This fact is behind the choice of values in Eq.~(\ref{Bm5-def-1}).

	For the $X_{5/3}\rightarrow t\ \phi^+$ channel, the dominant contribution comes from the left-handed coupling involving $\phi_1^+$ as the other couplings are suppressed by the misalignment angle.
	Here we consider a special situation where the two charged scalars are degenerate and, thus, indistinguishable. In a more general scenario, the mixing between them will share the coupling of $\phi^+_1$ to the two mass eigenstates, thus potentially reduce the rate into this channel. Furthermore, if the masses are one below and one above the $t\bar{b}$ threshold, an interesting situation occurs where both $X_{5/3} \to t W^+ \gamma$ and $tt\bar{b}$ final states are present.

	\begin{table}[tb]
		\renewcommand{\arraystretch}{1.}
		\scriptsize
		\begin{center}
			\begin{tabular}{ c|ccc c| c c c| c  c}
				& $b(t)\ \phi_0^\pm $ & $B(T_1)\ \phi_0^\pm $  & $b(t)\ \phi_1^\pm $ & $B(T_1)\ \phi_1^\pm $  & $t(b)\ \phi_0^0$ & $\phi_0^0T_1$ &$T_2\ \phi_0^0$  & $t\ Re[\phi_1^0]$ & $t\ Im[\phi_1^0]$ 
				\\
				\hline
				$T_1$ & $4 \cdot 10^{-4}$ & $-$ & $4 \cdot 10^{-4}$ & $-$ & $6 \cdot 10^{-3}$ & $-$ & $-$ & $0.21$ & $0.16$
				\\
				$T_2$ & $0.23$ & $-$ & $0.23$ & $-$ & $0.11$ & $2 \cdot 10^{-4}$ & $-$ & $4 \cdot 10^{-3}$ & $5 \cdot 10^{-3}$
				\\
				$T_3$ & $5 \cdot 10^{-4}$ & $4 \cdot 10^{-3}$ & $5 \cdot 10^{-4}$ & $5 \cdot 10^{-3}$ & $7 \cdot 10^{-4}$ & $2 \cdot 10^{-3}$ & $7 \cdot 10^{-4}$ & $4 \cdot 10^{-4}$ & $4 \cdot 10^{-5}$
				\\
				$B$ & $8 \cdot 10^{-3}$ & $1 \cdot 10^{-4}$ & $9 \cdot 10^{-3}$ & $1 \cdot 10^{-4}$ & $0.15$ & $-$ & $-$ & $-$ & $-$
				\\
				\hline
				\hline
				& $T_1\ Re[\phi_1^0]$ & $T_1\ Im[\phi_1^0]$ &$T_2\ Re[\phi_1^0]$ & $T_2\ Im[\phi_1^0]$ & $at(b)$ & $T_1\ a$ & $T_2\ a$ & $t(b)\ \eta$ & $ T_1\ \eta$
				\\
				\hline
				$T_1$ & $-$ & $-$ & $-$ & $-$ & $7\cdot 10^{-33}$ & $-$ & $-$ & $0.03$ & $-$
				\\
				$T_2$ & $0.03$ & $0.03$ & $-$ & $-$ & $4 \cdot 10^{-3}$ & $4 \cdot 10^{-33}$ & $-$ & $0.09$ & $9 \cdot 10^{-4}$
				\\
				$T_3$ & $9 \cdot 10^{-3}$ & $6 \cdot 10^{-4}$ & $3 \cdot 10^{-3}$ & $4 \cdot 10^{-3}$ & $2 \cdot 10^{-3}$ & $1 \cdot 10^{-34}$ & $5 \cdot 10^{-6}$ & $0.17$ & $0.01$
				\\
				$B$ & $-$ & $-$ & $-$ & $-$ & $4 \cdot 10^{-3}$ &  $-$ & $-$  & $0.14$ & $-$
				\\
			\end{tabular}
			\begin{tabular}{ c|c|ccc|ccc|ccc}
				\hline\hline
				&  $ T_2\ \eta$ & $b(t)\ W$ & $B(T_1)\ W$ & $X_{5/3}\ W(\phi_1^{--})$ &~$t(b)\ Z$ & ~$T_1\ Z$ & ~$T_2\ Z$ & $t\ h$ & $T_1\ h$ & $T_2\ h$
				\\
				\hline
				$T_1$ & $-$ & $0.03$ & $-$ & $-$ & $0.37$ & $-$ &$-$ &  $0.18$ & $-$ &$-$
				\\
				$T_2$ & $-$  & $7 \cdot 10^{-3}$ & $-$ & $0.01$ & $0.15$ & $0.02$ & $-$ & $0.07$ & $1 \cdot 10^{-4}$ & $-$
				\\
				$T_3$ &$1\cdot 10^{-3}$  & $3 \cdot 10^{-3}$ & $0.22$ & $0.23$ & $2 \cdot 10^{-3}$ & $0.09$ & $0.13$ & $6 \cdot 10^{-4}$ & $0.06$ & $0.05$
				\\
				$B$ & $-$  & $0.34$ & $0.01$ & $0.11$ & $0.22$ & $-$ & $-$ & $-$ & $-$ & $-$
				\\
			\end{tabular}
		\end{center}
		\caption{Branching ratios of the VLQs in the benchmark scenario Bm4 for  fixed values of the  pNGB triplet masses, $m_{\phi}=m_a=m_\eta=100$ GeV.}
		\label{table-Bm4}
	\end{table}

	\begin{figure}[tbh]
		\centering
		\begin{tabular}{cc}
			\includegraphics[width=0.48\textwidth]{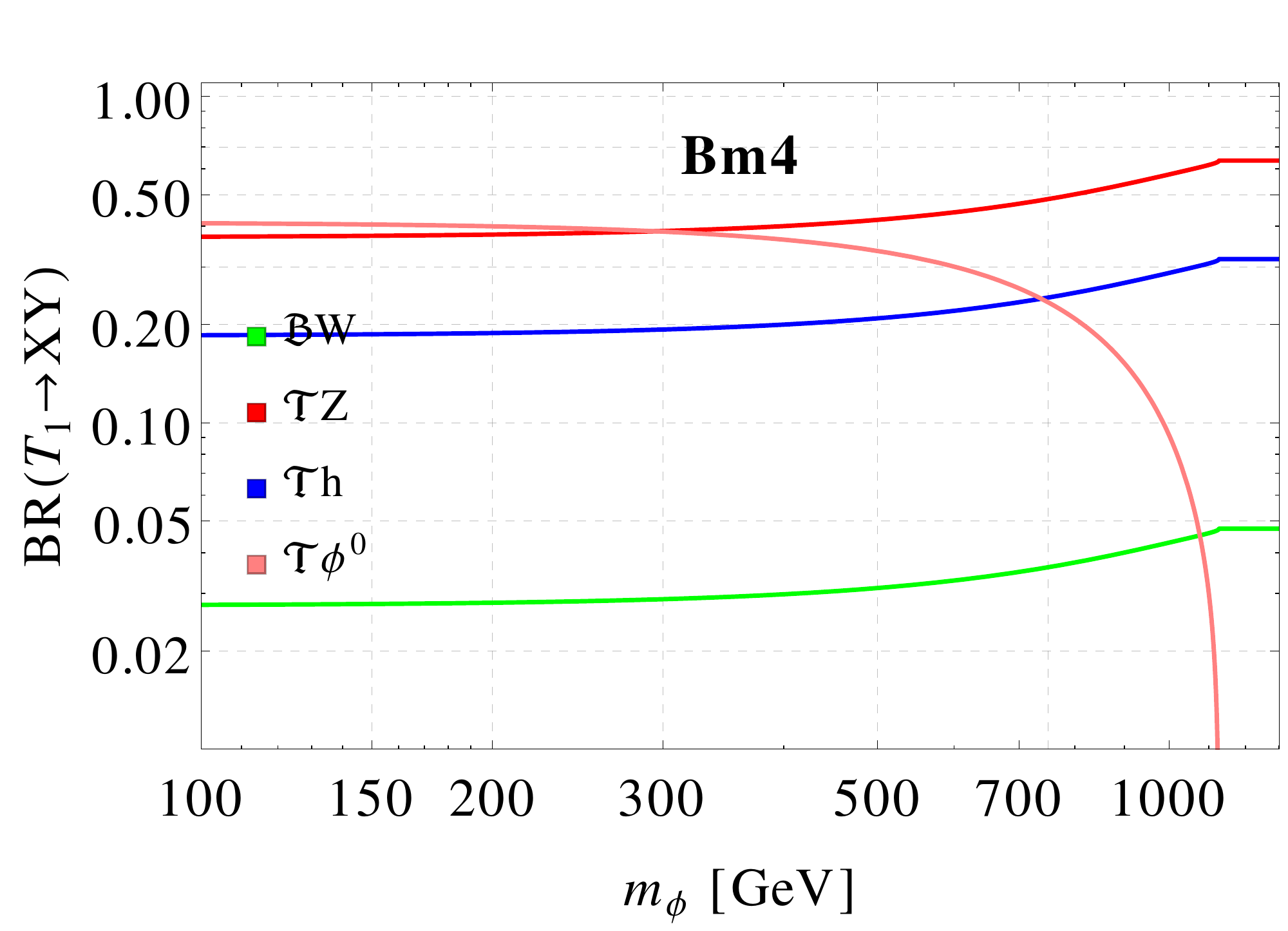} & \includegraphics[width=0.48\textwidth]{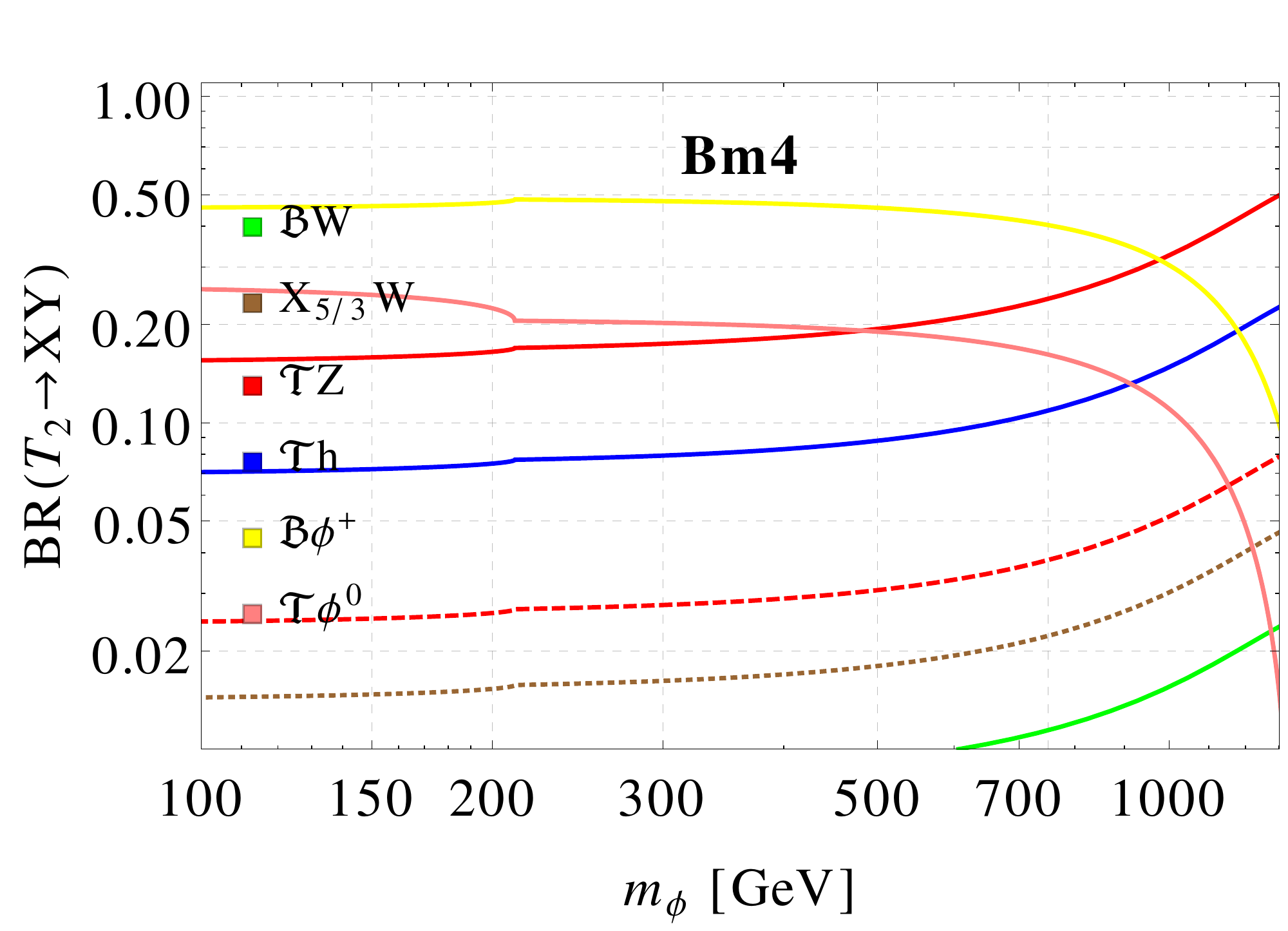} \\
			\includegraphics[width=0.48\textwidth]{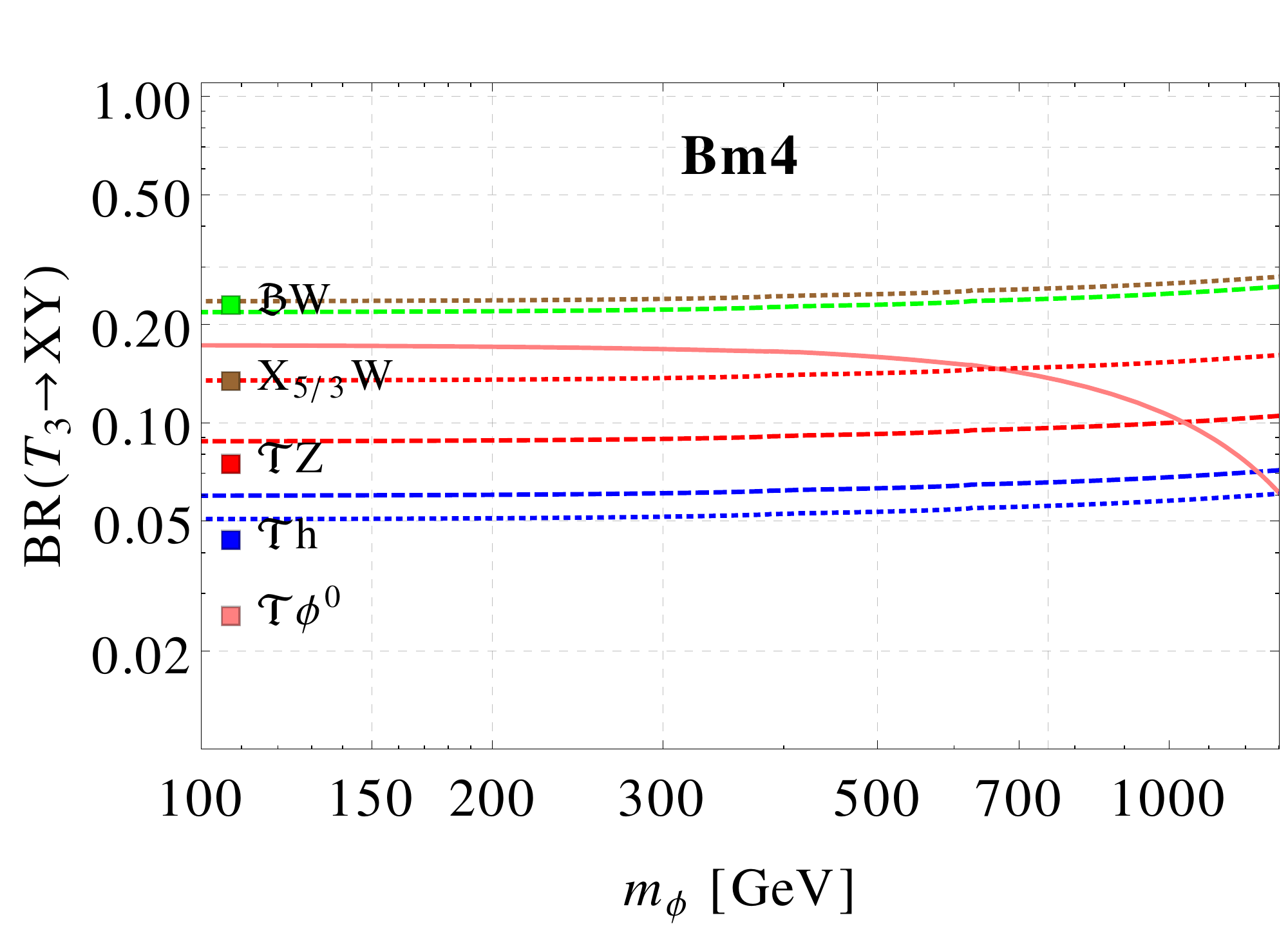} & \includegraphics[width=0.48\textwidth]{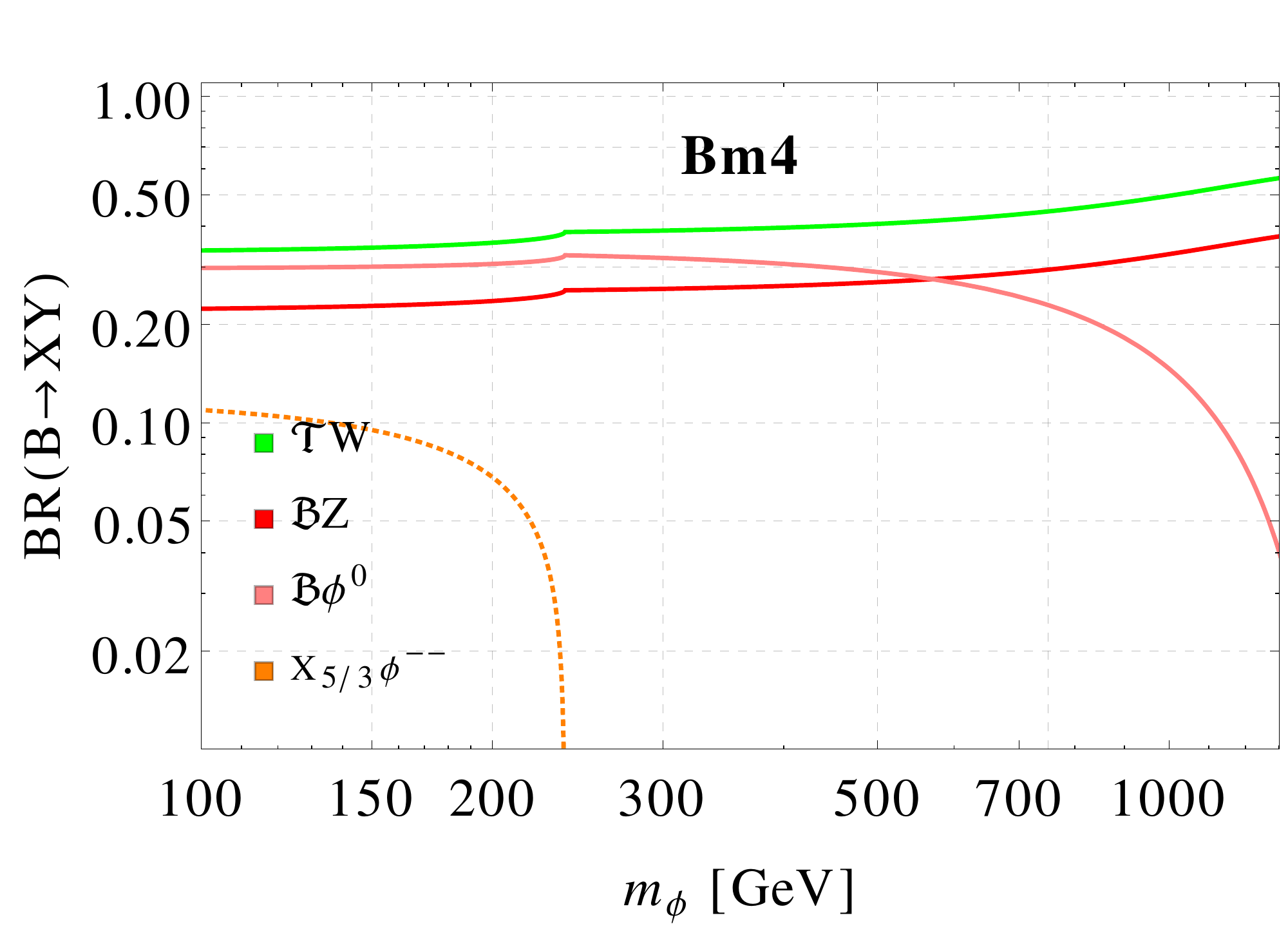}
		\end{tabular}
		\caption{Branching ratios of $T_1$, $T_2$, $T_3$ and $B$ as a function of the common mass of the  pNGB triplets $m_{\phi}$  for the benchmark model Bm4 defined in Eqs~(\ref{Bm5-def-1}) and (\ref{Bm5-def-2}).
			For simplicity, we assume that all components of the EW triplets have the same mass $m_\phi$.
			The continuous, dashed and dotted lines correspond respectively to $\mathfrak{T}=\{t,T_1,T_2\}$ and $\mathfrak{B}=\{b,B\}$ .
		}
		\label{branching-ratios-X53-phi}
	\end{figure}

	\item $\bf T \rightarrow b \phi^+$: 
	The charged pNGBs also couple to the charge $2/3$ VLQs $T_{1,2,3}$.
	Below  the $t\overline{b}$ threshold, the decay $T_i\rightarrow b\ \phi^+ \rightarrow b W^+ \gamma$ occurs and leads to a very interesting final state, which features a hard photon together with the standard charged-current final state. 
	Above the threshold, we have $T\rightarrow b\ \phi^+ \rightarrow b t \overline{b}$ with three bottom quarks in the final state when singly produced and six if pair-produced.
	
	As the parameter space is rather constrained due to the presence of few free parameters, we present results here for the same benchmark Bm4 defined above.
	The BRs for all the $T$-partners are shown in Fig.~\ref{branching-ratios-X53-phi}, while Table~\ref{table-Bm4} shows, as a references, the values of all channels for a fixed triplet mass of $100$~GeV.
	We see that the dominant decays are into standard channels, with a sizeable component into the neutral pNGBs. As they are taken degenerate here, we group them under a single channel $\phi_0 \to \eta/\phi_0^0/\phi_1^0/\phi_1^{0\ast}$: they will decay into a pair of gauge bosons via anomalies plus $t\bar{t}$ via the PC mixing, thus giving rise to exotic channels characterised by the simplified model of Section~\ref{sec:sm1} (except for the absence of decays into gluons).
	The intermediate state $T_2$, on the other hand, has sizeable decays into $b \phi^+$ (roughly 50\%), with the rest shared between standard channels and neutral pNGBs.
	For completeness we remark that the heaviest one, $T_3$, does not decay into any of the triplet pNGBs.

	Finally, we  report  in the bottom-right panel of Fig.~\ref{branching-ratios-X53-phi}  the BRs of the VLQ B (with mass $M_B=1.54$ TeV). Besides rather standard decays, it has a sizeable rate into $X_{5/3}\,\phi^{--}$. 
	The chain decay $X_{5/3} \to b \phi^{++}$ or $\to t W^+$, with $\phi^{\pm \pm} \to W^\pm W^\pm$, gives a final state with four $W$ bosons plus two $b$-quarks if $X_{5/3}$ is pair-produced.

\end{itemize}


\section{Conclusions}
\label{sec:concl}

The search for heavy VLQs, possibly partners of the top quark, continues to be one of the main physics
goals at the LHC. So far, however, the experimental efforts have been concentrated, and limited, to decays to one
massive electroweak boson ($W$, $Z$ and Higgs) plus a quark, mainly from the third generation.
The presence of additional decay channels would forcibly reduce the reach of these searches by adding different
final states for which searches are not optimised.

In this article we have proposed four simplified scenarios for additional decay modes for third generation partners,
focusing on fermions with charges $2/3$ ($T$), $-1/3$ ($B$) and $5/3$ ($X_{5/3}$).  
These modes are actually rather common in motivated underlying models for composite Higgs  with
partial compositeness.
First we consider decays into a light pseudo-scalar $a$ in addition to the standard ones, where $a$ can decay  
into a pair of gauge bosons via topological anomalies or into a pair of fermions via operators giving rise to the
fermion masses. In the underlying theories we consider, the pseudo-scalar is typically associated to an 
anomaly-free global $U(1)$ symmetry. Secondly, we consider a top partner decaying exclusively into a light
pseudo-scalar $\eta$, which further decays into electroweak gauge bosons. In the underlying models, $\eta$
originates as an additional pNGB of the Higgs coset. Thirdly, we consider decays of the charge $5/3$ partner
$X_{5/3}$ into charged coloured pNGBs. The latter originate from the sector of the underlying theory carrying
QCD colour. Finally, decays into additional un-coloured charged pNGBs are considered, also arising from the Higgs coset.

We provide simplified models that can be used for phenomenological studies or to design new searches, while
at the same time providing benchmark points coming from realistic underlying models.  In all cases, we show that
sizeable BRs in the new channels are a norm, rather than a tuned exception, as the parameters we
chose are generic. Thus, searches of these new modes are as justified as the ones in the standard channels.
The final states we highlight typically contain many top quarks or many electroweak gauge bosons, depending 
on the mass of the new scalars. They offer, therefore, a rich panorama of final states that can be easily detected
at the LHC. Another intriguing class of final states involves hard photons in association to more standard final
states: they can occur in the decays of $\eta \to Z \gamma$ below the $WW$ threshold, and decays of the
charged scalar $\phi^+ \to W^+ \gamma$ below the $t\bar{b}$ threshold.  We leave a detailed study of the
phenomenology of the new final states for further studies.

Finally, we provide a complete description of the underlying models we use for our benchmarks.
An additional interesting point that becomes apparent is that the heavier states also decay into the new light scalars, together with decay chains into the lighter VLQs, thus offering rich (but more complex)  signatures that deserve further investigation. Our study shows that the phenomenology of top partners is much richer than what can be described in the most minimal simplified models. The new final states are also a
remarkable  stamp at collider-accessible energies of the underlying model giving rise to the confining dynamics that may lurk behind the Higgs boson.


\section*{Acknowledgements}

NB and GC acknowledge partial support from the Labex-LIO (Lyon Institute of Origins) under grant ANR-10-LABX-66 and FRAMA (FR3127, F\'ed\'eration de Recherche ``Andr\'e Marie Amp\`ere'') and thank IBS for hospitality during the initial stages of this work and in part during the IBS focus meeting on ``Fundamental Composite Dynamics''. TF's work was supported by IBS under the project code, IBS-R018-D1. We also acknowledge support from the France Korea Particle Physics LIA (FKPPL).



\bibliography{VLQa}

\providecommand{\href}[2]{#2}\begingroup\raggedright\begin{thebibliography}{100}

\bibitem{Kaplan:1983fs}
D.~B. Kaplan and H.~Georgi, ``{SU(2) x U(1) Breaking by Vacuum
  Misalignment},''\href{http://dx.doi.org/10.1016/0370-2693(84)91177-8}{\emph{Phys.
  Lett.} {\bf 136B} (1984) 183--186}.

\bibitem{Dugan:1984hq}
M.~J. Dugan, H.~Georgi and D.~B. Kaplan, ``{Anatomy of a Composite Higgs
  Model},''\href{http://dx.doi.org/10.1016/0550-3213(85)90221-4}{\emph{Nucl.
  Phys.} {\bf B254} (1985) 299--326}.

\bibitem{Kaplan:1991dc}
D.~B. Kaplan, ``{Flavor at SSC energies: A New mechanism for dynamically
  generated fermion
  masses},''\href{http://dx.doi.org/10.1016/S0550-3213(05)80021-5}{\emph{Nucl.
  Phys.} {\bf B365} (1991) 259--278}.

\bibitem{Matsedonskyi:2012ym}
O.~Matsedonskyi, G.~Panico and A.~Wulzer, ``{Light Top Partners for a Light
  Composite
  Higgs},''\href{http://dx.doi.org/10.1007/JHEP01(2013)164}{\emph{JHEP} {\bf
  01} (2013) 164}, [\href{https://arxiv.org/abs/1204.6333}{{\tt 1204.6333}}].

\bibitem{Redi:2012ha}
M.~Redi and A.~Tesi, ``{Implications of a Light Higgs in Composite
  Models},''\href{http://dx.doi.org/10.1007/JHEP10(2012)166}{\emph{JHEP} {\bf
  10} (2012) 166}, [\href{https://arxiv.org/abs/1205.0232}{{\tt 1205.0232}}].

\bibitem{Contino:2011np}
R.~Contino, D.~Marzocca, D.~Pappadopulo and R.~Rattazzi, ``{On the effect of
  resonances in composite Higgs
  phenomenology},''\href{http://dx.doi.org/10.1007/JHEP10(2011)081}{\emph{JHEP}
  {\bf 10} (2011) 081}, [\href{https://arxiv.org/abs/1109.1570}{{\tt
  1109.1570}}].

\bibitem{Marzocca:2012zn}
D.~Marzocca, M.~Serone and J.~Shu, ``{General Composite Higgs
  Models},''\href{http://dx.doi.org/10.1007/JHEP08(2012)013}{\emph{JHEP} {\bf
  08} (2012) 013}, [\href{https://arxiv.org/abs/1205.0770}{{\tt 1205.0770}}].

\bibitem{Galloway:2010bp}
J.~Galloway, J.~A. Evans, M.~A. Luty and R.~A. Tacchi, ``{Minimal Conformal
  Technicolor and Precision Electroweak
  Tests},''\href{http://dx.doi.org/10.1007/JHEP10(2010)086}{\emph{JHEP} {\bf
  10} (2010) 086}, [\href{https://arxiv.org/abs/1001.1361}{{\tt 1001.1361}}].

\bibitem{Cacciapaglia:2014uja}
G.~Cacciapaglia and F.~Sannino, ``{Fundamental Composite (Goldstone) Higgs
  Dynamics},''\href{http://dx.doi.org/10.1007/JHEP04(2014)111}{\emph{JHEP} {\bf
  04} (2014) 111}, [\href{https://arxiv.org/abs/1402.0233}{{\tt 1402.0233}}].

\bibitem{Ibrahim:2008gg}
T.~Ibrahim and P.~Nath, ``{An MSSM Extension with a Mirror Fourth Generation,
  Neutrino Magnetic Moments and LHC
  Signatures},''\href{http://dx.doi.org/10.1103/PhysRevD.78.075013}{\emph{Phys.
  Rev.} {\bf D78} (2008) 075013}, [\href{https://arxiv.org/abs/0806.3880}{{\tt
  0806.3880}}].

\bibitem{Liu:2009cc}
C.~Liu, ``{Supersymmetry and Vector-like Extra
  Generation},''\href{http://dx.doi.org/10.1103/PhysRevD.80.035004}{\emph{Phys.
  Rev.} {\bf D80} (2009) 035004}, [\href{https://arxiv.org/abs/0907.3011}{{\tt
  0907.3011}}].

\bibitem{Martin:2009bg}
S.~P. Martin, ``{Extra vector-like matter and the lightest Higgs scalar boson
  mass in low-energy
  supersymmetry},''\href{http://dx.doi.org/10.1103/PhysRevD.81.035004}{\emph{Phys.
  Rev.} {\bf D81} (2010) 035004}, [\href{https://arxiv.org/abs/0910.2732}{{\tt
  0910.2732}}].

\bibitem{delAguila:2000rc}
F.~del Aguila, M.~Perez-Victoria and J.~Santiago, ``{Observable contributions
  of new exotic quarks to quark
  mixing},''\href{http://dx.doi.org/10.1088/1126-6708/2000/09/011}{\emph{JHEP}
  {\bf 09} (2000) 011}, [\href{https://arxiv.org/abs/hep-ph/0007316}{{\tt
  hep-ph/0007316}}].

\bibitem{AguilarSaavedra:2009es}
J.~A. Aguilar-Saavedra, ``{Identifying top partners at
  LHC},''\href{http://dx.doi.org/10.1088/1126-6708/2009/11/030}{\emph{JHEP}
  {\bf 11} (2009) 030}, [\href{https://arxiv.org/abs/0907.3155}{{\tt
  0907.3155}}].

\bibitem{Cacciapaglia:2010vn}
G.~Cacciapaglia, A.~Deandrea, D.~Harada and Y.~Okada, ``{Bounds and Decays of
  New Heavy Vector-like Top
  Partners},''\href{http://dx.doi.org/10.1007/JHEP11(2010)159}{\emph{JHEP} {\bf
  11} (2010) 159}, [\href{https://arxiv.org/abs/1007.2933}{{\tt 1007.2933}}].

\bibitem{Okada:2012gy}
Y.~Okada and L.~Panizzi, ``{LHC signatures of vector-like
  quarks},''\href{http://dx.doi.org/10.1155/2013/364936}{\emph{Adv. High Energy
  Phys.} {\bf 2013} (2013) 364936},
  [\href{https://arxiv.org/abs/1207.5607}{{\tt 1207.5607}}].

\bibitem{Garberson:2013jz}
F.~Garberson and T.~Golling, ``{Generalization of exotic quark
  searches},''\href{http://dx.doi.org/10.1103/PhysRevD.87.072007}{\emph{Phys.
  Rev.} {\bf D87} (2013) 072007}, [\href{https://arxiv.org/abs/1301.4454}{{\tt
  1301.4454}}].

\bibitem{Buchkremer:2013bha}
M.~Buchkremer, G.~Cacciapaglia, A.~Deandrea and L.~Panizzi, ``{Model
  Independent Framework for Searches of Top
  Partners},''\href{http://dx.doi.org/10.1016/j.nuclphysb.2013.08.010}{\emph{Nucl.
  Phys.} {\bf B876} (2013) 376--417},
  [\href{https://arxiv.org/abs/1305.4172}{{\tt 1305.4172}}].

\bibitem{DeSimone:2012fs}
A.~De~Simone, O.~Matsedonskyi, R.~Rattazzi and A.~Wulzer, ``{A First Top
  Partner Hunter's
  Guide},''\href{http://dx.doi.org/10.1007/JHEP04(2013)004}{\emph{JHEP} {\bf
  04} (2013) 004}, [\href{https://arxiv.org/abs/1211.5663}{{\tt 1211.5663}}].

\bibitem{Barnard:2013zea}
J.~Barnard, T.~Gherghetta and T.~S. Ray, ``{UV descriptions of composite Higgs
  models without elementary
  scalars},''\href{http://dx.doi.org/10.1007/JHEP02(2014)002}{\emph{JHEP} {\bf
  02} (2014) 002}, [\href{https://arxiv.org/abs/1311.6562}{{\tt 1311.6562}}].

\bibitem{Ferretti:2013kya}
G.~Ferretti and D.~Karateev, ``{Fermionic UV completions of Composite Higgs
  models},''\href{http://dx.doi.org/10.1007/JHEP03(2014)077}{\emph{JHEP} {\bf
  03} (2014) 077}, [\href{https://arxiv.org/abs/1312.5330}{{\tt 1312.5330}}].

\bibitem{Ferretti:2014qta}
G.~Ferretti, ``{UV Completions of Partial Compositeness: The Case for a SU(4)
  Gauge Group},''\href{http://dx.doi.org/10.1007/JHEP06(2014)142}{\emph{JHEP}
  {\bf 06} (2014) 142}, [\href{https://arxiv.org/abs/1404.7137}{{\tt
  1404.7137}}].

\bibitem{Vecchi:2015fma}
L.~Vecchi, ``{A dangerous irrelevant UV-completion of the composite Higgs},''
  \href{https://arxiv.org/abs/1506.00623}{{\tt 1506.00623}}.

\bibitem{Sannino:2016sfx}
F.~Sannino, A.~Strumia, A.~Tesi and E.~Vigiani, ``{Fundamental partial
  compositeness},''\href{http://dx.doi.org/10.1007/JHEP11(2016)029}{\emph{JHEP}
  {\bf 11} (2016) 029}, [\href{https://arxiv.org/abs/1607.01659}{{\tt
  1607.01659}}].

\bibitem{Cacciapaglia:2017cdi}
G.~Cacciapaglia, H.~Gertov, F.~Sannino and A.~E. Thomsen, ``{Minimal
  Fundamental Partial Compositeness},''
  \href{https://arxiv.org/abs/1704.07845}{{\tt 1704.07845}}.

\bibitem{Cai:2015bss}
H.~Cai, T.~Flacke and M.~Lespinasse, ``{A composite scalar hint from di-boson
  resonances?},'' \href{https://arxiv.org/abs/1512.04508}{{\tt 1512.04508}}.

\bibitem{Belyaev:2015hgo}
A.~Belyaev, G.~Cacciapaglia, H.~Cai, T.~Flacke, A.~Parolini and H.~Ser\^odio,
  ``{Singlets in composite Higgs models in light of the LHC 750 GeV diphoton
  excess},''\href{http://dx.doi.org/10.1103/PhysRevD.94.015004}{\emph{Phys.
  Rev.} {\bf D94} (2016) 015004}, [\href{https://arxiv.org/abs/1512.07242}{{\tt
  1512.07242}}].

\bibitem{DeGrand:2016pgq}
T.~DeGrand, M.~Golterman, E.~T. Neil and Y.~Shamir, ``{One-loop Chiral
  Perturbation Theory with two fermion
  representations},''\href{http://dx.doi.org/10.1103/PhysRevD.94.025020}{\emph{Phys.
  Rev.} {\bf D94} (2016) 025020}, [\href{https://arxiv.org/abs/1605.07738}{{\tt
  1605.07738}}].

\bibitem{Ferretti:2016upr}
G.~Ferretti, ``{Gauge theories of Partial Compositeness: Scenarios for Run-II
  of the LHC},''\href{http://dx.doi.org/10.1007/JHEP06(2016)107}{\emph{JHEP}
  {\bf 06} (2016) 107}, [\href{https://arxiv.org/abs/1604.06467}{{\tt
  1604.06467}}].

\bibitem{Belyaev:2016ftv}
A.~Belyaev, G.~Cacciapaglia, H.~Cai, G.~Ferretti, T.~Flacke, A.~Parolini
  et~al., ``{Di-boson signatures as Standard Candles for Partial
  Compositeness},''\href{http://dx.doi.org/10.1007/JHEP01(2017)094}{\emph{JHEP}
  {\bf 01} (2017) 094}, [\href{https://arxiv.org/abs/1610.06591}{{\tt
  1610.06591}}].

\bibitem{Gripaios:2009pe}
B.~Gripaios, A.~Pomarol, F.~Riva and J.~Serra, ``{Beyond the Minimal Composite
  Higgs
  Model},''\href{http://dx.doi.org/10.1088/1126-6708/2009/04/070}{\emph{JHEP}
  {\bf 04} (2009) 070}, [\href{https://arxiv.org/abs/0902.1483}{{\tt
  0902.1483}}].

\bibitem{Serra:2015xfa}
J.~Serra, ``{Beyond the Minimal Top Partner
  Decay},''\href{http://dx.doi.org/10.1007/JHEP09(2015)176}{\emph{JHEP} {\bf
  09} (2015) 176}, [\href{https://arxiv.org/abs/1506.05110}{{\tt 1506.05110}}].

\bibitem{Chala:2017sjk}
M.~Chala, G.~Durieux, C.~Grojean, L.~de~Lima and O.~Matsedonskyi, ``{Minimally
  extended SILH},''\href{http://dx.doi.org/10.1007/JHEP06(2017)088}{\emph{JHEP}
  {\bf 06} (2017) 088}, [\href{https://arxiv.org/abs/1703.10624}{{\tt
  1703.10624}}].

\bibitem{Cacciapaglia:2015eqa}
G.~Cacciapaglia, H.~Cai, A.~Deandrea, T.~Flacke, S.~J. Lee and A.~Parolini,
  ``{Composite scalars at the LHC: the Higgs, the Sextet and the
  Octet},''\href{http://dx.doi.org/10.1007/JHEP11(2015)201}{\emph{JHEP} {\bf
  11} (2015) 201}, [\href{https://arxiv.org/abs/1507.02283}{{\tt 1507.02283}}].

\bibitem{Deandrea:2017rqp}
A.~Deandrea and A.~M. Iyer, ``{Vector-like quarks and heavy coloured bosons at
  the LHC},'' \href{https://arxiv.org/abs/1710.01515}{{\tt 1710.01515}}.

\bibitem{Anandakrishnan:2015yfa}
A.~Anandakrishnan, J.~H. Collins, M.~Farina, E.~Kuflik and M.~Perelstein,
  ``{Odd Top Partners at the
  LHC},''\href{http://dx.doi.org/10.1103/PhysRevD.93.075009}{\emph{Phys. Rev.}
  {\bf D93} (2016) 075009}, [\href{https://arxiv.org/abs/1506.05130}{{\tt
  1506.05130}}].

\bibitem{Balkin:2017aep}
R.~Balkin, M.~Ruhdorfer, E.~Salvioni and A.~Weiler, ``{Charged Composite Scalar
  Dark Matter},'' \href{https://arxiv.org/abs/1707.07685}{{\tt 1707.07685}}.

\bibitem{Chala:2018qdf}
M.~Chala, R.~Gr{\"o}ber and M.~Spannowsky, ``{Searches for vector-like quarks
  at future colliders and implications for composite Higgs models with dark
  matter},'' \href{https://arxiv.org/abs/1801.06537}{{\tt 1801.06537}}.

\bibitem{Kraml:2016eti}
S.~Kraml, U.~Laa, L.~Panizzi and H.~Prager, ``{Scalar versus fermionic top
  partner interpretations of $t\bar t + E_T^{\rm miss}$ searches at the
  LHC},''\href{http://dx.doi.org/10.1007/JHEP11(2016)107}{\emph{JHEP} {\bf 11}
  (2016) 107}, [\href{https://arxiv.org/abs/1607.02050}{{\tt 1607.02050}}].

\bibitem{Chala:2017xgc}
M.~Chala, ``{Direct bounds on heavy toplike quarks with standard and exotic
  decays},''\href{http://dx.doi.org/10.1103/PhysRevD.96.015028}{\emph{Phys.
  Rev.} {\bf D96} (2017) 015028}, [\href{https://arxiv.org/abs/1705.03013}{{\tt
  1705.03013}}].

\bibitem{Aguilar-Saavedra:2017giu}
J.~A. Aguilar-Saavedra, D.~E. L\'opez-Fogliani and C.~Mu\~noz, ``{Novel
  signatures for vector-like
  quarks},''\href{http://dx.doi.org/10.1007/JHEP06(2017)095}{\emph{JHEP} {\bf
  06} (2017) 095}, [\href{https://arxiv.org/abs/1705.02526}{{\tt 1705.02526}}].

\bibitem{Brooijmans:2016vro}
G.~Brooijmans et~al., ``{Les Houches 2015: Physics at TeV colliders - new
  physics working group report},'' in \emph{{9th Les Houches Workshop on
  Physics at TeV Colliders (PhysTeV 2015) Les Houches, France, June 1-19,
  2015}}, 2016.
\newblock \href{https://arxiv.org/abs/1605.02684}{{\tt 1605.02684}}.

\bibitem{Sirunyan:2017pks}
{\scshape CMS} collaboration, A.~M. Sirunyan et~al., ``{Search for pair
  production of vector-like quarks in the bW$\overline{\mathrm{b}}$W channel
  from proton-proton collisions at $\sqrt{s} =$ 13 TeV},''
  \href{https://arxiv.org/abs/1710.01539}{{\tt 1710.01539}}.

\bibitem{CMS:2016dmr}
{\scshape CMS} collaboration, ``{Search for vector-like quark pair production
  in final states with leptons and boosted Higgs bosons at
  $\sqrt{s}=13~\mathrm{TeV}$},'' CMS-PAS-B2G-16-011.

\bibitem{Aaboud:2017zfn}
{\scshape ATLAS} collaboration, M.~Aaboud et~al., ``{Search for pair production
  of heavy vector-like quarks decaying to high-p$_{T}$ W bosons and b quarks in
  the lepton-plus-jets final state in pp collisions at $ \sqrt{s}=13 $ TeV with
  the ATLAS
  detector},''\href{http://dx.doi.org/10.1007/JHEP10(2017)141}{\emph{JHEP} {\bf
  10} (2017) 141}, [\href{https://arxiv.org/abs/1707.03347}{{\tt 1707.03347}}].

\bibitem{Aaboud:2017qpr}
{\scshape ATLAS} collaboration, M.~Aaboud et~al., ``{Search for pair production
  of vector-like top quarks in events with one lepton, jets, and missing
  transverse momentum in $ \sqrt{s}=13 $ TeV $pp$ collisions with the ATLAS
  detector},''\href{http://dx.doi.org/10.1007/JHEP08(2017)052}{\emph{JHEP} {\bf
  08} (2017) 052}, [\href{https://arxiv.org/abs/1705.10751}{{\tt 1705.10751}}].

\bibitem{TheATLAScollaboration:2016gxs}
{\scshape ATLAS} collaboration, ``{Search for production of vector-like top
  quark pairs and of four top quarks in the lepton-plus-jets final state in
  $pp$ collisions at $\sqrt{s}=13$ TeV with the ATLAS detector},''
  ATLAS-CONF-2016-013.

\bibitem{Aaboud:2016lwz}
{\scshape ATLAS} collaboration, M.~Aaboud et~al., ``{Search for top squarks in
  final states with one isolated lepton, jets, and missing transverse momentum
  in $\sqrt{s}=13$ TeV $pp$ collisions with the ATLAS
  detector},''\href{http://dx.doi.org/10.1103/PhysRevD.94.052009}{\emph{Phys.
  Rev.} {\bf D94} (2016) 052009}, [\href{https://arxiv.org/abs/1606.03903}{{\tt
  1606.03903}}].

\bibitem{ATLAS:2016btu}
{\scshape ATLAS} collaboration, ``{Search for new phenomena in $t\bar{t}$ final
  states with additional heavy-flavour jets in $pp$ collisions at $\sqrt{s}=13$
  TeV with the ATLAS detector},'' ATLAS-CONF-2016-104.

\bibitem{Sirunyan:2017ynj}
{\scshape CMS} collaboration, A.~M. Sirunyan et~al., ``{Search for single
  production of a vector-like T quark decaying to a Z boson and a top quark in
  proton-proton collisions at sqrt(s) = 13 TeV},''
  \href{https://arxiv.org/abs/1708.01062}{{\tt 1708.01062}}.

\bibitem{Sirunyan:2017tfc}
{\scshape CMS} collaboration, A.~M. Sirunyan et~al., ``{Search for single
  production of vector-like quarks decaying into a b quark and a W boson in
  proton-proton collisions at $\sqrt s =$ 13
  TeV},''\href{http://dx.doi.org/10.1016/j.physletb.2017.07.022}{\emph{Phys.
  Lett.} {\bf B772} (2017) 634--656},
  [\href{https://arxiv.org/abs/1701.08328}{{\tt 1701.08328}}].

\bibitem{Sirunyan:2017ezy}
{\scshape CMS} collaboration, A.~M. Sirunyan et~al., ``{Search for single
  production of vector-like quarks decaying to a Z boson and a top or a bottom
  quark in proton-proton collisions at $ \sqrt{s}=13 $
  TeV},''\href{http://dx.doi.org/10.1007/JHEP05(2017)029}{\emph{JHEP} {\bf 05}
  (2017) 029}, [\href{https://arxiv.org/abs/1701.07409}{{\tt 1701.07409}}].

\bibitem{Sirunyan:2016ipo}
{\scshape CMS} collaboration, A.~M. Sirunyan et~al., ``{Search for electroweak
  production of a vector-like quark decaying to a top quark and a Higgs boson
  using boosted topologies in fully hadronic final
  states},''\href{http://dx.doi.org/10.1007/JHEP04(2017)136}{\emph{JHEP} {\bf
  04} (2017) 136}, [\href{https://arxiv.org/abs/1612.05336}{{\tt 1612.05336}}].

\bibitem{Khachatryan:2016vph}
{\scshape CMS} collaboration, V.~Khachatryan et~al., ``{Search for single
  production of a heavy vector-like T quark decaying to a Higgs boson and a top
  quark with a lepton and jets in the final
  state},''\href{http://dx.doi.org/10.1016/j.physletb.2017.05.019}{\emph{Phys.
  Lett.} {\bf B771} (2017) 80--105},
  [\href{https://arxiv.org/abs/1612.00999}{{\tt 1612.00999}}].

\bibitem{ATLAS:2016ovj}
{\scshape ATLAS} collaboration, ``{Search for single production of vector-like
  quarks decaying into $Wb$ in $pp$ collisions at $\sqrt{s} =$ 13 TeV with the
  ATLAS detector},'' ATLAS-CONF-2016-072.

\bibitem{Aad:2016qpo}
{\scshape ATLAS} collaboration, G.~Aad et~al., ``{Search for single production
  of vector-like quarks decaying into Wb in pp collisions at $\sqrt{s} = 8$ TeV
  with the ATLAS
  detector},''\href{http://dx.doi.org/10.1140/epjc/s10052-016-4281-8}{\emph{Eur.
  Phys. J.} {\bf C76} (2016) 442},
  [\href{https://arxiv.org/abs/1602.05606}{{\tt 1602.05606}}].

\bibitem{Atre:2011ae}
A.~Atre, G.~Azuelos, M.~Carena, T.~Han, E.~Ozcan, J.~Santiago et~al.,
  ``{Model-Independent Searches for New Quarks at the
  LHC},''\href{http://dx.doi.org/10.1007/JHEP08(2011)080}{\emph{JHEP} {\bf 08}
  (2011) 080}, [\href{https://arxiv.org/abs/1102.1987}{{\tt 1102.1987}}].

\bibitem{Cacciapaglia:2017iws}
G.~Cacciapaglia, G.~Ferretti, T.~Flacke and H.~Serodio, ``{Revealing timid
  pseudo-scalars with taus at the LHC},''
  \href{https://arxiv.org/abs/1710.11142}{{\tt 1710.11142}}.

\bibitem{Casolino:2015cza}
M.~Casolino, T.~Farooque, A.~Juste, T.~Liu and M.~Spannowsky, ``{Probing a
  light CP-odd scalar in di-top-associated production at the
  LHC},''\href{http://dx.doi.org/10.1140/epjc/s10052-015-3708-y}{\emph{Eur.
  Phys. J.} {\bf C75} (2015) 498},
  [\href{https://arxiv.org/abs/1507.07004}{{\tt 1507.07004}}].

\bibitem{Bauer:2017ris}
M.~Bauer, M.~Neubert and A.~Thamm, ``{Collider Probes of Axion-Like
  Particles},''\href{http://dx.doi.org/10.1007/JHEP12(2017)044}{\emph{JHEP}
  {\bf 12} (2017) 044}, [\href{https://arxiv.org/abs/1708.00443}{{\tt
  1708.00443}}].

\bibitem{Sirunyan:2016iap}
{\scshape CMS} collaboration, A.~M. Sirunyan et~al., ``{Search for dijet
  resonances in proton-proton collisions at $\sqrt{s}$ = 13 TeV and constraints
  on dark matter and other
  models},''\href{http://dx.doi.org/10.1016/j.physletb.2017.09.029,
  10.1016/j.physletb.2017.02.012}{\emph{Phys. Lett.} {\bf B769} (2017)
  520--542}, [\href{https://arxiv.org/abs/1611.03568}{{\tt 1611.03568}}].

\bibitem{CMS:2017xrr}
{\scshape CMS} collaboration, ``{Searches for dijet resonances in pp collisions
  at $\sqrt{s}=13~\mathrm{TeV}$ using data collected in 2016.},''
  CMS-PAS-EXO-16-056.

\bibitem{Aaboud:2017yvp}
{\scshape ATLAS} collaboration, M.~Aaboud et~al., ``{Search for new phenomena
  in dijet events using 37 fb$^{-1}$ of $pp$ collision data collected at
  $\sqrt{s}=$13 TeV with the ATLAS
  detector},''\href{http://dx.doi.org/10.1103/PhysRevD.96.052004}{\emph{Phys.
  Rev.} {\bf D96} (2017) 052004}, [\href{https://arxiv.org/abs/1703.09127}{{\tt
  1703.09127}}].

\bibitem{Sirunyan:2017uhk}
{\scshape CMS} collaboration, A.~M. Sirunyan et~al., ``{Search for $
  \mathrm{t}\overline{\mathrm{t}} $ resonances in highly boosted lepton+jets
  and fully hadronic final states in proton-proton collisions at $ \sqrt{s}=13
  $ TeV},''\href{http://dx.doi.org/10.1007/JHEP07(2017)001}{\emph{JHEP} {\bf
  07} (2017) 001}, [\href{https://arxiv.org/abs/1704.03366}{{\tt 1704.03366}}].

\bibitem{Khachatryan:2015sma}
{\scshape CMS} collaboration, V.~Khachatryan et~al., ``{Search for resonant $t
  \bar t$ production in proton-proton collisions at $\sqrt s=$ 8
  TeV},''\href{http://dx.doi.org/10.1103/PhysRevD.93.012001}{\emph{Phys. Rev.}
  {\bf D93} (2016) 012001}, [\href{https://arxiv.org/abs/1506.03062}{{\tt
  1506.03062}}].

\bibitem{Aad:2015fna}
{\scshape ATLAS} collaboration, G.~Aad et~al., ``{A search for $ t\overline{t}
  $ resonances using lepton-plus-jets events in proton-proton collisions at $
  \sqrt{s}=8 $ TeV with the ATLAS
  detector},''\href{http://dx.doi.org/10.1007/JHEP08(2015)148}{\emph{JHEP} {\bf
  08} (2015) 148}, [\href{https://arxiv.org/abs/1505.07018}{{\tt 1505.07018}}].

\bibitem{ATLAS:2016fol}
{\scshape ATLAS} collaboration, ``{Search for resonances below 1.2 TeV from the
  mass distribution of $b$-jet pairs in proton-proton collisions at
  $\sqrt{s}$=13 TeV with the ATLAS detector},'' ATLAS-CONF-2016-031.

\bibitem{Khachatryan:2016qkc}
{\scshape CMS} collaboration, V.~Khachatryan et~al., ``{Search for heavy
  resonances decaying to tau lepton pairs in proton-proton collisions at $
  \sqrt{s}=13 $
  TeV},''\href{http://dx.doi.org/10.1007/JHEP02(2017)048}{\emph{JHEP} {\bf 02}
  (2017) 048}, [\href{https://arxiv.org/abs/1611.06594}{{\tt 1611.06594}}].

\bibitem{CMS:2017epy}
{\scshape CMS} collaboration, ``{Search for additional neutral MSSM Higgs
  bosons in the di-tau final state in $pp$ collisions at $\sqrt{s}=13$ TeV},''
  CMS-PAS-HIG-17-020.

\bibitem{Aaboud:2016cre}
{\scshape ATLAS} collaboration, M.~Aaboud et~al., ``{Search for Minimal
  Supersymmetric Standard Model Higgs bosons $H/A$ and for a $Z^{\prime}$ boson
  in the $\tau \tau$ final state produced in $pp$ collisions at $\sqrt{s}=13$
  TeV with the ATLAS
  Detector},''\href{http://dx.doi.org/10.1140/epjc/s10052-016-4400-6}{\emph{Eur.
  Phys. J.} {\bf C76} (2016) 585},
  [\href{https://arxiv.org/abs/1608.00890}{{\tt 1608.00890}}].

\bibitem{Aaboud:2017sjh}
{\scshape ATLAS} collaboration, M.~Aaboud et~al., ``{Search for additional
  heavy neutral Higgs and gauge bosons in the ditau final state produced in 36
  fb$^{-1}$ of pp collisions at $ \sqrt{s}=13 $ TeV with the ATLAS
  detector},''\href{http://dx.doi.org/10.1007/JHEP01(2018)055}{\emph{JHEP} {\bf
  01} (2018) 055}, [\href{https://arxiv.org/abs/1709.07242}{{\tt 1709.07242}}].

\bibitem{Chatrchyan:2012am}
{\scshape CMS} collaboration, S.~Chatrchyan et~al., ``{Search for a light
  pseudoscalar Higgs boson in the dimuon decay channel in $pp$ collisions at
  $\sqrt{s}=7$
  TeV},''\href{http://dx.doi.org/10.1103/PhysRevLett.109.121801}{\emph{Phys.
  Rev. Lett.} {\bf 109} (2012) 121801},
  [\href{https://arxiv.org/abs/1206.6326}{{\tt 1206.6326}}].

\bibitem{Sirunyan:2017acf}
{\scshape CMS} collaboration, A.~M. Sirunyan et~al., ``{Search for massive
  resonances decaying into WW, WZ, ZZ, qW, and qZ with dijet final states at
  $\sqrt(s) = 13$~TeV},'' \href{https://arxiv.org/abs/1708.05379}{{\tt
  1708.05379}}.

\bibitem{Aaboud:2017eta}
{\scshape ATLAS} collaboration, M.~Aaboud et~al., ``{Search for diboson
  resonances with boson-tagged jets in $pp$ collisions at $\sqrt{s}=13$ TeV
  with the ATLAS
  detector},''\href{http://dx.doi.org/10.1016/j.physletb.2017.12.011}{\emph{Phys.
  Lett.} {\bf B777} (2018) 91--113},
  [\href{https://arxiv.org/abs/1708.04445}{{\tt 1708.04445}}].

\bibitem{Sirunyan:2016cao}
{\scshape CMS} collaboration, A.~M. Sirunyan et~al., ``{Search for massive
  resonances decaying into WW, WZ or ZZ bosons in proton-proton collisions at
  $\sqrt{s} = $ 13
  TeV},''\href{http://dx.doi.org/10.1007/JHEP03(2017)162}{\emph{JHEP} {\bf 03}
  (2017) 162}, [\href{https://arxiv.org/abs/1612.09159}{{\tt 1612.09159}}].

\bibitem{CMS:2017mjm}
{\scshape CMS} collaboration, ``{Search for heavy resonances decaying to pairs
  of vector bosons in the $\ell \nu q \bar{q}$ final state with the CMS
  detector in proton-proton collisions at $\sqrt{s} = 13$~TeV},''
  CMS-PAS-B2G-16-029.

\bibitem{CMS:2016pfl}
{\scshape CMS} collaboration, ``{Search for new resonances decaying to
  $\mathrm{WW}/\mathrm{WZ} \to \ell\nu \mathrm{qq}$},'' CMS-PAS-B2G-16-020.

\bibitem{Aaboud:2017gsl}
{\scshape ATLAS} collaboration, M.~Aaboud et~al., ``{Search for heavy
  resonances decaying into $WW$ in the $e\nu\mu\nu$ final state in $pp$
  collisions at $\sqrt{s}=13$ TeV with the ATLAS
  detector},''\href{http://dx.doi.org/10.1140/epjc/s10052-017-5491-4}{\emph{Eur.
  Phys. J.} {\bf C78} (2018) 24}, [\href{https://arxiv.org/abs/1710.01123}{{\tt
  1710.01123}}].

\bibitem{Aaboud:2017fgj}
{\scshape ATLAS} collaboration, M.~Aaboud et~al., ``{Search for $WW/WZ$
  resonance production in $\ell \nu qq$ final states in $pp$ collisions at
  $\sqrt{s} =$ 13 TeV with the ATLAS detector},''
  \href{https://arxiv.org/abs/1710.07235}{{\tt 1710.07235}}.

\bibitem{CMS:2017xyz}
{\scshape CMS} collaboration, ``{Search for a new scalar resonance decaying to
  a pair of Z bosons in proton-proton collisions at $\sqrt{s}=13$~TeV},''
  CMS-PAS-HIG-17-012.

\bibitem{CMS:2017sbi}
{\scshape CMS} collaboration, ``{Search for new diboson resonances in the
  dilepton + jets final state at $\sqrt{s} = 13~\mathrm{TeV}$ with 2016
  data},'' CMS-PAS-HIG-16-034.

\bibitem{Sirunyan:2017jtu}
{\scshape CMS} collaboration, A.~M. Sirunyan et~al., ``{Search for diboson
  resonances in the 2$\ell$2$\nu$ final state},''
  \href{https://arxiv.org/abs/1711.04370}{{\tt 1711.04370}}.

\bibitem{CMS:2017pfj}
{\scshape CMS} collaboration, ``{Search for heavy resonances decaying into a Z
  boson and a vector boson in the $\nu \nu$ $q\bar{q}$ final state},''
  CMS-PAS-B2G-17-005.

\bibitem{Aaboud:2017rel}
{\scshape ATLAS} collaboration, M.~Aaboud et~al., ``{Search for heavy $ZZ$
  resonances in the $\ell^+\ell^-\ell^+\ell^-$ and $\ell^+\ell^-\nu\bar\nu$
  final states using proton proton collisions at $\sqrt{s}= 13$ TeV with the
  ATLAS detector},'' \href{https://arxiv.org/abs/1712.06386}{{\tt 1712.06386}}.

\bibitem{Aaboud:2017itg}
{\scshape ATLAS} collaboration, M.~Aaboud et~al., ``{Searches for heavy $ZZ$
  and $ZW$ resonances in the $\ell\ell qq$ and $\nu\nu qq$ final states in $pp$
  collisions at $\sqrt{s}=13$ TeV with the ATLAS detector},''
  \href{https://arxiv.org/abs/1708.09638}{{\tt 1708.09638}}.

\bibitem{Sirunyan:2017hsb}
{\scshape CMS} collaboration, A.~M. Sirunyan et~al., ``{Search for Z$\gamma$
  resonances using leptonic and hadronic final states in proton-proton
  collisions at $\sqrt{s}=$ 13 TeV},''
  \href{https://arxiv.org/abs/1712.03143}{{\tt 1712.03143}}.

\bibitem{Khachatryan:2016odk}
{\scshape CMS} collaboration, V.~Khachatryan et~al., ``{Search for high-mass
  Z$\gamma$ resonances in $\mathrm{ e }^{+}\mathrm{ e }^{-}\gamma $ and $
  \mu^{+}\mu^{-}\gamma$ final states in proton-proton collisions at $\sqrt{s}
  =$ 8 and 13
  TeV},''\href{http://dx.doi.org/10.1007/JHEP01(2017)076}{\emph{JHEP} {\bf 01}
  (2017) 076}, [\href{https://arxiv.org/abs/1610.02960}{{\tt 1610.02960}}].

\bibitem{Aaboud:2017uhw}
{\scshape ATLAS} collaboration, M.~Aaboud et~al., ``{Searches for the $Z\gamma$
  decay mode of the Higgs boson and for new high-mass resonances in $pp$
  collisions at $\sqrt{s} = 13$ TeV with the ATLAS
  detector},''\href{http://dx.doi.org/10.1007/JHEP10(2017)112}{\emph{JHEP} {\bf
  10} (2017) 112}, [\href{https://arxiv.org/abs/1708.00212}{{\tt 1708.00212}}].

\bibitem{Khachatryan:2016yec}
{\scshape CMS} collaboration, V.~Khachatryan et~al., ``{Search for high-mass
  diphoton resonances in proton-proton collisions at 13 TeV and combination
  with 8 TeV
  search},''\href{http://dx.doi.org/10.1016/j.physletb.2017.01.027}{\emph{Phys.
  Lett.} {\bf B767} (2017) 147--170},
  [\href{https://arxiv.org/abs/1609.02507}{{\tt 1609.02507}}].

\bibitem{Khachatryan:2016hje}
{\scshape CMS} collaboration, V.~Khachatryan et~al., ``{Search for Resonant
  Production of High-Mass Photon Pairs in Proton-Proton Collisions at $\sqrt s$
  =8 and 13
  TeV},''\href{http://dx.doi.org/10.1103/PhysRevLett.117.051802}{\emph{Phys.
  Rev. Lett.} {\bf 117} (2016) 051802},
  [\href{https://arxiv.org/abs/1606.04093}{{\tt 1606.04093}}].

\bibitem{CMS:2017yta}
{\scshape CMS} collaboration, ``{Search for new resonances in the diphoton
  final state in the mass range between 70 and 110 GeV in pp collisions at
  $\sqrt{s}=$ 8 and 13 TeV},'' CMS-PAS-HIG-17-013.

\bibitem{Aaboud:2017yyg}
{\scshape ATLAS} collaboration, M.~Aaboud et~al., ``{Search for new phenomena
  in high-mass diphoton final states using 37 fb$^{-1}$ of proton--proton
  collisions collected at $\sqrt{s}=13$ TeV with the ATLAS
  detector},''\href{http://dx.doi.org/10.1016/j.physletb.2017.10.039}{\emph{Phys.
  Lett.} {\bf B775} (2017) 105--125},
  [\href{https://arxiv.org/abs/1707.04147}{{\tt 1707.04147}}].

\bibitem{Deandrea:2014raa}
A.~Deandrea and N.~Deutschmann, ``{Multi-tops at the
  LHC},''\href{http://dx.doi.org/10.1007/JHEP08(2014)134}{\emph{JHEP} {\bf 08}
  (2014) 134}, [\href{https://arxiv.org/abs/1405.6119}{{\tt 1405.6119}}].

\bibitem{Katz:2010iq}
A.~Katz, M.~Son and B.~Tweedie, ``{Ditau-Jet Tagging and Boosted Higgses from a
  Multi-TeV
  Resonance},''\href{http://dx.doi.org/10.1103/PhysRevD.83.114033}{\emph{Phys.
  Rev.} {\bf D83} (2011) 114033}, [\href{https://arxiv.org/abs/1011.4523}{{\tt
  1011.4523}}].

\bibitem{Conte:2016zjp}
E.~Conte, B.~Fuks, J.~Guo, J.~Li and A.~G. Williams, ``{Investigating light
  NMSSM pseudoscalar states with boosted ditau
  tagging},''\href{http://dx.doi.org/10.1007/JHEP05(2016)100}{\emph{JHEP} {\bf
  05} (2016) 100}, [\href{https://arxiv.org/abs/1604.05394}{{\tt 1604.05394}}].

\bibitem{Alanne:2018wtp}
T.~Alanne, N.~Bizot, G.~Cacciapaglia and F.~Sannino, ``{Classification of NLO
  operators for composite-Higgs models},''
  \href{https://arxiv.org/abs/1801.05444}{{\tt 1801.05444}}.

\bibitem{Arbey:2015exa}
A.~Arbey, G.~Cacciapaglia, H.~Cai, A.~Deandrea, S.~Le~Corre and F.~Sannino,
  ``{Fundamental Composite Electroweak Dynamics: Status at the
  LHC},''\href{http://dx.doi.org/10.1103/PhysRevD.95.015028}{\emph{Phys. Rev.}
  {\bf D95} (2017) 015028}, [\href{https://arxiv.org/abs/1502.04718}{{\tt
  1502.04718}}].

\bibitem{Contino:2008hi}
R.~Contino and G.~Servant, ``{Discovering the top partners at the LHC using
  same-sign dilepton final
  states},''\href{http://dx.doi.org/10.1088/1126-6708/2008/06/026}{\emph{JHEP}
  {\bf 06} (2008) 026}, [\href{https://arxiv.org/abs/0801.1679}{{\tt
  0801.1679}}].

\bibitem{CMS:2017wwc}
{\scshape CMS} collaboration, ``{Search for top quark partners with charge 5/3
  in the single-lepton final state at $\sqrt{s}=13~\mathrm{TeV}$},''
  CMS-PAS-B2G-17-008.

\bibitem{CMS:2017jfv}
{\scshape CMS} collaboration, ``{Search for heavy vector-like quarks decaying
  to same-sign dileptons},'' CMS-PAS-B2G-16-019.

\bibitem{Sirunyan:2017jin}
{\scshape CMS} collaboration, A.~M. Sirunyan et~al., ``{Search for top quark
  partners with charge 5/3 in proton-proton collisions at $ \sqrt{s}=13 $
  TeV},''\href{http://dx.doi.org/10.1007/JHEP08(2017)073}{\emph{JHEP} {\bf 08}
  (2017) 073}, [\href{https://arxiv.org/abs/1705.10967}{{\tt 1705.10967}}].

\bibitem{Backovic:2014uma}
M.~Backovic, T.~Flacke, S.~J. Lee and G.~Perez, ``{LHC Top Partner Searches
  Beyond the 2 TeV Mass
  Region},''\href{http://dx.doi.org/10.1007/JHEP09(2015)022}{\emph{JHEP} {\bf
  09} (2015) 022}, [\href{https://arxiv.org/abs/1409.0409}{{\tt 1409.0409}}].

\bibitem{Kanemura:2014ipa}
S.~Kanemura, M.~Kikuchi, H.~Yokoya and K.~Yagyu, ``{LHC Run-I constraint on the
  mass of doubly charged Higgs bosons in the same-sign diboson decay
  scenario},''\href{http://dx.doi.org/10.1093/ptep/ptv071}{\emph{PTEP} {\bf
  2015} (2015) 051B02}, [\href{https://arxiv.org/abs/1412.7603}{{\tt
  1412.7603}}].

\bibitem{Degrande:2017naf}
C.~Degrande, K.~Hartling and H.~E. Logan, ``{Scalar decays to $\gamma\gamma$,
  $Z\gamma$, and $W\gamma$ in the Georgi-Machacek
  model},''\href{http://dx.doi.org/10.1103/PhysRevD.96.075013}{\emph{Phys.
  Rev.} {\bf D96} (2017) 075013}, [\href{https://arxiv.org/abs/1708.08753}{{\tt
  1708.08753}}].

\bibitem{DeGrand:2016mxr}
T.~A. DeGrand, D.~Hackett, W.~I. Jay, E.~T. Neil, Y.~Shamir and B.~Svetitsky,
  ``{Towards Partial Compositeness on the Lattice: Baryons with Fermions in
  Multiple Representations},''{\emph{PoS} {\bf LATTICE2016} (2016) 219},
  [\href{https://arxiv.org/abs/1610.06465}{{\tt 1610.06465}}].

\bibitem{Cacciapaglia:2015dsa}
G.~Cacciapaglia, H.~Cai, T.~Flacke, S.~J. Lee, A.~Parolini and H.~Ser{\^o}dio,
  ``{Anarchic Yukawas and top partial compositeness: the flavour of a
  successful
  marriage},''\href{http://dx.doi.org/10.1007/JHEP06(2015)085}{\emph{JHEP} {\bf
  06} (2015) 085}, [\href{https://arxiv.org/abs/1501.03818}{{\tt 1501.03818}}].

\bibitem{Golterman:2017vdj}
M.~Golterman and Y.~Shamir, ``{Effective potential in ultraviolet completions
  for composite Higgs models},'' \href{https://arxiv.org/abs/1707.06033}{{\tt
  1707.06033}}.

\bibitem{Ayyar:2017qdf}
V.~Ayyar, T.~DeGrand, M.~Golterman, D.~C. Hackett, W.~I. Jay, E.~T. Neil
  et~al., ``{Spectroscopy of SU(4) composite Higgs theory with two distinct
  fermion representations},'' \href{https://arxiv.org/abs/1710.00806}{{\tt
  1710.00806}}.

\bibitem{Bennett:2017kga}
E.~Bennett, D.~K. Hong, J.-W. Lee, C.~J.~D. Lin, B.~Lucini, M.~Piai et~al.,
  ``{Sp(4) gauge theory on the lattice: towards SU(4)/Sp(4) composite Higgs
  (and beyond)},'' \href{https://arxiv.org/abs/1712.04220}{{\tt 1712.04220}}.

\bibitem{Raby:1979my}
S.~Raby, S.~Dimopoulos and L.~Susskind, ``{Tumbling Gauge
  Theories},''\href{http://dx.doi.org/10.1016/0550-3213(80)90093-0}{\emph{Nucl.
  Phys.} {\bf B169} (1980) 373--383}.

\end{thebibliography}\endgroup
\bibliographystyle{JHEP-2-2.bst}

\end{document}